\documentclass[pdflatex,sn-mathphys-num]{sn-jnl-arxiv}

\usepackage{graphicx}%
\usepackage{multirow}%
\usepackage{amsmath,amssymb,amsfonts}%
\usepackage{amsthm}%
\usepackage{mathrsfs}%
\usepackage[title]{appendix}%
\usepackage{xcolor}%
\usepackage{textcomp}%
\usepackage{manyfoot}%
\usepackage{booktabs}%
\usepackage{algorithm}%
\usepackage{algorithmicx}%
\usepackage{algpseudocode}%
\usepackage{listings}%
\usepackage[utf8]{inputenc} 
\usepackage[T1]{fontenc}    
\usepackage{hyperref}       
\usepackage{url}            
\usepackage{nicefrac}       
\usepackage[expansion=false]{microtype}      
\usepackage{natbib}
\usepackage{subcaption}
\usepackage{threeparttable}
\usepackage{adjustbox}
\usepackage{siunitx}
\usepackage{rotating}

\theoremstyle{thmstyleone}%
\theoremstyle{thmstyletwo}%

\theoremstyle{thmstylethree}%

\begin{document}

\title{A Priori Sampling of Transition States with Guided Diffusion}


\author[1]{\fnm{Hyukjun} \sur{Lim}}\email{hyukjunlim@snu.ac.kr}
\equalcont{These authors contributed equally to this work.}

\author*[2]{\fnm{Soojung} \sur{Yang}}\email{soojungy@mit.edu}
\equalcont{These authors contributed equally to this work.}

\author[3]{\fnm{Lucas} \sur{Pinède}}\email{lucas.pinede@etu.chimieparistech.psl.eu}
\equalcont{These authors contributed equally to this work.}

\author[4]{\fnm{Miguel} \sur{Steiner}}\email{steinmig@mit.edu}

\author[5]{\fnm{Yuanqi} \sur{Du}}\email{yd392@cornell.edu}

\author*[4]{\fnm{Rafael} \sur{G\'omez-Bombarelli}}\email{rafagb@mit.edu}

\affil[1]{\orgdiv{Department of Materials Science and Engineering}, \orgname{Seoul National University}, \orgaddress{\street{1 Gwanak-ro, Gwanak-gu}, \city{Seoul}, \postcode{08826}, \country{Republic of Korea}}}

\affil[2]{\orgdiv{Computational and Systems Biology}, \orgname{MIT}, \orgaddress{\street{77 Massachusetts Avenue}, \city{Cambridge}, \postcode{02139}, \state{Massachusetts}, \country{USA}}}

\affil[3]{\orgname{Chimie ParisTech, PSL University}, \orgaddress{\street{11 rue Pierre et Marie Curie}, \city{Paris}, \postcode{75005}, \country{France}}}

\affil[4]{\orgdiv{Department of Materials Science and Engineering}, \orgname{MIT}, \orgaddress{\street{77 Massachusetts Avenue}, \city{Cambridge}, \postcode{02139}, \state{Massachusetts}, \country{USA}}}

\affil[5]{\orgdiv{Department of Computer Science}, \orgname{Cornell University}, \orgaddress{\street{127 Hoy Road}, \city{Ithaca}, \postcode{14850}, \state{NY}, \country{USA}}}

\abstract{
Transition states, the first-order saddle points on the potential energy surfaces, govern the kinetics and mechanisms of chemical reactions and conformational changes. 
Locating them is challenging because transition pathways are topologically complex and can proceed via an ensemble of diverse routes. 
Existing methods address these challenges by introducing heuristic assumptions about the pathway or reaction coordinates, which limits their applicability when a good initial guess is unavailable or when the guess precludes alternative, potentially relevant pathways.
We propose to bypass such heuristic limitations by introducing ASTRA, \textbf{A} Priori \textbf{S}ampling of \textbf{TRA}nsition States with Guided Diffusion, which reframes the transition state search as an inference-time scaling problem for generative models.
ASTRA trains a score-based diffusion model on configurations from known metastable states.
Then, ASTRA guides inference toward the isodensity surface separating the basins of metastable states via a principled composition of conditional scores. 
A Score-Aligned Ascent (SAA) process then approximates a reaction coordinate from the difference between conditioned scores and combines it with physical forces to drive convergence onto first-order transition states.
Validated on benchmarks ranging from 2D potentials to biomolecular conformational changes and a chemical reaction, ASTRA locates transition states with high precision and discovers multiple reaction pathways, enabling mechanistic studies of complex molecular systems.
}




\maketitle

\section{Introduction}\label{sec1}


Understanding the dynamics of molecular systems, from chemical reactions to protein folding, requires characterizing their transition states (TSs)~\cite{eyring1935activated, wigner1938transition}. 
Defined as first-order saddle points on the potential energy surface (PES), TSs represent the highest energy configurations along a minimum energy path and act as kinetic bottlenecks that determine reaction rates and mechanisms. 
Despite their central importance, locating TSs remains notoriously difficult: they are too short-lived for atomistic experimental characterization, and too rarely visited for unbiased molecular dynamics simulation~\cite{lindorff2011fast}. 
Various enhanced sampling techniques have therefore been developed to circumvent these timescale limitations~\cite{tiwary2016review, yang2019enhanced, henin2022enhanced, shen2023enhanced}, but their efficacy relies on defining collective variables (CVs) that accurately describe the relevant slow degrees of freedom. Identifying appropriate CVs is challenging when the transition mechanism is unknown, requiring iterative refinement from an initial guess that can become computationally expensive.
Time-independent TS optimization (TSO) methods offer an alternative by finding individual TS structures and describing a series of transformations within the framework of TS theory~\cite{eyring1935activated, wigner1938transition}, although their per-structure operation has historically restricted application to systems with low conformational flexibility.
Over the past decade, these methods have been increasingly automated and scaled up to model complex reaction networks across diverse chemical systems~\cite{dewyer2017finding, simm2018exploration, unsleber2020exploration, baiardi2022expansive, ismail2022graph, steiner2022autonomous}.
However, TSO methods remain computationally intensive, and require chemically informed initial guesses that are difficult to generalize across chemical systems~\cite{steiner2024human}. 
The reliance of both enhanced sampling MD and TSO methods on heuristics makes the TS search tedious and difficult to scale to complex systems.  

In recent years, data-driven strategies have aimed to accelerate MD and TSO approaches by automatically learning CVs~\cite{mehdi2024enhanced, mendels2018collective, wang2019past, kang2025committors, trizio2025everything, das2025machine, zhang2024descriptor, tan2025enhanced}, but the transferability of learned manifolds remains limited for systems outside the training distribution -- a limitation shared across data-driven approaches.
A prominent example is the committor function — the probability of reaching the product state before the reactant from a given configuration — which can be learned from transition path data via maximum-likelihood methods~\cite{jung2019artificial, sun2022multitask, jung2023machine}, or from equilibrium and short-trajectory MD data via variational~\cite{khoo2019solving, li2019computing} and self-consistency formulations~\cite{li2022semigroup, strahan2023predicting, mitchell2024committor}. 
In either case, success requires either sufficient transition path data or substantial computational investment in iterative sampling~\cite{megias2025iterative, arredondo2025atoms, talmazan2025static, giuseppe2026following} and active learning~\cite{rotskoff2022active, kang2024computing}, and the need for iterative enhanced sampling to cover the transition region persists regardless.  

Generative ML models offer another alternative: conditional models sample TSs given the reactant and product states~\cite{pattanaik2020generating, duan2023accurate, kim2024diffusion, duan2025optimal}, while unconditional models target the equilibrium distribution~\cite{lewis2025scalable, jing2024generative, costa2025accelerating, thiemann2025force} and require post-hoc identification of TSs. Both remain limited by their training data distribution.
Recent works have partially addressed this by using geometric optimization and energy evaluations during training to find transition paths and extract TSs without a pre-curated TS dataset~\cite{nam2025transferable,hait2025locating} or by leveraging MD-derived metrics~\cite{wang2024generalized} and the Onsager--Machlup functional with a force field approximated from a generative model trained on equilibrium distribution to find transition paths~\cite{raja2025action}.
Nevertheless, these approaches still rely on either accurately learning the PES near the TS or constructing mappings to pre-computed TS distributions~\cite{raja2025action, tuo2025flow}, limiting their utility to underexplored systems.  

To go beyond sampling the learned distribution, inference-time control in diffusion models offers a promising avenue for steering generative processes toward desired regions of chemical and conformational space. 
Steering the sampling of generative models based on objectives, constraints, or rewards has been an active area of research, spanning earlier approaches such as classifier-free guidance~\cite{ho2022classifier} to reward-tiltedd, annealed, and equal-density distribution sampling~\cite{du2023reduce,skreta2024superposition,skretafeynman}. 
However, steered sampling outside the training PES distribution has not been extensively explored. 
Recent attempts to integrate enhanced sampling with generative models remain limited by poor transferability to unseen systems or reliance on predefined CVs~\cite{nam2025enhancing}. Similarly, while flow models can be steered at inference time to upsample the transition states distribution~\cite{kolloff2025minimum}, the generalization of this approach to novel systems is limited by the reliance on \textit{a priori} samples from the target ensemble.
Nevertheless, given that diffusion models can learn to sample diverse states and that atomistic simulations provide ground truth energetics, combining them offers a natural framework for TS ensemble generation with low data requirements.

\begin{figure}[!htb]
    \centering
    \includegraphics[width=\textwidth]{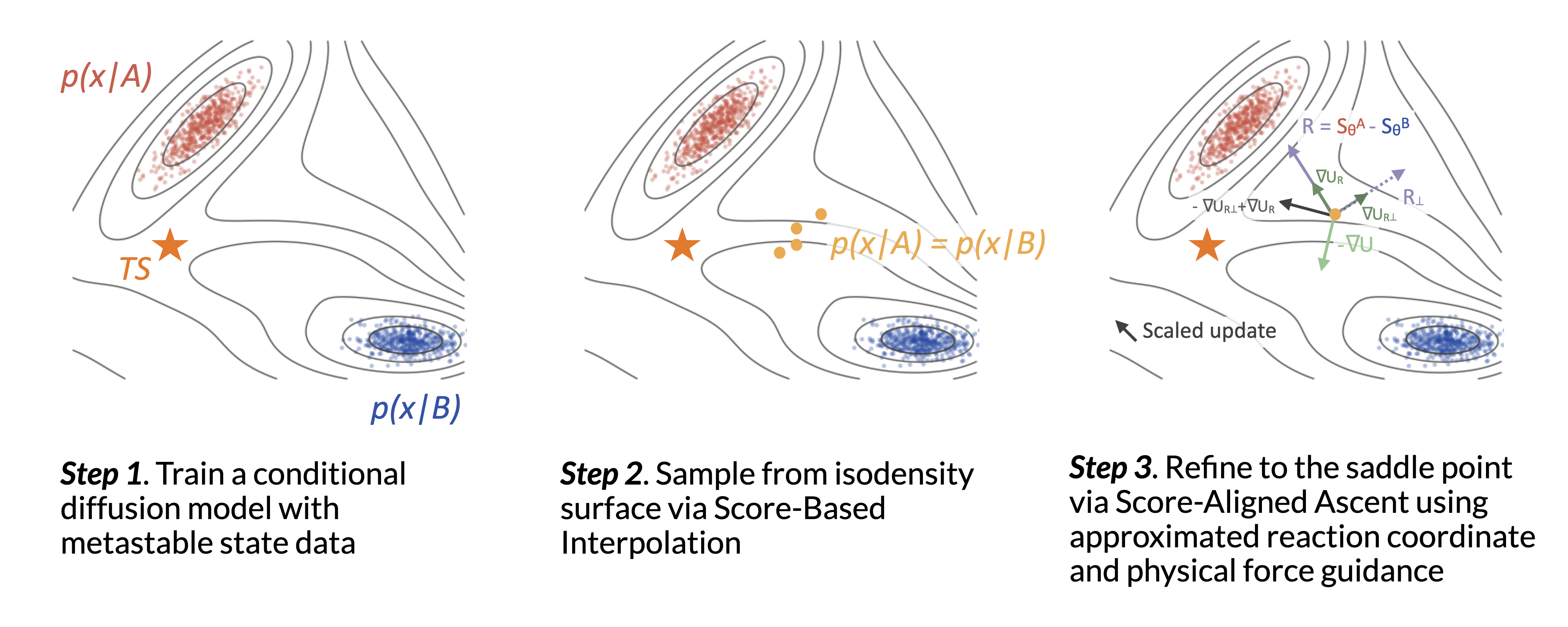}
    \caption{Overview of \textbf{ASTRA}. The method consists of three stages: (1) training a conditional generative model, (2) sampling from an isodensity surface, and (3) inference-time guidance that combines a reaction coordinate approximated from score differences of the two conditional models $\theta^A, \theta^B$ ($R = S_\theta^A - S_\theta^B$) with physical forces to rapidly sample transition states defined as first-order saddle points.}
    \label{fig:overview}
\end{figure}

We introduce ASTRA (\textbf{A} Priori \textbf{S}ampling of \textbf{TRA}nsition States with Guided Diffusion), a system-independent workflow to locate and optimize transition states of arbitrary molecular systems without iterative enhanced sampling, heuristics, or TS training data (see Figure~\ref{fig:overview}).
Our approach combines recent advances in generative ML with principles from computational chemistry.
The central idea is that a useful TS ensemble guess can be inferred from the learned probability distributions of the surrounding metastable states. Therefore, ASTRA employs a conditionally trained score-based diffusion model~\cite{song2020score, ho2022classifier} that learns the data manifold of reactant and product basins from short MD trajectories.
At inference, the reverse diffusion process is guided to sample configurations on the isodensity surface where the probability of belonging to either state is equal, via a principled composition of conditional scores~\cite{skreta2024superposition} -- which we term \textbf{Score-Based Interpolation (SBI)}.
These samples are then refined by force-based updates on the PES, ascending along the reaction coordinate and descending along orthogonal directions, leveraging our second key insight that the reaction coordinate can be effectively approximated from the conditional diffusion scores. 
This process, which we call \textbf{Score-Aligned Ascent (SAA)}, alleviates the need for the diffusion model to learn the PES perfectly.
Crucially, ASTRA operates entirely \textit{a priori}, requiring only samples from the reactant and product states with no knowledge of the transition region, yielding high-quality TS guesses ideally suited for rapid convergence with single-ended optimization methods. Thus, ASTRA uniquely offers a complementary alternative to conventional double-ended TS guess methods in standard hierarchical search protocols~\cite{del2008hierarchical,peters2004growing,zimmerman2013reliable,maiti2025benchmark}, providing a universal and \textit{a priori} approach to saddle point location.  

Our contributions are threefold: (1) We propose a workflow for direct TS sampling that leverages guided diffusion, eliminating the need for path-finding algorithms or prior TS data.
(2) We introduce the combination of Score-Based Interpolation and Score-Aligned Ascent as a principled mechanism to guide a diffusion process to sample first-order saddle points. 
(3) We show that ASTRA successfully identifies known transition states and competing pathways in systems ranging from two-dimensional potentials to high-dimensional chemical systems. 

\section{Results}\label{sec2}

To rigorously assess the performance of the ASTRA algorithm, we established a unified evaluation framework consistent across all test systems.
This is designed to probe the quality of the TS ensembles generated by our method. 
For each system, a score-based diffusion model is first trained on configurations, from known metastable states, obtained from short MD trajectories.
The ASTRA algorithm is then employed to generate a candidate TS ensemble.
Detailed experimental parameters are provided in Appendix~\ref{exp:explanation}, and the results of all ablation studies are presented in Appendix~\ref{abl:ib_saa}.

For the 2D analytical potential energy surfaces, the ASTRA-generated samples are directly compared to the known transition states using the L2 distance metric.
In the higher-dimensional chemical systems, we quantify the quality of the samples with committor values. 
The committor function $q(x)$ defines the probability that a trajectory initiated from a given state $x$ will reach the designated product state before returning to the reactant state.
By definition, the true TS ensemble corresponds to the $q(x)=0.5$ isosurface.
We compute this metric using a pretrained ML committor model or MD-based evaluations when a force field is available (i.e. for all-atom systems).
In the latter case, we approximate committor values by initiating multiple replicas of Langevin dynamics simulations from each structure with random initial velocities sampled from the Maxwell--Boltzmann distribution.  

We note that ASTRA is designed to locate first-order saddle points on the potential energy surface (PES), whereas the committor isosurface is governed by the underlying free energy surface; these two characterizations of the TS coincide for low-dimensional systems with a dominant barrier but can differ for flexible high-dimensional systems. We discuss this distinction and its implications for evaluation in Section~\ref{TS characterization reference}.  

Additionally, for chemical systems with a force field available, we assess the quality of the TS guesses provided by the different approaches with the convergence of the Dimer algorithm, a single-ended TS optimization method. This setup mirrors standard practice for obtaining an optimized saddle point structure, where double-ended methods, such as Nudged Elastic Band (NEB), typically provide a robust initial guess for the convergence of a subsequent single-ended TS optimization method. In that context, we evaluate ASTRA specifically as a TS guess generation method.

\subsection{Analytical Potentials}
\begin{figure}[!htb]
    \centering
    \begin{subfigure}[b]{0.325\textwidth}    
        \centering
        \includegraphics[width=\textwidth]{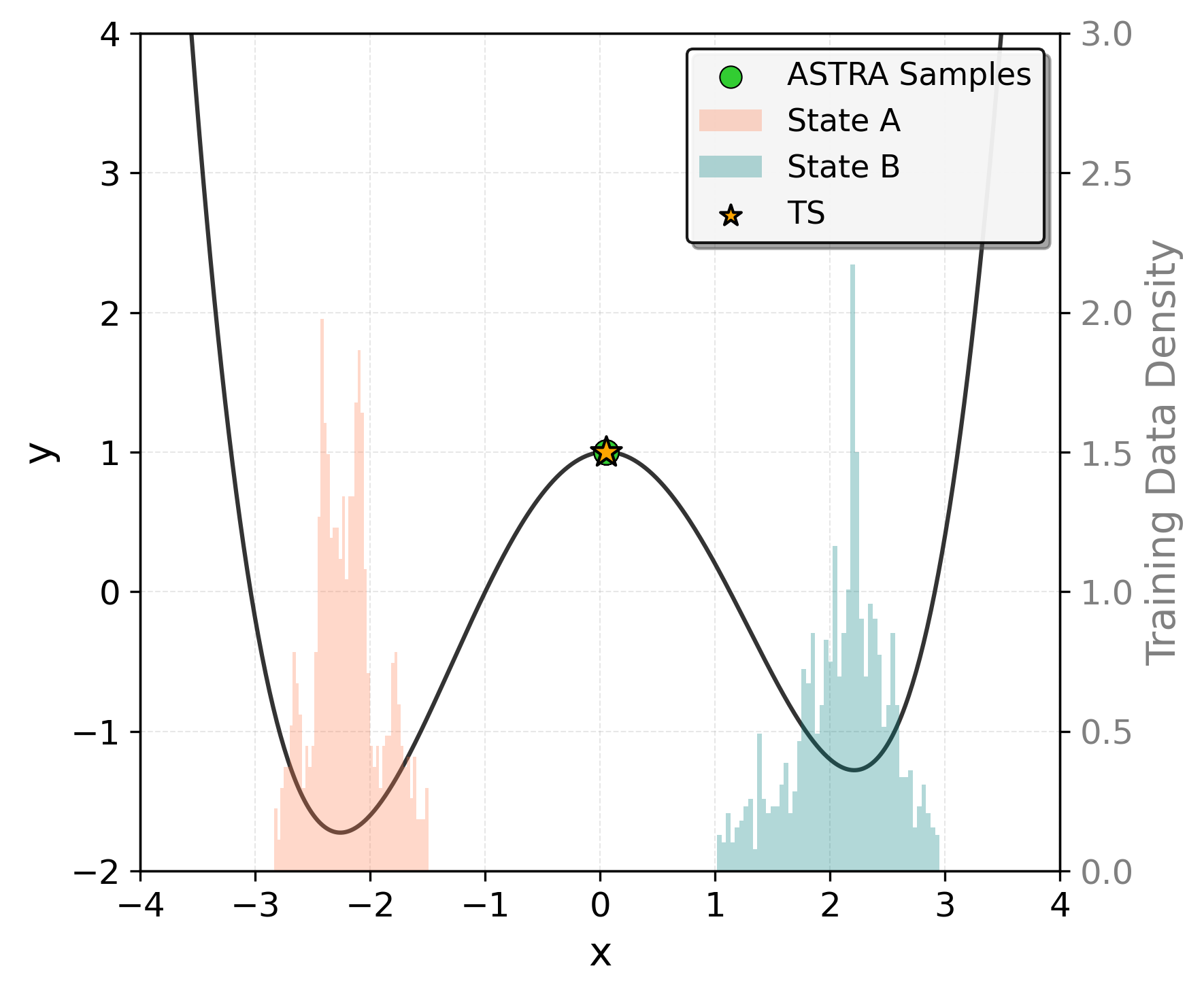}
        \caption{Double well potential.}
        \label{fig:double_well_samples}
    \end{subfigure}
    \hfill 
    \begin{subfigure}[b]{0.325\textwidth}
        \centering
        \includegraphics[width=\textwidth]{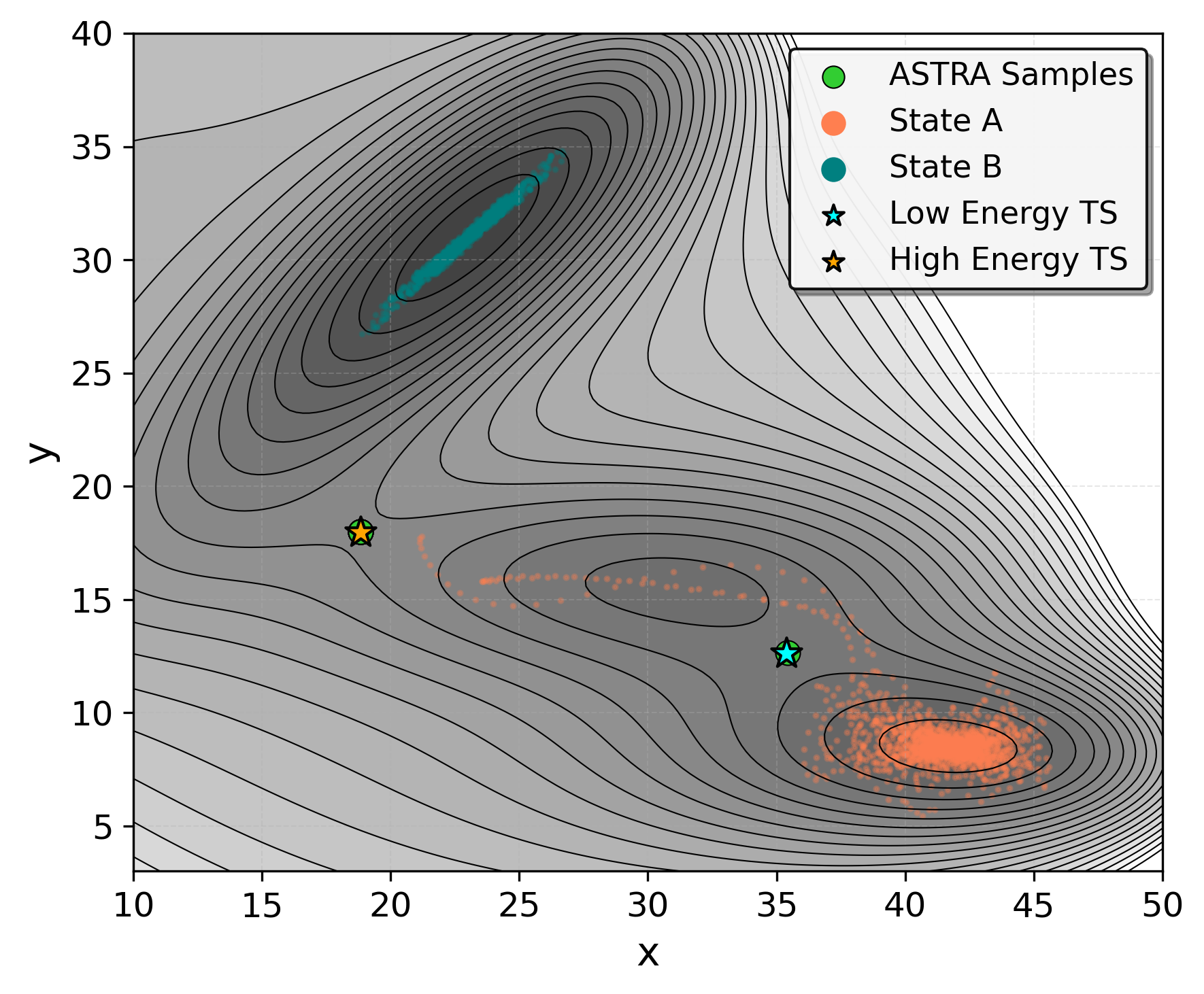}
        \caption{Müller-Brown potential.}
        \label{fig:muller_brown_samples}
    \end{subfigure}
    \hfill 
    \begin{subfigure}[b]{0.325\textwidth}
        \centering
        \includegraphics[width=\textwidth]{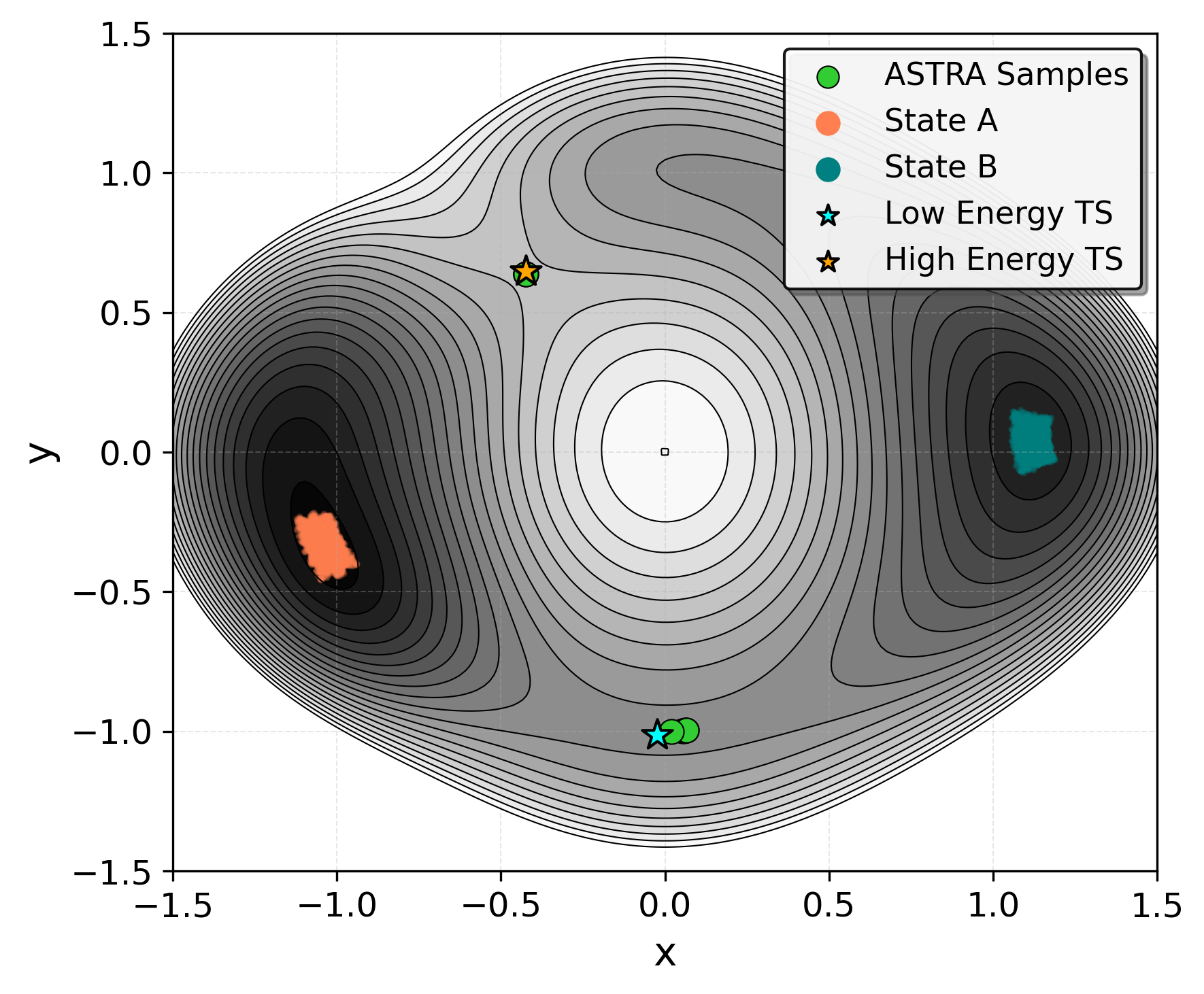}
        \caption{Double path potential.}
        \label{fig:double_path_samples}
    \end{subfigure}
    \caption{Our method discovers transition state regions with high precision. It finds a transition state for the (a) double well potential, and multiple transition states for the (b) Müller-Brown potential and (c) double path potential.}                      
    \label{fig:combined_samples}
\end{figure}

On a one-dimensional double well potential, we first demonstrate the fundamental capability of our method to sample TSs over a simple energy barrier. 
The model is trained on data from two distinct energy wells, designated as State A and State B.
As illustrated in Figure~\ref{fig:double_well_samples}, our algorithm identifies the TS by generating samples densely clustered at the peak of the potential barrier separating the two minima.

To further assess the performance of our method, we employ the Müller-Brown potential~\cite{muller1979location}, which is a common two-dimensional benchmark for evaluating TS search algorithms. 
Its energy surface features three local minima connected by two first-order saddle points, offering a controlled environment to assess a method's ability to identify non-linear pathways and distinct saddle points. 
For our analysis, we assign the two lowest-energy minima as the initial and final states (State A and State B in Figure~\ref{fig:muller_brown_samples}).
The ASTRA algorithm successfully locates configurations clustering around both accessible transition states connecting A and B.
Ablation studies shown in Figure~\ref{fig:abl:mb} indicate that this ability to discover multiple, topologically distinct saddle points is primarily driven by the Score-Aligned Ascent (SAA) component of our method.

A critical test for advanced sampling methods is the ability to identify multiple, topologically distinct pathways for a given transition. 
We evaluate this specific capability on a two-dimensional potential engineered to include two competing reaction channels connecting the same reactant and product states. 
As shown in Figure~\ref{fig:double_path_samples}, the ASTRA algorithm successfully populates selectively both TS regions corresponding to the two competing pathways. 
This simultaneous discovery of distinct reaction channels, achieved without any explicit path-based guidance, demonstrates a key advantage of our method. 
Whereas traditional algorithms are designed to converge to a single pathway, our approach can explore the entire TS manifold, thereby providing mechanistic insights into complex energy landscapes that can be easily overlooked with conventional techniques without extensive sampling, and avoids being trapped in local maxima that are irrelevant for the studied chemical process.

\subsection{Conformational Isomerization in Common Peptide Model Systems}
\begin{figure}[!htb]
    \centering
    \begin{subfigure}[b]{0.48\textwidth}
        \centering
        \includegraphics[width=\textwidth]{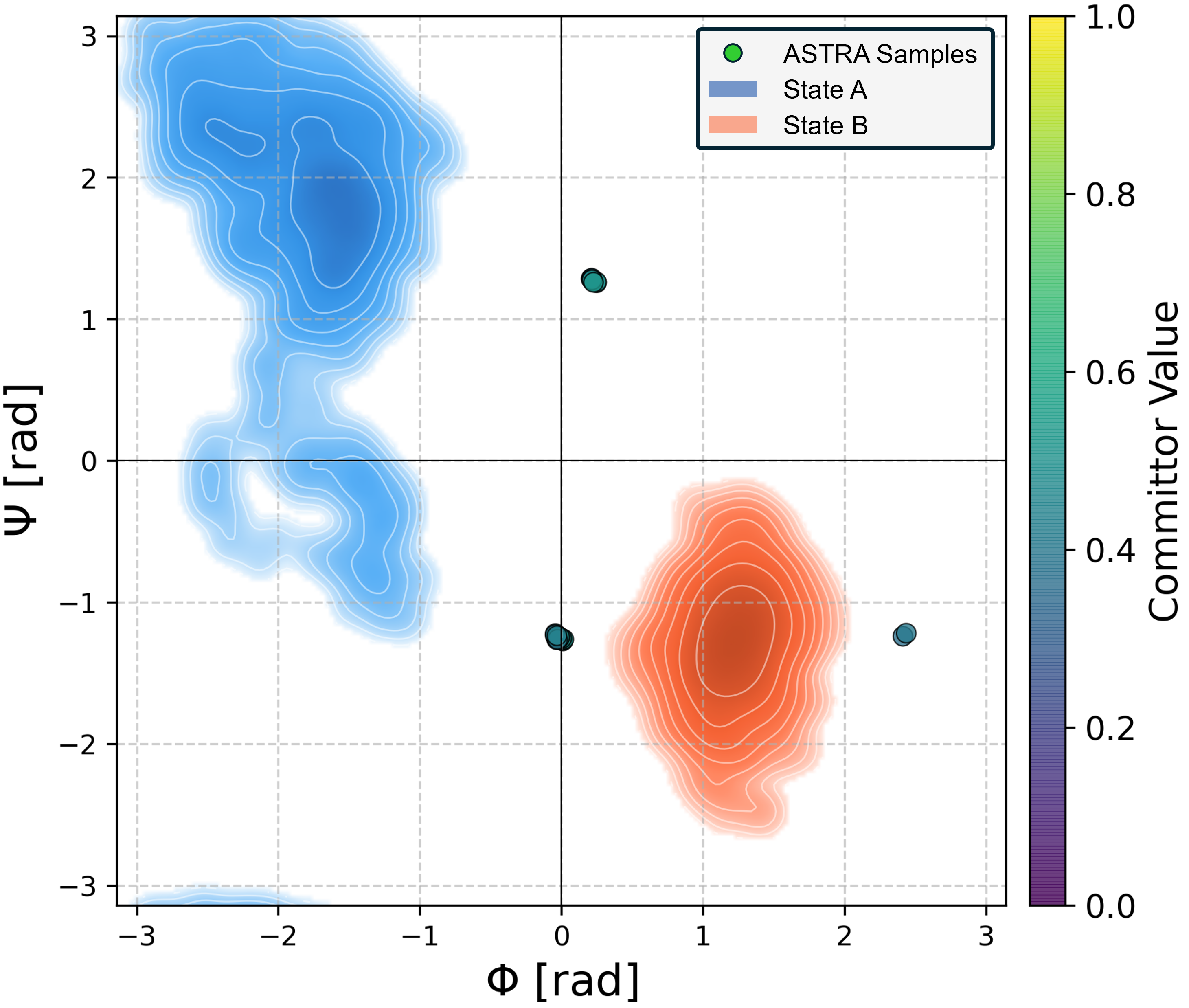}
        \caption{Alanine dipeptide.}
        \label{fig:alanine_ramachandran}
    \end{subfigure}
    \hfill 
    \begin{subfigure}[b]{0.48\textwidth}
        \centering
        \includegraphics[width=\textwidth]{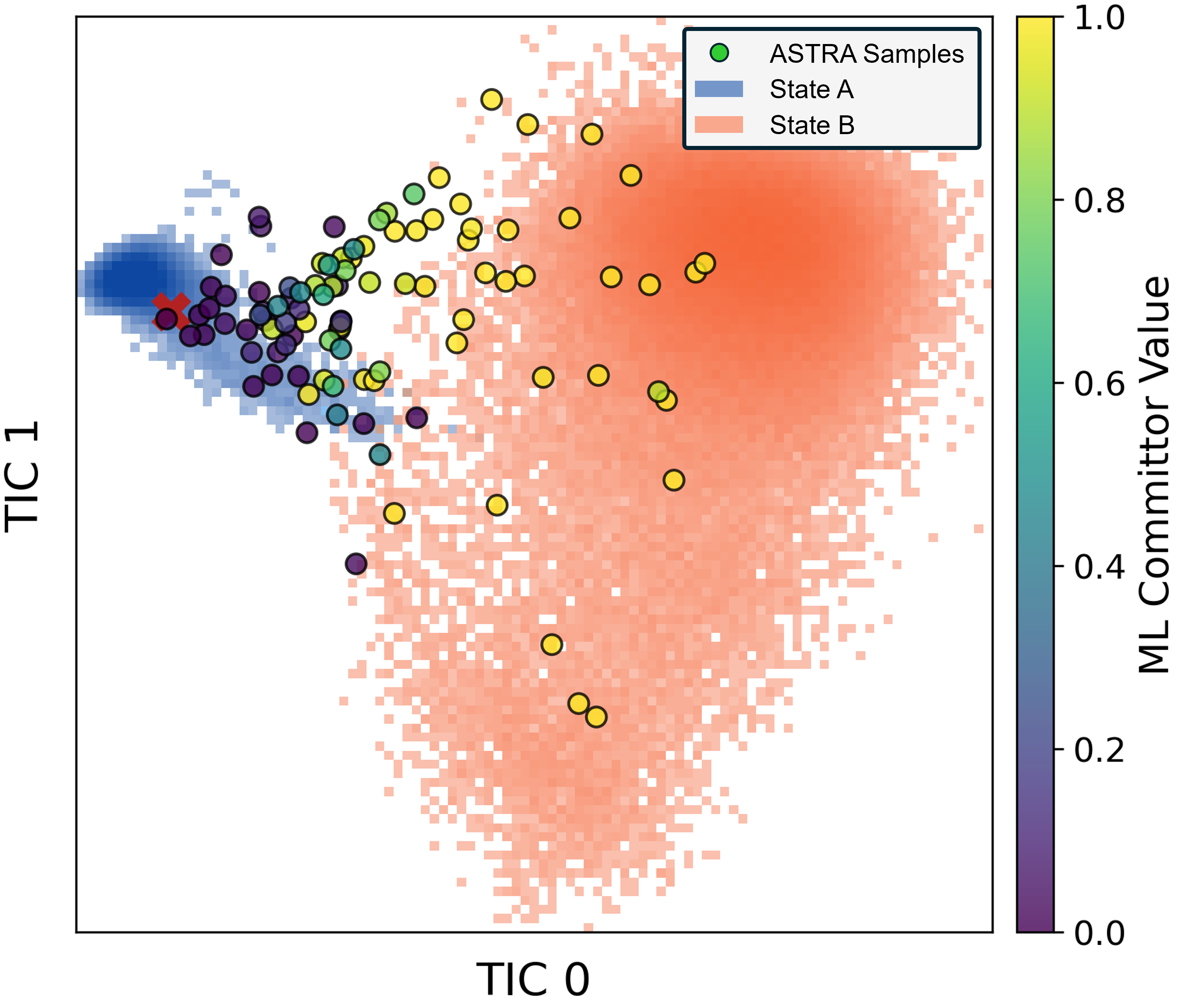}
        \caption{Chignolin.}
        \label{fig:chignolin_tica}
    \end{subfigure}
    \caption{Application of our method to chemical systems. (a) For the alanine dipeptide, the generated samples (circles) localize the transition region between two states. (b) For chignolin, the free energy landscape is projected onto the two slowest collective modes from time-lagged independent component analysis fitted from a converged simulation~\cite{lindorff2011fast}.
    The background densities in both plots correspond to the training data distribution where distinct colors indicate the defined State A and State B. The color scale for the alanine dipeptide plot represents the MD-based committor value while the scale for chignolin indicates the machine-learned committor value~\cite{kang2024computing}.
    }
    \label{fig:chemical_systems_ala2_chig}
\end{figure}

Having demonstrated proficiency on low-dimensional analytical potentials, we tested our algorithm's performance on molecular systems.
The conformational dynamics of alanine dipeptide, a 22-atom molecular system, provide a widely adopted model and benchmark for studying conformational ensembles of biomolecules.
The dynamics are conventionally analyzed through a Ramachandran plot, defined by the dihedral angles $\phi$ and $\psi$. 
We focus on the isomerization between the two most populated conformational states, the C5 and C7ax conformations, which represent distinct local minima separated by free energy barriers~\cite{chekmarev2004long}.

\begin{figure}[!htb]
    \centering
    \includegraphics[width=0.99\textwidth]{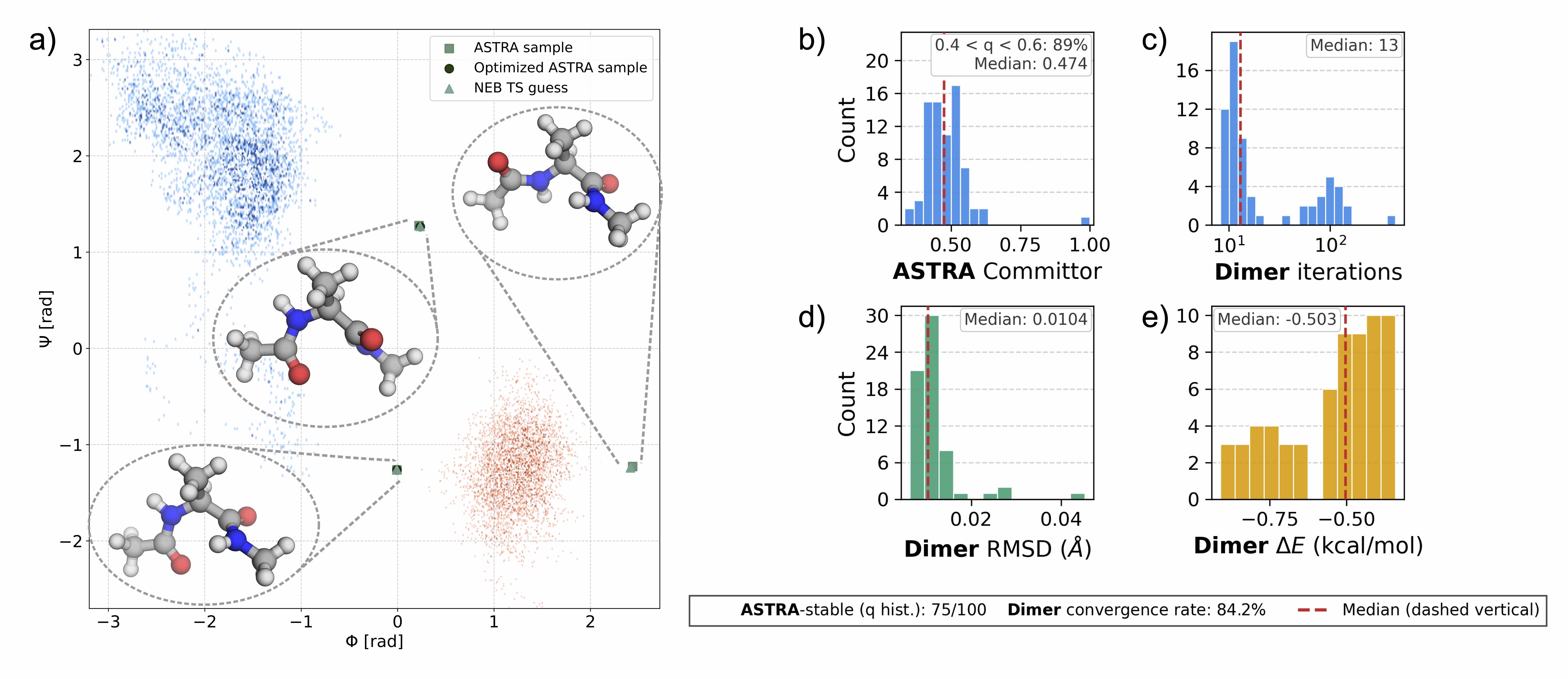}
    \caption{Detailed performance of ASTRA on alanine dipeptide. a) The ASTRA samples covering all three TSs are compared to the Nudged Elastic Band reference method. One ASTRA sample is selected randomly for each TS for readability. The background shows the training data distribution and colors represent the two defined states for the classifier free guidance training. The three-dimensional structures are overlayed for each identified TS. b) shows the distribution of committor value for the stable ASTRA samples peaked around 0.5. c-e) characterize how close ASTRA samples are from their Dimer-optimized counterpart. c) shows the distribution of number of iterations necessary to converge Dimer from the stable ASTRA samples. d) displays the Root Mean Square Deviation (RMSD) of ASTRA samples after Dimer optimization compared to before, while e) shows that difference in terms of the energy of the structures. We indicate the number of stable ASTRA samples, compared to the total number of structures drawn from our algorithm for that analysis, alongside the Dimer convergence rate from ASTRA stable samples. Stable ASTRA samples correspond to structures sampled by ASTRA for which running 2ps MD from is stable.}
    \label{fig:abl:ala2_TSE}
\end{figure}

We highlight the absence of model knowledge in the transition region in Figure~\ref{fig:alanine_ramachandran} a), showing the distribution of the ASTRA-generated structures overlaid on the distribution of the training set.
We observe that they are localized precisely within three narrow regions of the Ramachandran plot, which correspond to the known transition states for the C5-C7ax interconversion. 
This is validated by committor analysis, where short Langevin Dynamics simulations initiated from these configurations yield a distribution of committor values sharply peaked at 0.5 (Figure~\ref{fig:abl:ala2_TSE}) for all transition states. This highlights the key ability of ASTRA to identify distinct and structurally diverse pathways for a transition in a chemical system, without any input or prior knowledge of the mechanisms at play.

Additionally, we assess the quality of our samples with a single-ended TS optimization procedure, which typically requires a high-quality initial guess, geometrically close to the saddle point.
The objective is to evaluate how similar the ASTRA-generated samples are to the reference structures obtained by an optimization method.
Specifically, we report the convergence rate, the root mean square deviation (RMSD) of atomic positions, and the energy differences between the initial and optimized structures. 
As illustrated in Figure \ref{fig:abl:ala2_TSE}, and in greater detail in Table \ref{tab:hyperparameter_ablations} and Table \ref{tab:hyperparameter_ablations_committor}, we find that our method is able to provide a majority of samples that converge to a saddle point within a few optimization steps.
Additionally, final optimized structures remain close to the original ASTRA samples, as evidenced by the low RMSD values and small changes in energy barriers.
The validity of the transition state guesses generated by ASTRA is further confirmed by comparing them to structures obtained by NEB, a reference method for finding minimum energy paths. Transition configurations generated by SAA are nearly indistinguishable from those optimized by NEB (Figure~\ref{fig:abl:ala2_TSE}, Table~\ref{tab:geo_lin_neb}). Crucially, SAA exhibits superior robustness to initialization noise. For example, when initialized from a linearly interpolated path for alanine dipeptide, NEB fails to converge, whereas SAA successfully locates the transition state (Figure~\ref{fig:abl:geo_lin_opt_neb}, Table~\ref{tab:geo_lin_neb}). Furthermore, as an interpolation-based method, NEB requires precisely informed starting and ending configurations to identify all three transition states for alanine dipeptide (Figure \ref{fig:abl:geo_lin_opt_neb}). This strict dependence on prior mechanistic knowledge and extensive sampling of the potential energy surface remains a fundamental bottleneck for the high-throughput exploration of novel chemistries. By contrast, the synergetic use of SBI and SAA enables broad, heuristics-free approximate sampling of the Transition State Ensemble (TSE) alongside robust refinement to saddle points.

We also refer to Appendix~\ref{analysis_ASTRA} for complete analysis and details of the method and the key components of its success for this benchmark system.

\begin{figure}[!htb]
    \centering
    \begin{subfigure}[b]{0.48\textwidth}
        \centering
        \includegraphics[width=\textwidth]{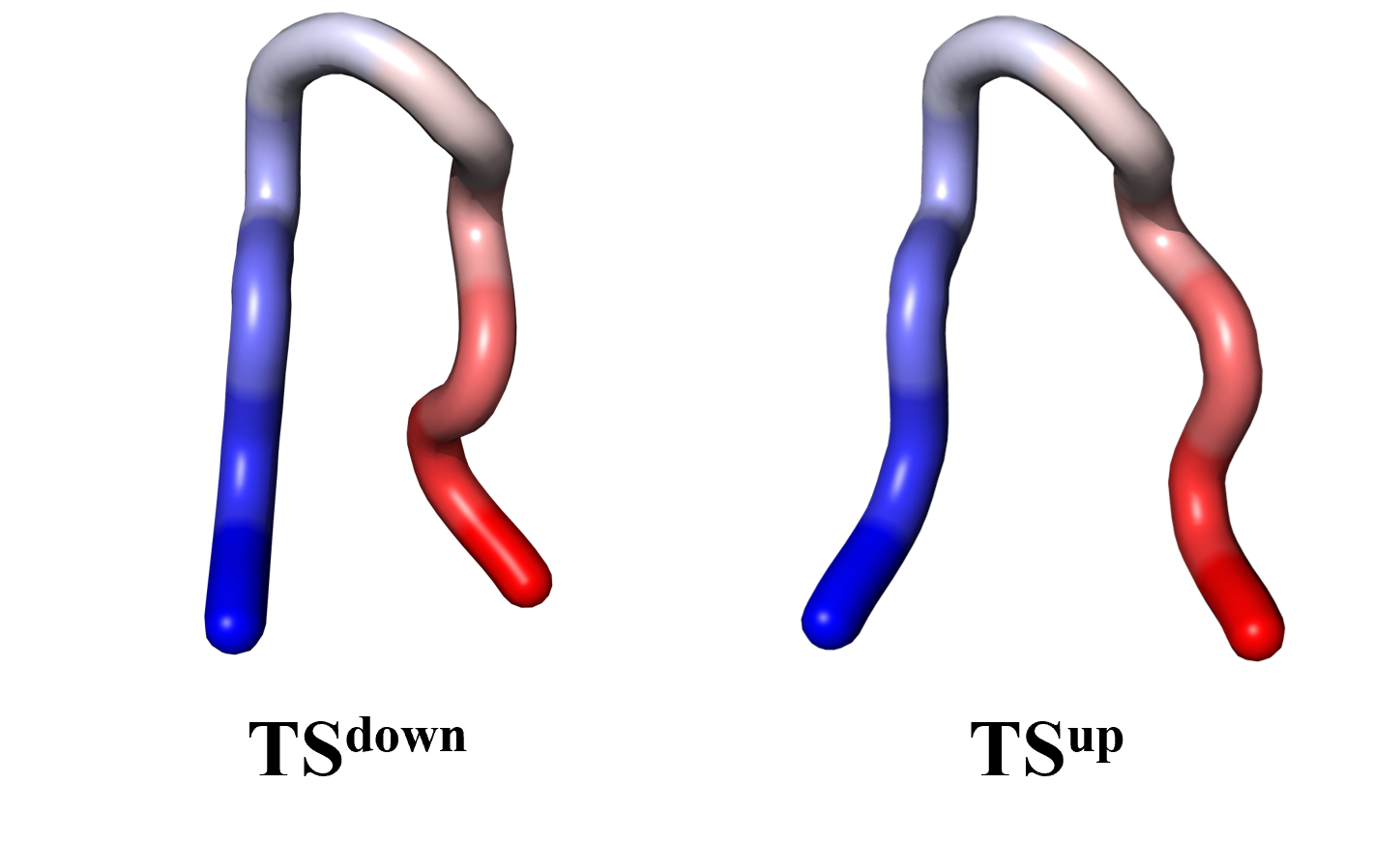}
        \caption{ASTRA-generated samples.}
        \label{fig:abl:chig_updown_ours}
    \end{subfigure}
    \hfill 
    \begin{subfigure}[b]{0.48\textwidth}
        \centering
        \includegraphics[width=\textwidth]{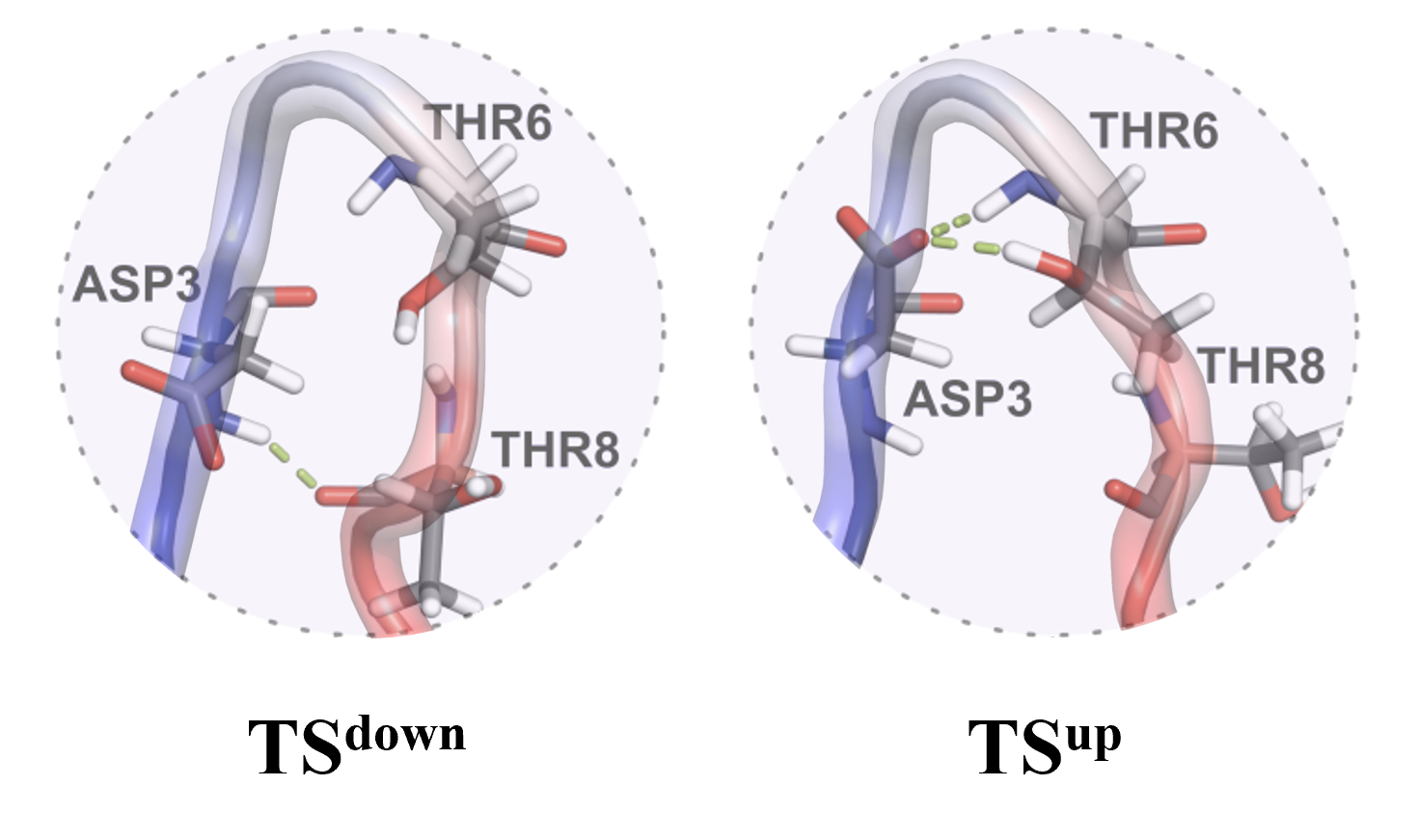}
        \caption{References.}
        \label{fig:abl:chig_updown_ref}
    \end{subfigure}
    \caption{Visualization of chignolin transition state mechanisms. Representative structures from ASTRA-sampled TS\textsuperscript{down} and TS\textsuperscript{up} (\textbf{left}) ensembles are overlaid with transparent tubes on the reference conformations from Ref.~\citenum{kang2024computing} (\textbf{right}). The structural agreement validates our method's ability to resolve distinct folding pathways.}
    \label{fig:abl:chig_updown} 
\end{figure}

We extend our analysis to chignolin, a 10-residue fast-folding protein with multiple competing folding pathways, and transition timescales spanning microseconds.
For this task, we initialized our score-based diffusion model from the publicly available checkpoint of Ref.~\citenum{arts2023two}, which was pretrained on coarse-grained chignolin trajectories.
We then fine-tuned the model using our training protocol to enable Classifier-Free Guidance on the MD trajectory dataset by D.~E.~Shaw Research~\cite{lindorff2011fast}.
To prevent direct supervision on the transition region during fine-tuning, all frames corresponding to the transition region were removed from the fine-tuning set. 
This region was identified using a pre-trained, machine learning-based committor model~\cite{kang2024computing}: frames with ML committor values $q$ in the range $0.0001 \le q \le 0.9999$ were excluded, retaining only configurations deep within either basin. 
The resulting configurations generated by our method were then projected onto a two-dimensional space defined by time-lagged independent component analysis (TICA) for visualization~\cite{molgedey1994separation, perez2013identification}.
The generated configurations are concentrated along the $q \approx 0.5$ isosurface, providing strong quantitative evidence that our method concentrates probability mass on the TS ensemble (Figures~\ref{fig:chignolin_tica} and \ref{fig:abl:com_chig}). 
Furthermore, these samples are not concentrated on one spot but are spread across the TICA space within the transition region. 
Structural analysis reveals that the generated samples can be classified into the two known competing folding pathways for chignolin, commonly referred to as TS\textsuperscript{down} and TS\textsuperscript{up} (Figure~\ref{fig:abl:chig_updown}). This confirms our method's ability to accurately describe transition state diversity in folding mechanisms. 

\subsection{Chemical Reaction}

Beyond conformational sampling of biomolecules, TS are of major importance in the study of chemical reaction mechanisms.
However, learning high energy structures governed by covalent bond formation and cleavage without prior data of such structures has been a major challenge.
We test our algorithm on a first-generation donor-acceptor Stenhouse adduct, which has already been studied in the context of learning collective variables~\cite{kang2024computing}.
The molecule consists of 42 atoms. It transitions from a linear, conjugated (open) state to a cyclic (closed) state through a multi-step reaction pathway that includes \textit{Z-E} isomerization, conformational rotation, and a thermally driven 4$\pi$-electrocyclization~\cite{reyes2024compartmentalizing}.
We focus on the TS sampling of the 4$\pi$-electrocyclization that involves a concerted formation of two and breaking of one covalent bond in two distinct regions of the molecule, demonstrating whether our algorithm can interpolate between reactant and product states involving multiple simultaneous bond changes.

Analogous to alanine dipeptide, we constructed a training dataset by sampling configurations from the metastable transient isomer and stable enol closed isomer with conventional MD simulations.
Both simulations remained within their respective basins, and we further ensured that our dataset excludes any state close to the transition region by using a pre-trained ML committor model~\cite{kang2024computing}; we retained only frames whose predicted committor values were either below 0.0001 or above 0.9999.

\begin{figure}[!htb]
    \centering
    \includegraphics[width=\textwidth]{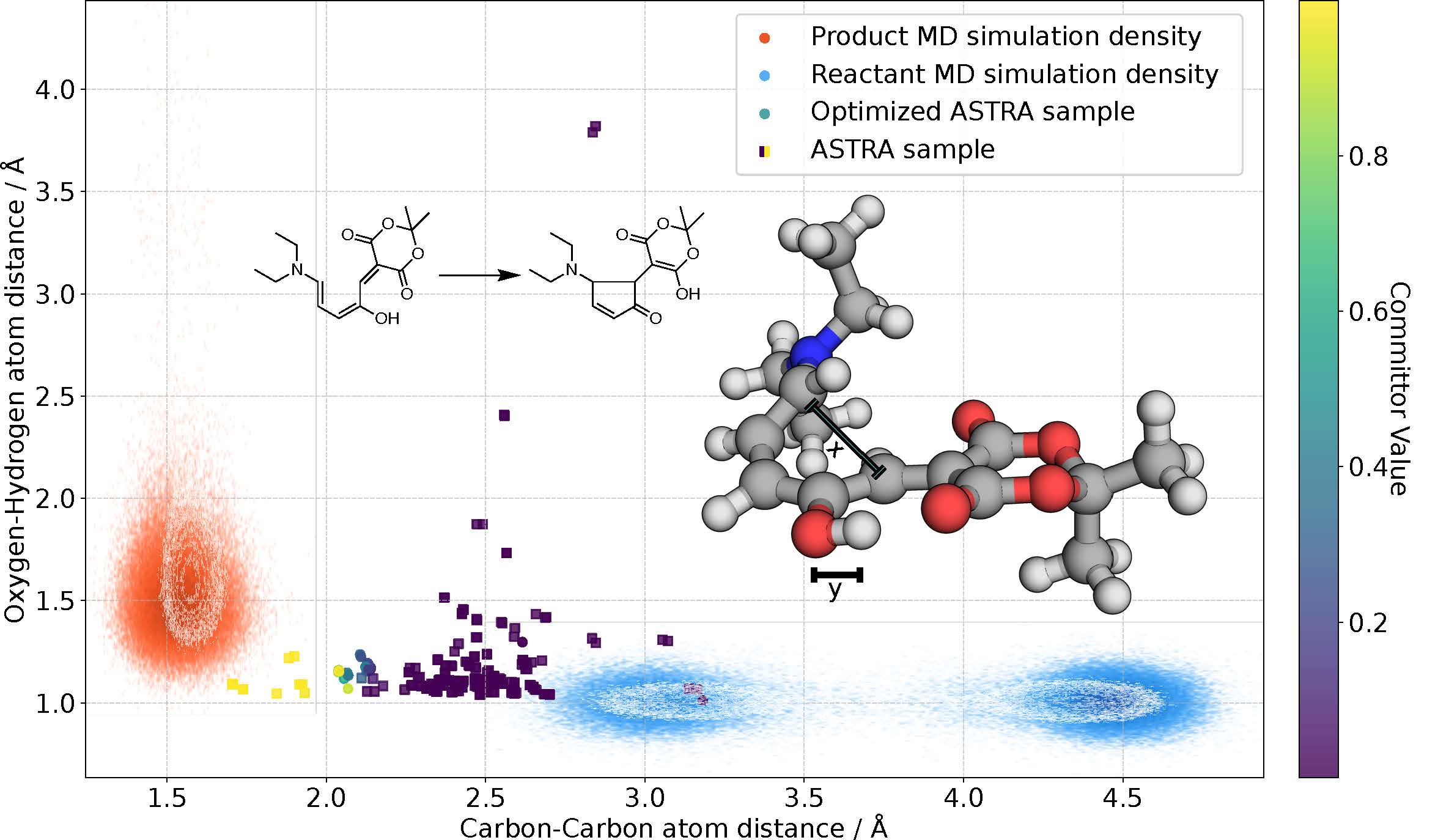}
    \caption{Lewis structure representation of the chemical reaction and three dimensional reactant structure with the two coordinates highlighted that span the plot. The two chosen degrees of freedom separate the reactant and product state well, but were only used for visualization and not included in any part of our algorithm. The MD simulation data points are plotted as densities. The structures generated by the ASTRA algorithm and the structures optimized with the Dimer algorithm and converged to a true transition state are shown as differently shaped points that are colored based on the committor value predicted by the pretrained model.}
    \label{fig:dasa_committor}
\end{figure}

At inference, the ASTRA algorithm consistently generated structures within the TS region (Figure~\ref{fig:dasa_committor}), as indicated by predicted committor values that differ substantially from 0 or 1.
However, the ML committor values do not peak at 0.5 for both the structures generated by ASTRA or structures optimized with the Dimer algorithm that are confirmed to be TS structures with a single imaginary normal mode.
A possible reason is that the ML committor model was not trained on optimized TS structures, but on frames of MD simulations.
Hence, we do not rely on the committor values in this analysis when comparing converged TS structures obtained with different methods and instead focus our analysis on the coverage of the various possible TS conformers and compare this coverage to other single- and double-ended TS search algorithms.
Various TS conformers are possible for this reaction due to i) the different out-of-plane positions of the geminal methyl groups as shown in a previous study~\cite{kang2024computing}, ii) the dihedral angle of the 6-membered ring with the formed 5-membered ring, and iii) the orientation of the ethyl groups of the amine group.

Generating various TS structures from baseline workflows and ASTRA reveals a clear clustering of data points on the two-dimensional subspace, which is depicted in Figure~\ref{fig:si:dasa_analysis}.
The different clusters reveal a large structural diversity of potential TS conformers that is successfully covered by the ASTRA samples and required a multi-level workflow of conformer generation and TS guess generation with existing approaches.

\section{Discussion}\label{sec4}

In this work, we introduce ASTRA, a workflow that reconceptualizes TS sampling as an inference-time guidance task of generative models by leveraging a pretrained score-based model conditioned on known metastable states and physical forces.
Our method composes learned distributions with Score-Based Interpolation (SBI) to target directly the dividing surface of equal density, while the Score-Aligned Ascent (SAA) mechanism steers the sampling process towards transition states. 
This approach successfully locates the TS ensemble regions with high precision and discovers competing reaction pathways in chemical systems, as demonstrated on alanine dipeptide, chignolin, and an electrocyclical reaction, without any prior knowledge of the underlying transition mechanisms.

The \textit{a priori} sampling of transition states is primarily enabled by the synergetic use of SBI and SAA.
The implicit probabilistic definition of the transition state region by SBI provides a clear advantage over classical interpolation methods, such as linear or geodesic interpolation, that require carefully selected starting structures to ensure diversity.
Yet, in the alanine dipeptide benchmark, only a guidance scale of 1.0 with Isodensity Interpolation enables sampling of all three transition states; larger guidance scales tend to collapse the samples onto high-density modes.
Enhancing the diversity of samples on the PES during SBI could further improve the robustness, particularly in systems exhibiting complex, multi-pathway reaction dynamics. 


For convergence to a reliable MEP and saddle point, traditional minimum energy path algorithms such as NEB rely on a good initial path interpolating reasonable structures between minimized endpoints. When this assumption is met and the interpolated path passes close to the relevant saddle, NEB is highly accurate. In our alanine dipeptide benchmark, NEB initialized from minimized geodesic interpolation recovers both main TSs with TS structures of comparable RMSD to ASTRA (Table~\ref{tab:geo_lin_neb}). However, starting from linear Cartesian interpolation between the same minimized endpoints, NEB converges along only one of the two main pathways and misses the other TS (Figure~\ref{fig:abl:lin_min_neb}), because the linear path passes through high-energy, sterically clashing intermediates that bias the optimization toward a single basin of attraction. NEB performance degrades further when endpoints are not energy-minimized: from non-minimized geodesic interpolation, almost all images become unstable during NEB optimization (Table~\ref{tab:geo_lin_neb}).  

In contrast, SAA leverages the difference of conditional scores, emulating a reaction coordinate that connects the two basins at a low latent level. Because the optimization is performed in latent space and the resulting samples are projected back onto the data manifold via the remaining denoising steps, SAA is less constrained by the physicality of the initial guess structures. 
This robustness allows SAA to recover both main TSs in the alanine dipeptide system from linearly interpolated paths, where NEB recovers only one, and from non-minimized geodesic paths, where NEB fails entirely. We refer to Appendix~\ref{analysis_ASTRA} for an extended analysis and methodological details.


While ASTRA is not yet able to substitute all existing TS exploration workflows with a drop-in replacement, we show that (i) diffusion models can infer reliable TS structures from adjacent states and (ii) our inference-time guidance with SBI and SAA enables massive shortcuts in the generation of out-of-distribution samples.
The concurrent work by Tang et al.~\cite{tang2025breaking} learns the committor function of a given system by launching simulations from structures generated with a diffusion model that was also trained only on metastable states at first.
They do not modify the reverse-diffusion process and hence require a computationally intensive iterative approach to converge on a reliable model of the committor function.
In contrast, ASTRA acts as a ``single-shot'' generator.
While it remains to be seen if one-shot generation can fully replace iterative enhanced sampling in large biological systems, it is straightforward to integrate our ASTRA approach into existing iterative approaches as a powerful initialization strategy.
This could potentially eliminate the reliance on heuristic collective variables and reduce the number of iterations needed to uncover complex reaction mechanisms.
Conversely, in the application of ASTRA as a TS guess generation model in time-independent TS optimization tasks, our generative workflow is computationally more expensive than existing traditional computational chemistry workflows because of the system-focused training data generation and training overhead.
However, existing methods can often become trapped in single pathways due to heuristic initialization strategies without massive scale-up and automation of the TS search, while ASTRA enables direct broad and diverse TS guess generation free from heuristic assumptions.
This diversity is key in conformational flexible systems, where ASTRA's computational overhead could be offset by the increased sampling efficiency at inference.

A key advantage of our approach is its compatibility with increasingly powerful pretrained generative models for molecular conformational sampling. For instance, models such as BioEmu~\cite{lewis2025scalable} generate protein ensembles consistent with the Boltzmann distribution and are therefore biased toward low-energy states, with potentially limited accuracy near transition regions. When combined with our inference-time, force-guided scaling strategy, however, such models can still serve as strong priors. Starting from these pretrained priors rather than from scratch further improves the data efficiency of the search. 

Future work will extend ASTRA toward a general-purpose inference-time guidance framework for rare event discovery across diverse chemical systems. Although this study employs classical force fields and semi-empirical quantum chemistry methods, the framework easily generalizes to employing higher-fidelity force predictions such as machine learning interatomic potentials (MLIPs). Beyond transition state discovery based on saddle point definitions, ASTRA can be adapted to explore rare event regions more broadly, for example through uncertainty-aware guidance~\cite{tan2025enhanced, koulischer2025feedback} that targets configurations with high predictive uncertainty under the generative model or MLIPs. Such a strategy can be used for autonomous data acquisition and self-refinement steps in the active learning framework to improve the models.
More broadly, the principle of SBI and SAA in generative modeling may extend beyond chemistry and offer a foundation for rare-event discovery in complex dynamical systems.

\section{Methods}\label{sec3}
\subsection{Score-Based Interpolation of Diffusion Models for Initial Guess}
Our objective is to infer the TS, given short MD data of each metastable state. 
In particular, our aim was to sample the TS from a generative model that learned only the respective distribution of each metastable state, without any knowledge of the TS region and without further training on TS structures. 
We model the metastable distributions with a single system-specific diffusion model.
In particular, following the denoising score matching framework~\cite{song2020score}, we formulate a mapping between the target data distribution $p_0(\mathbf{x}_0)$ and a simple Gaussian prior $p_T(\mathbf{x}_T)\sim\mathcal{N}(0,\mathbf{I})$, through a Stochastic Differential equation:
\begin{equation}
  d\mathbf{x} = f(\mathbf{x}, t)dt + g(t)d{\mathbf{w}},
\end{equation}
where $d\mathbf{w}$ is a Wiener process, $f(\mathbf{x}, t)$ is the drift coefficient, and $g(t)$ is the diffusion coefficient. The reverse process can be written
\begin{equation} \label{eq:reverse}
  d\mathbf{x} = [f(\mathbf{x}, \tau) - g(\tau)^2 \nabla_{\mathbf{x}} \log p_\tau(\mathbf{x})]d\tau + g(\tau)d\bar{\mathbf{w}},
\end{equation}
where $\bar{\mathbf{w}}$ is a reverse Wiener process, and the reverse time $\tau \in [T, 0]$~\cite{song2020score}.
The model learns $s_\theta(\mathbf{x}, t)\approx\nabla_\textbf{x}\text{log}p_\tau(\textbf{x})$ so that by initializing the reverse SDE with noise $\mathbf{x}_\tau \sim \mathcal{N}(0, I)$ and solving it numerically, we can sample from the data distribution after training.
The training is generally self-supervised and requires only nuclear positions.
However, we require the diffusion model to additionally be conditioned on each metastable state to sample from the transition region between each state.
Hence, we annotate each data point with its metastable state class $c$ and train a denoising score matching model with Classifier-Free Guidance (CFG)~\cite{ho2022classifier}.
The labeling is trivial as the training data is generated from two separate molecular dynamics simulations for systems without any pre-existing trajectory data.
The scores are then parametrized by the CFG-based score $s_{\theta}^{\text{c}}(\mathbf{x}, t)\approx \nabla \log p_t(\mathbf{x} | c)$, given by the extrapolation
\begin{equation} \label{eq:cfg}
s_{\theta}^{\text{c}}(\mathbf{x}, t) = s_\theta(\mathbf{x}, t, \emptyset) + \gamma (s_\theta(\mathbf{x}, t, c) - s_\theta(\mathbf{x}, t, \emptyset)),
\end{equation}
where $c$ is the condition, replaced by a null token $\emptyset$ for the unconditional score, and $\gamma$ is the hyperparameter of the guidance scale that controls the conditioning strength. 
We trained the diffusion models with the Denoising Diffusion Probabilistic Models\cite{ho2020denoising} (DDPM) framework.

Recent methods have proposed various approaches to steer the diffusion sampling process at inference time\cite{skreta2025feynman,singhal2025general}. We are particularly interested in sampling an intermediate region between two metastable states while producing structures lying on the data manifold. The Superposition of Diffusion models \cite{skreta2024superposition} (SuperDiff) proposed a principled way to sample the surface of equal density under two conditional models denoted $A$ and $B$, where the equal density surface is sampled by following the reverse SDE while respecting
\begin{equation}\label{eq:iso}
    d\log p^A_t(\mathbf{x}) = d\log p^B_t(\mathbf{x}).
\end{equation}

Our key intuition is that the TS region overlaps with this equal density region of two conditional generative models trained on the respective metastable states MD sampling. We formulate this inference steering following the SuperDiff AND operator. The reverse diffusion process is biased to follow along the target equal probability surface by interpolating between the two conditional scores
\begin{equation} \label{eq:interp}
  s_{\text{SBI}}(\mathbf{x}, t) = s_{\theta}^{B}(\mathbf{x}, t) + \kappa (s_{\theta}^{A}(\mathbf{x}, t) - s_{\theta}^{B}(\mathbf{x}, t)),
\end{equation}
where the interpolation weight $\kappa$ controls the contribution of each state's score.
To obtain the $\kappa$ that follows the path dictated by equation \ref{eq:iso}, the change of probability densities can be estimated with an Itô density estimator. One can then obtain the value of $\kappa$ at each time step by solving a simple system of linear equations. We refer to the original SuperDiff paper for details on the derivation and implementation \cite{skreta2024superposition}.

We get our initial guesses following the reverse diffusion process of equation \ref{eq:reverse} with the interpolated score of equation \ref{eq:interp}. We evaluate this method against Score Averaging (SA) that provides a simpler baseline that directly averages the two conditional scores, which corresponds to setting $\kappa=0.5$~\cite{liu2022compositional}.

\subsection{Score-Aligned Ascent for Saddle Point Search}
In transition state theory, TSs are first-order saddle points on a multi-dimensional PES.
Locating a TS corresponds to an optimization problem, where the energy needs to be maximized along the given transformation, often referred to as the reaction coordinate $\mathbf{r}$, and minimized in all orthogonal directions.
This is achieved by decomposing the negative gradient of the energy with respect to all nuclear coordinates $-\nabla U(\mathbf{x})$ into components parallel and perpendicular to the reaction coordinate.
The parallel component is inverted to create an ascent force that pushes the system energetically upward along $\mathbf{r}$, while preserving the original components that minimize the energy in all orthogonal directions. 
This operation yields the ideal ascent force
\begin{equation}
    \mathbf{F} = -\nabla U(\mathbf{x}) + 2(\nabla U(\mathbf{x}) \cdot \mathbf{r})\mathbf{r}/\|\mathbf{r}\|_2^2.
\end{equation}
The key challenge of this method lies in the definition of $\mathbf{r}$. For example, existing approaches such as the Dimer algorithm~\cite{henkelman1999dimer, olsen2004comparison, heyden2005efficient, kastner2008superlinearly, shang2010constrained}, find local approximations to $\mathbf{r}$ by locating the direction of least curvature with gradient calculations of two close-lying configurations on the PES. We detail existing methods reproducing this kind of dynamics in section \ref{TS characterization reference} that serve as baselines for our approach.

We propose to leverage the ability of diffusion models to emulate these dynamics with the Score-Aligned Ascent (SAA). In practice, towards the end of the reverse diffusion process steered with SBI, we pause the denoising and optimize with SAA. We find that, at this low noise level, an approximate reaction coordinate $\mathbf{r}_\text{SAA}$ can be dynamically formulated at each step by taking the difference of the scores conditioned on each state $A$ and $B$
\begin{equation}\label{saa_rc}
    \mathbf{r}_\text{SAA}=s_{\theta}^B(\mathbf{x}, t) - s_{\theta}^A(\mathbf{x}, t).
\end{equation}
This vector points from one metastable state to the other (see Appendix~\ref{abl:val_sda}) and provides an efficient ascent direction.
Then, by calculating the physical force $-\nabla U(\mathbf{x})$ that produced the training data, one computes the SAA force
\begin{equation}\label{saa_equation_force}
    \mathbf{F}_{\text{SAA}} = -\nabla U(\mathbf{x}) + 2(\nabla U(\mathbf{x}) \cdot \mathbf{r}_\text{SAA})\mathbf{r}_\text{SAA}/\|\mathbf{r}_\text{SAA}\|_2^2.
\end{equation}

There are two main objectives when performing the SAA dynamics at a low latent level.
First, because of its sensitivity to small topological changes, the force field must be evaluated on clean structures, i.e. configurations on the data manifold. 
Therefore, we estimate the denoised data point $\hat{\mathbf{x}}_t$ for a configuration $\mathbf{x}_t$ from the posterior mean $\hat{\mathbf{x}}_t\equiv\mathop{\mathbb{E}}_{p(\mathbf{x}_0|\mathbf{x}_t)}[\mathbf{x}_0]$ with
\begin{equation}
    \hat{\mathbf{x}}_t = \frac{1}{\sqrt{\bar{\alpha}_t}} \left( \mathbf{x}_t - \sqrt{1 - \bar{\alpha}_t} \mathbf{\epsilon}_\theta(\mathbf{x}_t, t) \right),
\end{equation}
where we follow the standard DDPM notations and write $\hat{\mathbf{x}}_t$ to emphasize the dependency of that estimate on time. 
However, this estimate deteriorates rapidly as the diffusion timestep increases, particularly for molecular systems. We address this limitation by performing SAA at a low noise level. Moreover, the score difference used as an approximation for the ascent direction is computed at the selected latent level. Therefore, there is a discrepancy between $\mathbf{r}_\text{SAA}(\mathbf{x}_t)$ and the forces $-\nabla U(\hat{\mathbf{x}}_t)$, as these two directions are not computed at the same latent level. To effectively minimize the impact of this structural mismatch in equation \ref{saa_equation_force}, we chose a lower noise level for SAA dynamics.

A second, equally important reason for pausing the denoising and not operating SAA directly on the data manifold is central to the success of the method. The ascent direction derived from equation \ref{saa_rc} can lead to non-physical structures, because the score difference provides no guarantee of movement along the true physical structure manifold. Nonetheless, this direction offers an efficient and meaningful pathway to connect two states, which can subsequently be projected back onto the manifold by denoising to $\tau=0$. By performing the dynamics at an intermediate latent level, we can correct such non-physical structures after SAA optimization by completing the denoising reverse process.
\subsection{ASTRA}

Summarizing the workflow, we first solve the reverse-time SDE with interpolated score up to a chosen fixed timestep close to the SDE endpoint to ensure both reliable force and score calculations. We then pause the reverse SDE process, and optimize with $\mathbf{\hat{F}}_{\text{SAA}}=-\nabla U(\mathbf{x}) + 2(\nabla U(\mathbf{x}) \cdot \mathbf{r_\text{SAA}})\mathbf{r_\text{SAA}}/\|\mathbf{r_\text{SAA}}\|_2^2$ to steer the configuration towards a TS. 
We then resume the reverse SDE sampling, which transfers the sample onto the data manifold and corrects non-physical configurations stemming from the non-physical ascent direction. We refer to Algorithm~\ref{alg:ASTRA_sampling} for the complete details of the method and its key hyperparameters.

\subsection{Evaluation and baselines}

Except for coarse-grained chignolin, all baseline and benchmark methods were evaluated with the same classical force field or electronic structure method that was used to generate the dataset and to perform SAA. A common force field reference ensures that the transition states and other reference structures targeted by each algorithm are consistent with the underlying PES, so that observed differences reflect methodological performance rather than differences between potentials. In that regard, we emphasize that the notion of 'true' TS is relative to the given force field used.

\subsubsection{Transition state characterization} \label{TS characterization reference}

There are two related but distinct definitions of a TS. In transition state theory, it is a first-order saddle point on the PES: a stationary point whose Hessian has exactly one negative eigenvalue, lying along the minimum energy path (MEP) as its maximum and a minimum in all orthogonal directions. The committor-based TSE, by contrast, is the $q=0.5$ isosurface, which is the set of configurations from which dynamics is equally likely to commit to either metastable state, and is governed by the free energy surface, reflecting both energetic and entropic contributions. The two coincide when a single dominant barrier captures the transition and entropic contributions are small, but can diverge in flexible high-dimensional systems. While ASTRA targets first-order saddle points, we evaluate the resulting samples in two complementary ways.

\paragraph{Saddle-point evaluation via Dimer.} For systems where evaluating the PES is tractable (e.g., alanine dipeptide and the DASA reaction), we run the dimer method~\cite{henkelman1999dimer} as implemented in \textsc{SCINE} ReaDuct~\cite{weymuth2024scine, readuct2024} from each ASTRA sample, and from baseline guesses obtained by linear and geodesic interpolation. The dimer is a Hessian-free, minimum-mode-following method that iteratively identifies the lowest-curvature mode from forces alone and climbs along it while relaxing all orthogonal degrees of freedom; several variants improve convergence and robustness~\cite{olsen2004comparison, heyden2005efficient, kastner2008superlinearly, shang2010constrained}. We initialize the dimer direction from a Hessian calculation rather than a random guess as in \textsc{SCINE} ReaDuct implementation~\cite{weymuth2024scine, readuct2024}. 
For alanine dipeptide we report convergence rate, average steps to convergence, and the average energy and RMSD differences between the initial guess and the final converged structure; for DASA, Dimer is used to confirm that converged structures are true first-order saddle points, and we focus our analysis on the structural coverage of the resulting TS conformers.  

\paragraph{Committor evaluation.} Across all chemical systems, we report estimated committor values $q(x)$, the probability that a trajectory starting from $x$ reaches one metastable state before the other. A TS satisfies $q(x)=0.5$. For alanine dipeptide and DASA we estimate $q(x)$ from multiple replicas of Langevin dynamics with velocities drawn from the Maxwell--Boltzmann distribution, after defining basin outlines in a dimensionality-reduced space and discarding configurations that fail a 2~ps stability check. For chignolin, the high dimensionality makes verification via saddle point intractable and ambiguous, as the PES contains many saddle points that are dynamically irrelevant. In this case, the committor is closer to the ground truth, and we use the pretrained ML committor model of Ref.~\cite{kang2024computing} to estimate committor values.

\subsubsection{TS sampling baselines}
We benchmark the quality of our TS guesses against traditional approaches. As a primary baseline, we employed the NEB~\cite{henkelman2000climbing,henkelman2000improved} method, a standard approach for locating transition states and minimum energy paths (MEPs) between known reactant and product states. NEB represents the pathway as a chain of discrete images connected by virtual springs.
The springs prevent all images from converging to one of the two minima while the energy of all images is minimized. This “nudging” drives the images towards the MEP, enabling accurate saddle point localization without explicit knowledge of the reaction coordinate. NEB has become widely applied in computational chemistry and materials science for exploring energy barriers and reaction mechanisms.
\backmatter

\bmhead{Acknowledgements}
H.L. acknowledges the Korea Institute for Advancement of Technology (KIAT). This paper was supported by a KIAT grant funded by the Korea Government (Ministry of Education)(P0025681-G02P22450002201-10054408, Semiconductor-Specialized University).
M.S. gratefully acknowledges the Mobility fellowship P500PN\_225736 from the Swiss National Science Foundation. Y.D. acknowledges the support from Cornell University.
The authors acknowledge MIT SuperCloud and Lincoln Laboratory Supercomputing Center for providing (HPC, database, consultation) resources that have contributed to the research results reported within this paper.

\newpage
\begin{appendices}
\section{Related Works}
\subsection{Machine Learning for Transition State Search}
Numerous machine learning approaches have been developed to study rare events and in particular transition search. Early works leveraged a dataset of accumulated transition states to train generative models~\cite{pattanaik2020generating} with diffusion models~\cite{duan2023accurate,kim2024diffusion} and flow matching~\cite{duan2025optimal}. However, to alleviate the requirement of creating a transition state dataset, later work aimed to leverage geometric optimization and energy evaluations during training to find transition paths and extract transition states from it~\cite{nam2025transferable,hait2025locating}. Ref.~\cite{wang2024generalized} also leveraged the metric induced by local molecular dynamics simulation to learn a generalized flow matching model, while Ref.~\cite{raja2025action} relied on the learned force field from pre-trained diffusion models with the Onsager--Machlup functional to approximate the transition path. 

Additionally, a large body of literature tackled the committor function estimation problem and estimate reaction rates directly. When transition path data are available, maximum-likelihood methods are straightforward to learn the committor function \cite{jung2019artificial,sun2022multitask,jung2023machine}. Ref.~\cite{khoo2019solving,li2019computing} directly solved the variational formulation of the Kolmogorov backward equation (KBE) with equilibrium data. Ref.~\cite{rotskoff2022active, kang2024computing} proposed active approaches to use the current estimate of committor function to improve sampling of transition state regions when solving the KBE equation. Ref.~\cite{li2022semigroup, strahan2023predicting, mitchell2024committor} instead considered the Feynman–-Kac formulation of the KBE and used self-consistency objectives to match the committor values across the domain and their expected values after time evolution. 


\subsection{Inference-time Control for Diffusion Models}
Diffusion models excel at generating high-quality data, spanning images, texts to molecular structures. Beyond sampling from the learned distribution, they can be guided to generate from modified target distributions that encode design objectives, constraints, or rewards. The early and well-known examples are classifier guidance and classifier-free guidance to sample from conditional distributions~\cite{ho2022classifier}. Later on, inference-time control has been extended to compositional, reward-tilted, annealed, equal density distributions and more~\cite{du2023reduce,skreta2024superposition,skretafeynman}. One popular branch of methods develop upon the heuristics and approximate guidance~\cite{chung2022diffusion}. Another branch of methods study exact guidance where expensive Monte Carlo (MC) estimations are required~\cite{lu2023contrastive}. Recently, sequential Monte Carlo methods have become relevant to reweight on path space to reduce the variance of inference-time control~\cite{wu2023practical,skretafeynman,he2025rne}.

\section{Background}
\subsection{Score-Based Diffusion Models}
Given a forward process that transforms a data distribution $p_0(\mathbf{x}_0)$ over a continuous time variable $t \in [0, T]$ into a known prior distribution $p_T(\mathbf{x}_T)$, typically the standard Normal distribution $\mathcal{N}(0, I)$, a diffusion model $s_\theta(\mathbf{x}, t)$ can be trained such that a reverse process transforms samples from the prior distribution to the data distribution~\cite{sohl2015deep}.
The model can learn the added noise in each time step $t$~\cite{ho2020denoising} or the gradient of the log-probability of the data distribution $\nabla_{\mathbf{x}} \log p_t(\mathbf{x})$, the so-called score function~\cite{song2019generative}.
Both training strategies can be related as discretizing the same continuous reverse process described by the stochastic differential equation (SDE)
\begin{equation}
  d\mathbf{x} = [f(\mathbf{x}, \tau) - g(\tau)^2 \nabla_{\mathbf{x}} \log p_\tau(\mathbf{x})]d\tau + g(\tau)d\bar{\mathbf{w}},
\end{equation}
where $f(\mathbf{x}, t)$ is the drift coefficient, $g(t)$ is the diffusion coefficient, $\bar{\mathbf{w}}$ is a reverse Wiener process, and the reverse time $\tau \in [T, 0]$~\cite{song2020score}.
Once the model $s_\theta(\mathbf{x}, t)$ is trained, it can be applied as a generative model by initializing the reverse SDE with noise $\mathbf{x}_\tau \sim \mathcal{N}(0, I)$ and solving it numerically.



\subsection{Classifier-Free Guidance}
Classifier-Free Guidance (CFG)~\cite{ho2022classifier} is a technique for conditional generation that trains a single diffusion model to handle both conditional and unconditional generation scenarios. 
The model is trained to predict the score of the data distribution, conditioned on the label $c \in \{C_1, C_2, ..., \emptyset\}$.
During training, the conditioning information is randomly dropped with a given probability $p_{drop}$.
This technique enables the single model to learn both the conditional score functions, $\nabla \log p_t(\mathbf{x} | c)$, for each label, and the unconditional score function, $\nabla \log p_t(\mathbf{x})$, over the entire data manifold.

During inference, the CFG-based score $s_{\theta}^{\text{c}}(\mathbf{x}, t)$ is given by interpolation
\begin{equation}
s_{\theta}^{\text{c}}(\mathbf{x}, t) = s_\theta(\mathbf{x}, t, \emptyset) + \gamma (s_\theta(\mathbf{x}, t, c) - s_\theta(\mathbf{x}, t, \emptyset)),
\end{equation}
where $s_\theta(\mathbf{x}, t, c)$ is the conditional score, $s_\theta(\mathbf{x}, t, \emptyset)$ is the unconditional score, where the condition $c$ is replaced by a null token $\emptyset$, and $\gamma$ is the guidance scale that controls the strength of conditioning.
This approach provides fine-grained control over conditioning strength without requiring separately trained conditional models.

\subsection{Superposition of Diffusion Models}
The Superposition of Diffusion Models (SuperDiff)~\cite{skreta2024superposition} provides a principled way to combine the score functions from multiple models, each trained on a different data distribution. 
Given two score models, $s_A$ and $s_B$, trained on distributions $p_A$ and $p_B$ respectively, a composite score function can be constructed by a weighted average:
\begin{equation}
  s_{\text{SBI}}(\mathbf{x}, t) = s_{\theta}^{\text{B}}(\mathbf{x}, t) + \kappa (s_{\theta}^{\text{A}}(\mathbf{x}, t) - s_{\theta}^{\text{B}}(\mathbf{x}, t)).
\end{equation}
The interpolation factor $\kappa(t)$ determines the nature of the composition, enabling the generation of samples from distributions that represent logical combinations of the base distributions. 
For example, setting $\kappa (t)$ based on the relative likelihoods of a sample under each model can produce samples from the union (OR) of the distributions, while setting $\kappa$ based on equal likelihoods can produce samples from the intersection (AND).
In the context of transition state sampling, we are interested in the latter and a $\kappa (t)$ that satisfies the equal density:
\begin{equation}
    d\log p^\text{A}_t(\mathbf{x}) = d\log p^\text{B}_t(\mathbf{x})
\end{equation}
can be found by solving a set of three linear equations~\cite{skreta2024superposition}.
As a comparison, we also include in our ablations Simple Averaging of the scores to interpolate and steer the diffusion process~\cite{liu2022compositional}.

\subsection{Local mode maximization}
Transition states are first-order saddle points on a multi-dimensional potential energy surface (PES).
Localizing a transition state corresponds to an optimization problem, where the energy needs to be maximized along the given transformation and minimized in all orthogonal directions.
If the mode of transformation $\mathbf{r}$, commonly referred to as reaction coordinate, is known, a gradient-based optimization to the transition state can be achieved with a sufficiently close start condition.
The optimization is driven by the step $\mathbf{\Delta x}_i$ that is determined by the gradient of the potential energy with respect to nuclear coordinates $\mathbf{g}(\mathbf{x})$
\begin{equation}
    \mathbf{\Delta x}_i = -\mathbf{g}(\mathbf{x}) + 2 \left(\mathbf{g}(\mathbf{x})^T \mathbf{r}\right) \mathbf{r}.
\end{equation}
However, the definition of $\mathbf{r}$ is non-trivial for a given chemical system.
Existing approaches, such as the Dimer algorithm~\cite{henkelman1999dimer, olsen2004comparison, heyden2005efficient, kastner2008superlinearly, shang2010constrained}, find local approximations to $\mathbf{r}$ by locating the direction of least curvature with gradient calculations for multiple close-lying configurations on the PES.

\section{Additional Details on Experiments}
\label{exp:explanation}
\subsection{Double Well Potential}
The double well potential is a simple potential with two distinct states. 
The analytical formula for this potential energy surface is given by: 
\begin{equation} 
    V(x) = 0.1x^4 - x^2 + 0.1x + 1.0 \label{eq:1d_pes} 
\end{equation}

\subsection{Müller-Brown Potential}
For the Müller-Brown potential, our methodology adheres to the training procedure detailed in Ref.~\citenum{raja2025action}. 
A score-based diffusion model was trained on the `tiny' subset of the provided dataset, which consists of 4,000 samples.
The two metastable states, A and B, are defined by the coordinate criteria y $<$ 20 and y $>$ 20, respectively.

The analytical surface of this system is defined as: 
\begin{equation} 
    V(x, y) = \sum_{i=1}^{4} A_i \exp\left[\alpha_i(x - a_i)^2 + \beta_i(x - a_i)(y - b_i) + \gamma_i(y - b_i)^2\right] \label{eq:muller_brown} \end{equation} with the parameters taken directly from the simulation code: 
\begin{align*} 
    A &= 10 \times \text{barrier} \times [-1.73, -0.87, -1.47, 0.13] \\ \alpha &= 10^{-2} \times [-0.39, -0.39, -2.54, 0.273] \\ a &= [48, 32, 24, 16] \\ \beta &= 10^{-2} \times [0, 0, 4.30, 0.23] \\ b &= [8, 16, 32, 24] \\ \gamma &= 10^{-2} \times [-3.91, -3.91, -2.54, 0.273] 
\end{align*} 
where `barrier` is a scaling factor for the height of the potential barrier, for which we used 1.0.

\subsection{Double Path Potential}
For the double path potential, the dataset was generated with Langevin Dynamics simulations, following the procedure outlined in Ref.~\citenum{raja2025action}. 
A score-based diffusion model was subsequently trained on a curated subset of these simulation data, with 4,000 samples.
State A and B are defined by the regions $x < 0$ and $x > 0$, respectively.

The analytical formula for the potential energy surface is given by: 
\begin{equation} 
    \begin{split} 
        V(x,y) = {} & 10 \left( 2 + \frac{4}{3}x^4 - 2y^2 + y^4 + \frac{10}{3}x^2(y^2 - 1) \right) \\ & + 7 \exp\left(-\frac{(x + 0.7)^2 + (y - 0.8)^2}{0.4^2}\right) \\ & + \exp\left(-\frac{(x - 1.0)^2 + (y + 0.3)^2}{0.4^2}\right) \\ & - 6 \exp\left(-\frac{(x + 1.0)^2 + (y + 0.6)^2}{0.4^2}\right) 
    \end{split} \label{eq:double_path} 
\end{equation}

\begin{figure}[!htb]
    \centering
    \begin{subfigure}[b]{0.48\textwidth}
        \centering
        \includegraphics[width=0.8\textwidth]{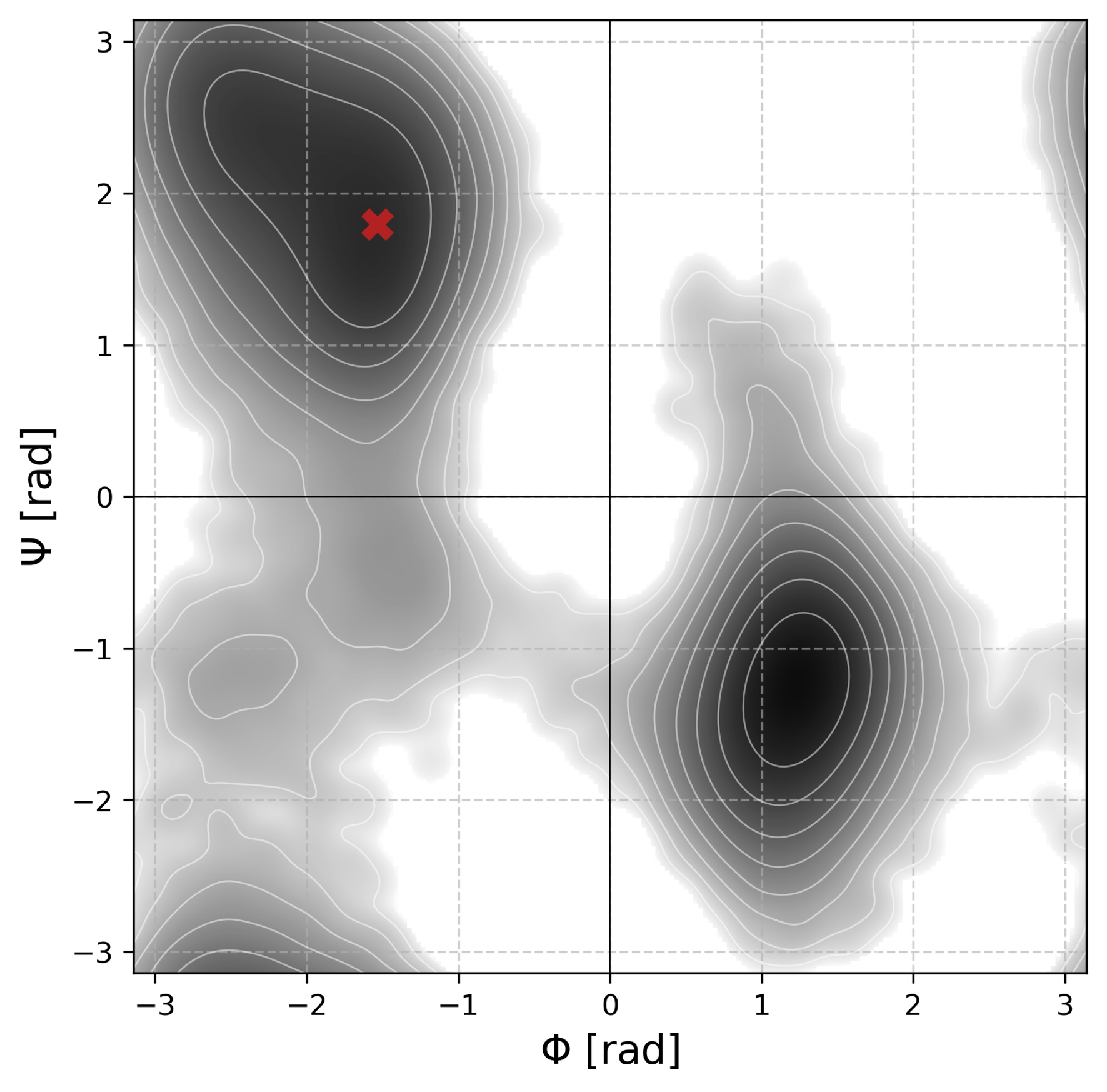}
        \caption{Alanine dipeptide.}
        \label{fig:abl:ala2_pes}
    \end{subfigure}
    \hfill 
    \begin{subfigure}[b]{0.48\textwidth}
        \centering
        \includegraphics[width=0.8\textwidth]{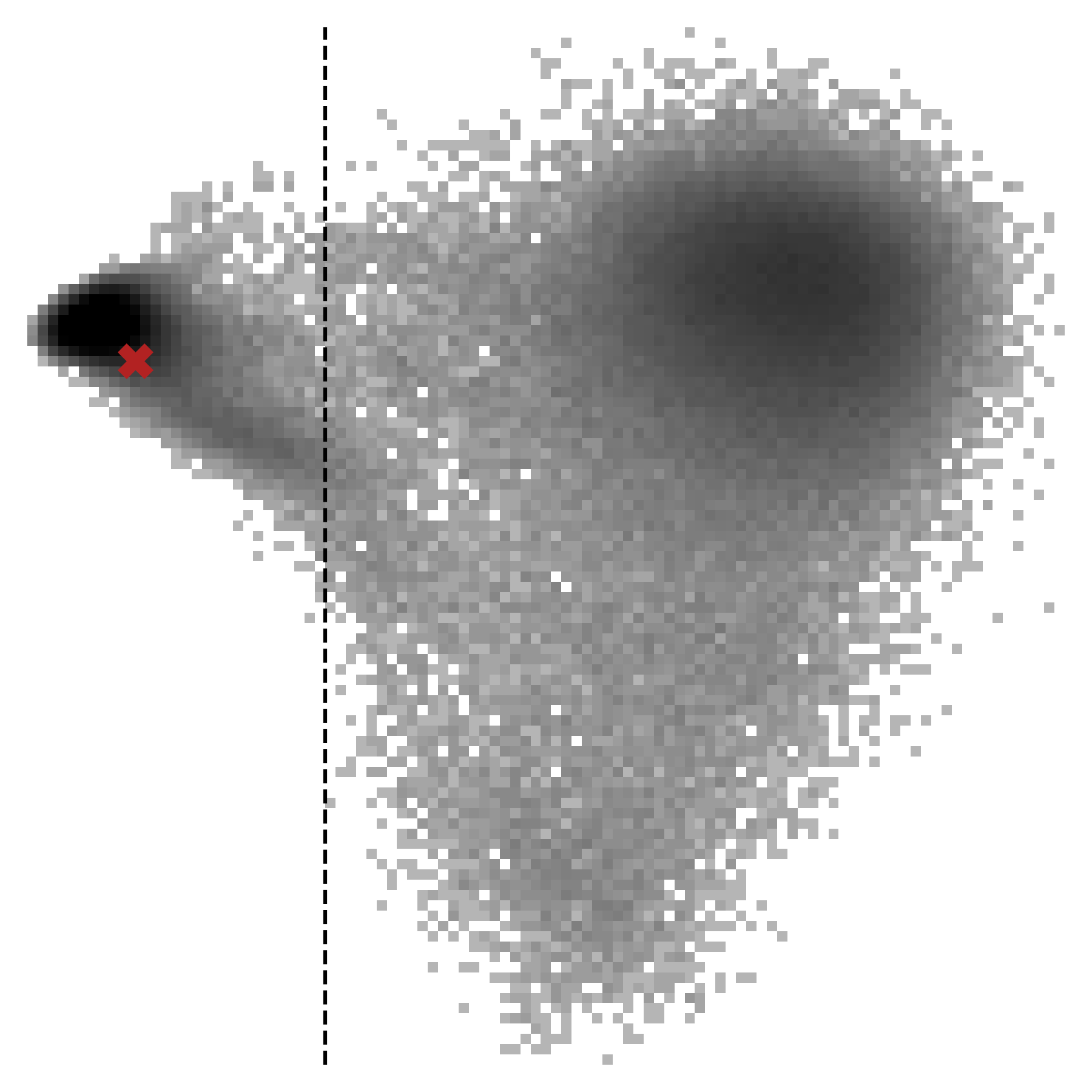}
        \caption{Chignolin.}
        \label{fig:abl:chig_pes}
    \end{subfigure}
    \caption{Potential Energy Surface (PES) of alanine dipeptide and chignolin.}
    \label{fig:abl:pes} 
\end{figure}

\subsection{Alanine Dipeptide}

\begin{figure}[!htb]
    \centering
    \begin{subfigure}[b]{0.32\textwidth}
        \centering
        \includegraphics[width=\textwidth]{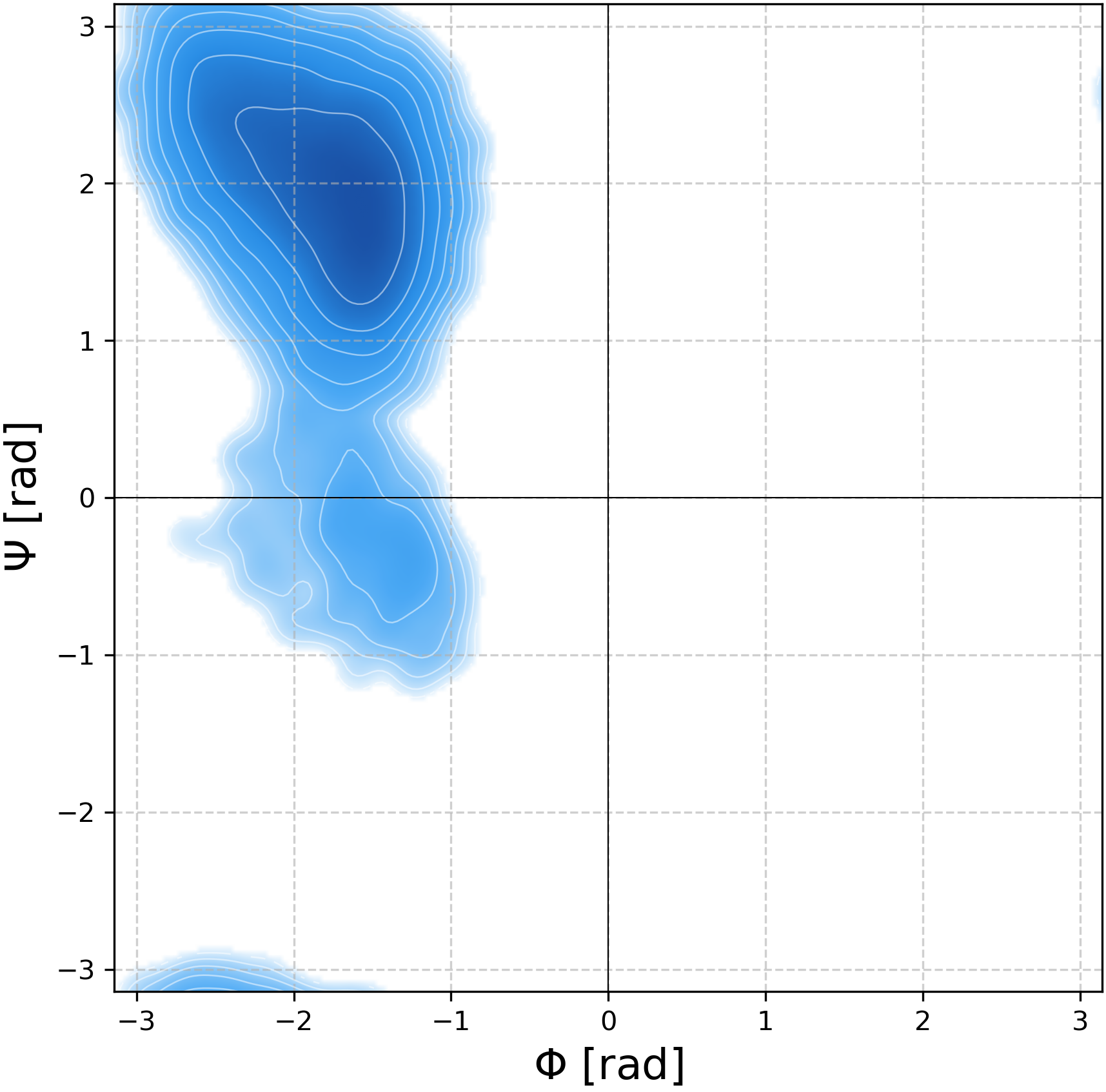}
        \caption{Conditioned on C5.}
        \label{fig:abl:ala2_pes_A}
    \end{subfigure}
    \hfill 
    \begin{subfigure}[b]{0.32\textwidth}
        \centering
        \includegraphics[width=\textwidth]{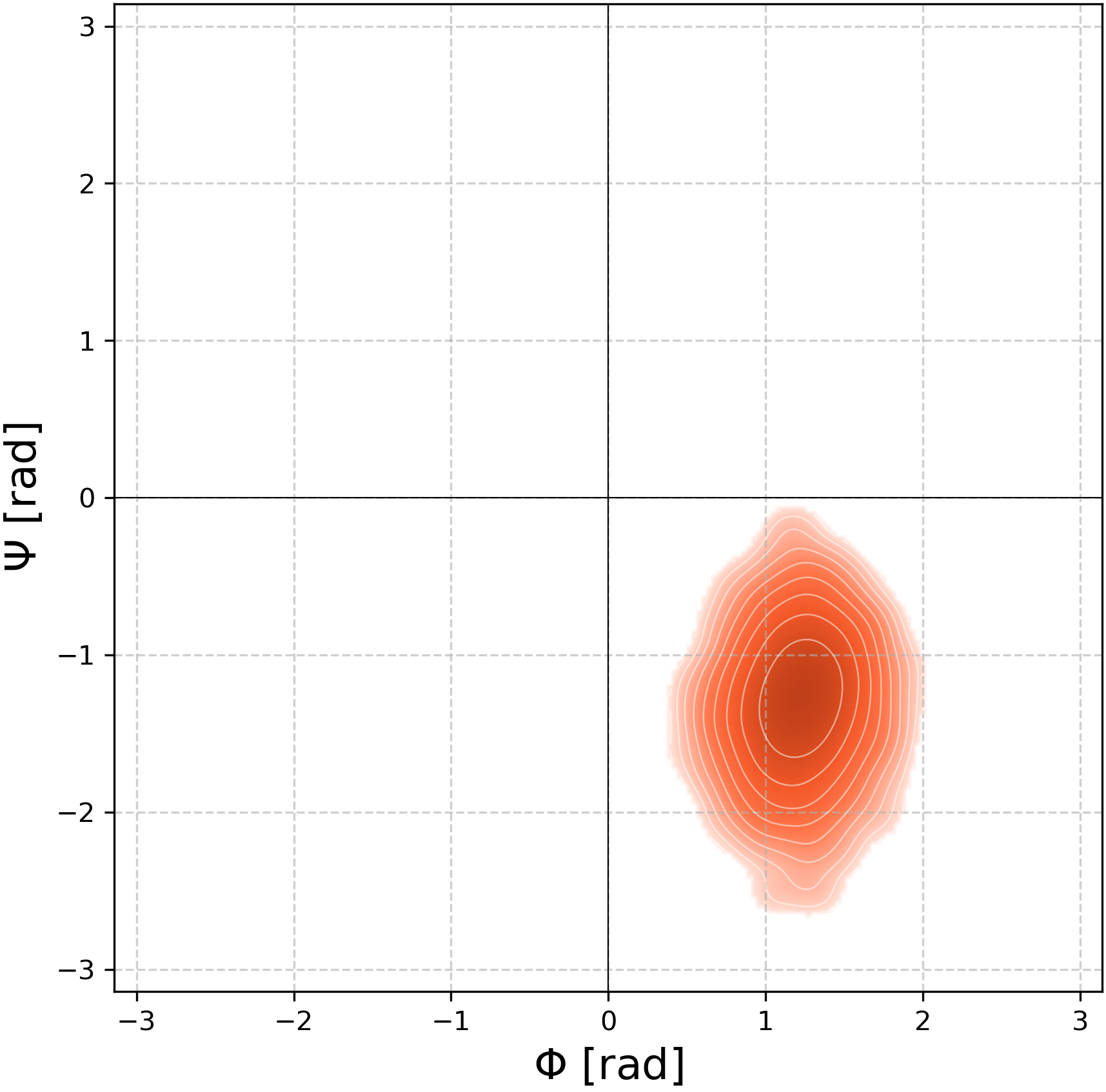}
        \caption{Conditioned on C7ax.}
        \label{fig:abl:ala2_pes_B}
    \end{subfigure}
    \begin{subfigure}[b]{0.32\textwidth}
        \centering
        \includegraphics[width=\textwidth]{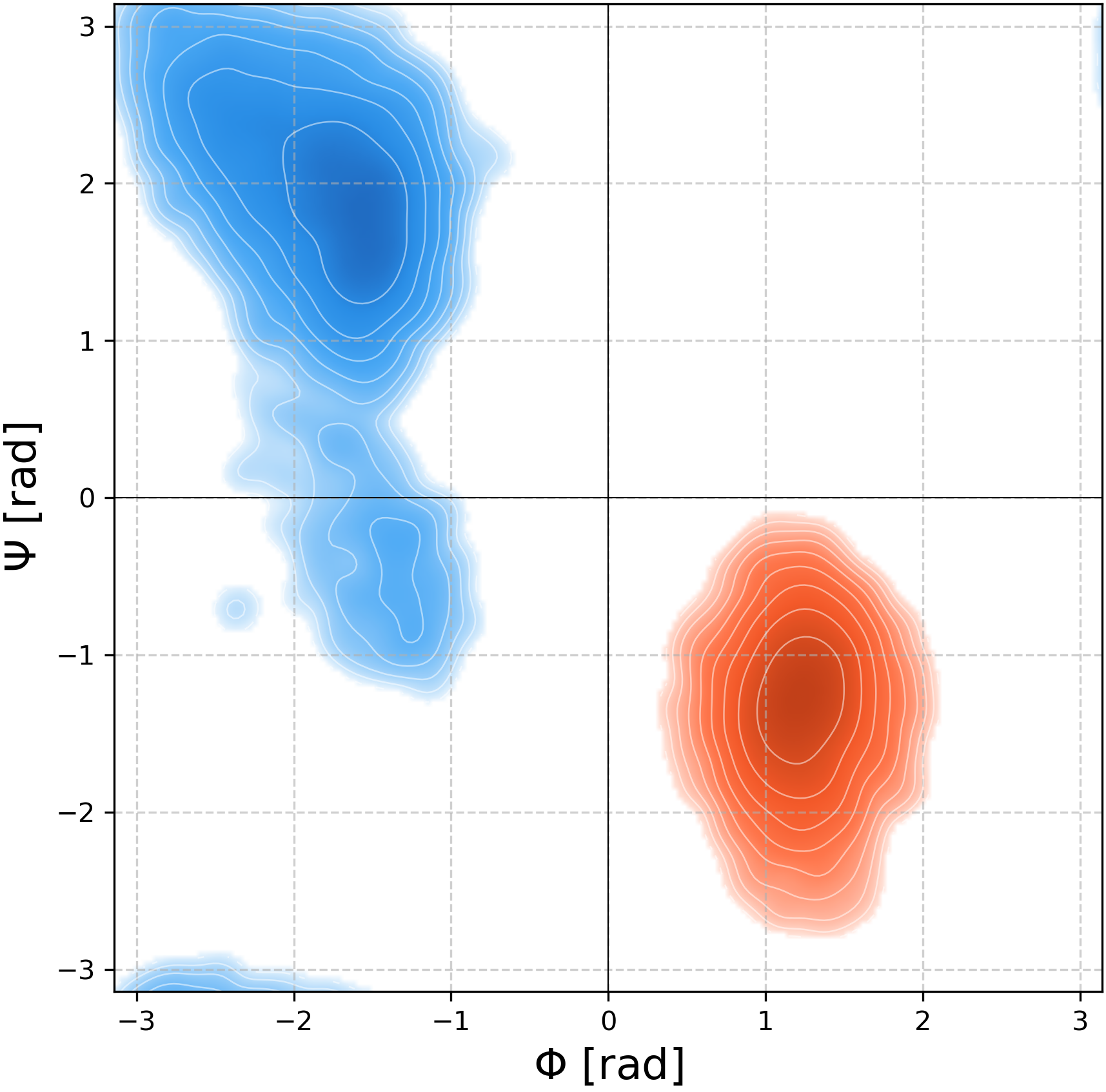}
        \caption{Unconditional.}
        \label{fig:abl:ala2_pes_uncond}
    \end{subfigure}
    \caption{Potential Energy Surface (PES) of alanine dipeptide sample from the trained diffusion model.}
    \label{fig:abl:pes_sample} 
\end{figure}

We generated the dataset by running Langevin Dynamics from each state (C5 and C7ax) for 500 ps with a 1 fs time step saving every 10 steps and a temperature of 300 K. 
We then sliced parts of the trajectory that covers the widest region in Ramachandran space $(\phi,\psi)$ and subsample randomly 2,500 configurations for each of the two states. 
In Figure~\ref{fig:abl:ala2_pes}, we show the PES of alanine dipeptide.
We propagated the dynamics with a custom force field taken from Ref.~\citenum{raja2025action}.
We then trained a diffusion model based on the EquiformerV2 architecture~\cite{liao2023equiformerv2} with up to $L=2$ representations, four attention heads and 64 channels. 
The radius graph is computed with $r_{cutoff}=5.0$ \AA. 
We use the AdamW optimizer with a constant learning rate $6.10^{-4}$, $0.001$ weight decay. 
We use an 0.999 Exponential Moving Average (EMA) decay with an effective batch size of $128$, and train for $500$ epochs.
For diffusion, we leverage the Denoising Diffusion Probabilistic Model framework (DDPM) with 1000 time steps and a cosine schedule~\cite{ho2020denoising}.
For SAA, we use 1000 optimization steps by Adam optimizer with a learning rate of 0.01, $\beta_1=0.0$, and $\beta_2=0.999$. 

The committor probabilities are computed from 100 configurations sampled from the method. 
We first run a 2 ps long simulation and filter out all the configurations leading to instability that are labelled as non physical. 
For all reported results, more than 98\% of samples were kept and used for the committor calculations. 
We compute the committor probabilities by running 100 replicas stochastic Langevin Dynamics with randomly initialized velocities using the same force field that generated the dataset to stay consistent with the PES distribution. 
The simulations are run for 1 ps with 1 fs steps at T=300 K. We define each state limits and record the region first reach for each replica to compute the probability of reaching one state or the other. 

\subsection{Chignolin}
For our analysis of Chignolin, which is a small fast-folding protein, we employed the D.~E.~Shaw Research coarse-grained (CG) dataset~\cite{lindorff2011fast}. 
In Figure~\ref{fig:abl:chig_pes}, we show the PES of the chignolin training dataset. 
We removed all the conformations with a ML committor value between 0.0001 and 0.9999 in the training dataset.
Our conditional diffusion model is an adaptation of the pre-trained architecture from Ref.~\citenum{arts2023two}. 
We finetuned this base model to enable conditional generation by incorporating a group embedding that specifies the state, i.e., folded or unfolded, at the beginning of the forward process. 
Sampling was then performed using SAA hyperparameters with 100 optimization steps, a learning rate of 0.0005, and a pause ratio of 0.05. 
Finally, the generated configurations were validated by computing their committor probabilities with the machine-learned committor predictor model from Ref.~\citenum{kang2024computing}.

\subsection{Electrocyclical reaction}

We generated the dataset by running Langevin Dynamics from the reactant and product structure each for 500 ps with a 0.5 fs time step saving every 10 steps and a temperature of 300 K.
We propagated the dynamics with the semi-empirical PM6 method~\cite{stewart2007optimization} calculated with \textsc{SCINE Sparrow}~\cite{Sparrow2024, bosia2023ultra}.
We then trained a diffusion model with the exact same procedure as before with alanine dipeptide.



\newpage
\section{Conformational diversity of the transition state ensemble for the electrocyclical reaction}

\begin{figure}[!htb]
    \centering
    \includegraphics[width=\textwidth]{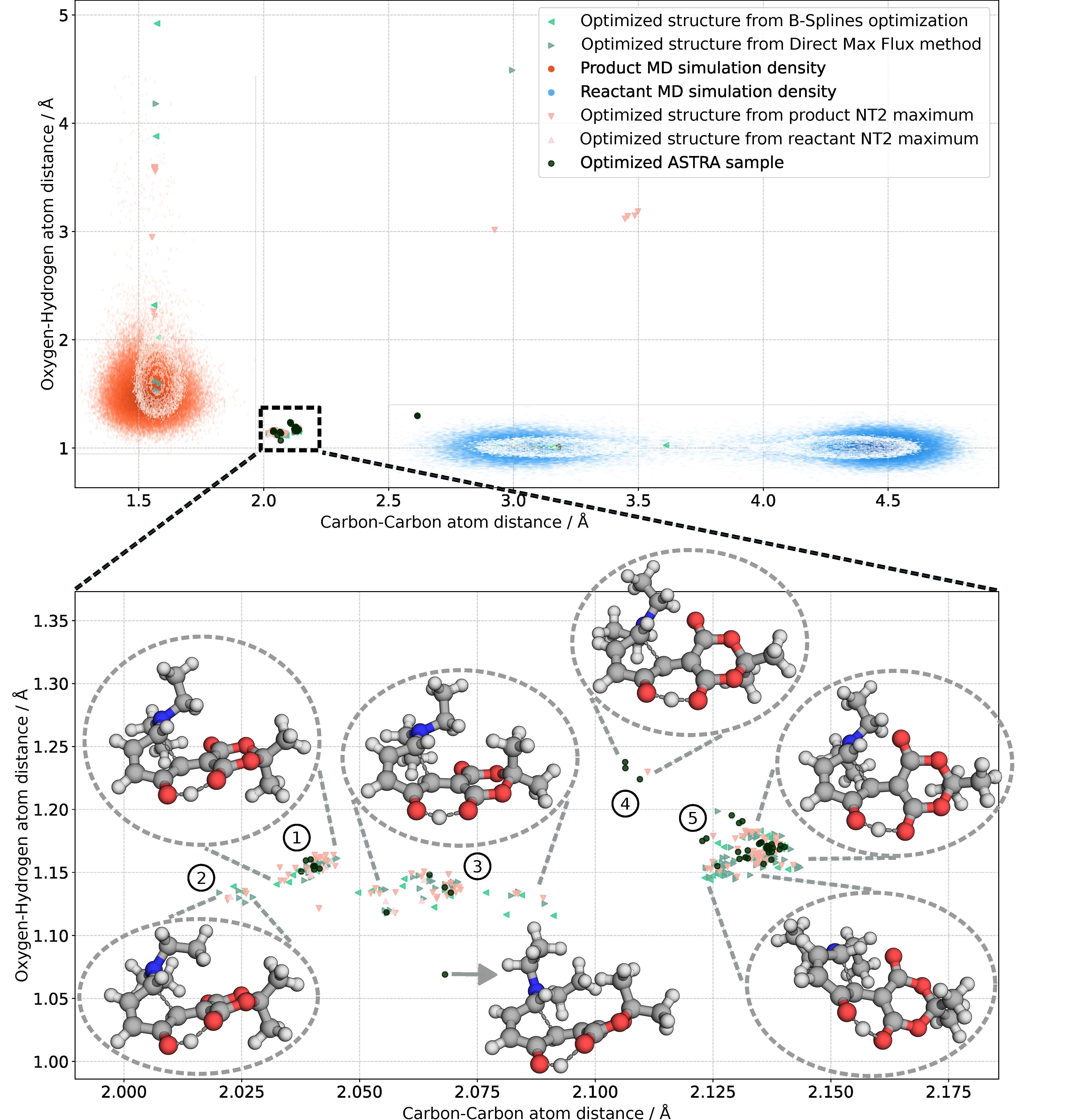}
    \caption{Two dimensional representation of the reactive potential energy surface (PES) spanned by two important internuclear distances. The MD simulation data points are plotted as densities. The various points represent structures optimized with the Dimer algorithm where the initial structure was generated with different TS guess structure generation algorithms. The various baseline methods are detailed in section \ref{sec:dasa_baselines}. The zoomed-in version of the PES features representative structures for each cluster of points that are numbered from 1 to 5.}
    \label{fig:si:dasa_analysis}
\end{figure}

To validate the structural diversity of ASTRA-generated samples for the DASA system (Figure \ref{fig:si:dasa_analysis}), we generated a large set of reference structures using multiple transition-state optimization methods with varying initializations. By comparing the ASTRA samples directly against these resulting reference clusters, we characterize the coverage and validity of our generative approach.

Notably, the only cluster missed by ASTRA is 2). It contains structures that continue a structural trend of an increasing dihedral angle and hence structural distortion going from the top right to the bottom left of cluster 1), hence, cluster 2) is an extreme version of cluster 1).
The lack of cluster 2) structures indicates that the distorted structure is disfavored in the diffusion process.
Due to this distortion, the chemical relevance of this cluster might be low as all structures in this cluster are higher in energy than the ASTRA generated structures in cluster 1). 
Beyond the coverage of each cluster, ASTRA can also capture a finer grained diversity within a certain cluster.
Cluster 5) mixes different methyl group orientations, with the right side of the cluster including structures with downward methyl orientation, while the left side of the cluster includes upward orientations.
We observe that ASTRA successfully captures this internal diversity of structures.
Similarly, cluster 3) has a large extent of possible carbon-carbon distances while preserving the other structural features, which is covered by ASTRA samples.
This demonstrates that our algorithm can recover physically valid variability of transitory configurations without prior knowledge or sampling of such ensembles due to our SBI and SAA sampling process.
Our ASTRA algorithm produced an additional structure outside of cluster 3) that exhibits similar structural features, but includes an asymmetrical position of the transferred proton.
This is missed by the other methods, but it also includes a strong distortion of the 6-membered ring and is high in energy.
The ASTRA algorithm also produced three out of four structures in cluster 4).
The structures of this cluster show an upward tilted left oxygen atom while the right 6-membered ring stays flat.
Structures within all clusters generally show different amine ethyl group orientations that aim at reducing steric hindrance.
Based on these findings, we are confident that the structural diversity of ASTRA samples observed in the smaller test systems before also extends to more complex chemical reactions.

For examples, the structures in cluster 1) exhibit a small dihedral angle of the transferred proton and the 6-membered ring and the geminal methyl groups are positioned downward. In comparison, cluster 5), the largest set of points, shows a strong tilt of the oxygen atoms transferring the proton with the left one tilted upwards and the right one downwards. Sampling with ASTRA encompasses this diversity of structural features typically encountered in traditional approaches with a diverse initialization.

\newpage
\section{Validation of Score Difference Approximation to Reaction Coordinate}
\label{abl:val_sda}
\begin{figure}[!htb]
    \centering
    \begin{subfigure}[b]{0.48\textwidth}
        \centering
        \includegraphics[width=\textwidth]{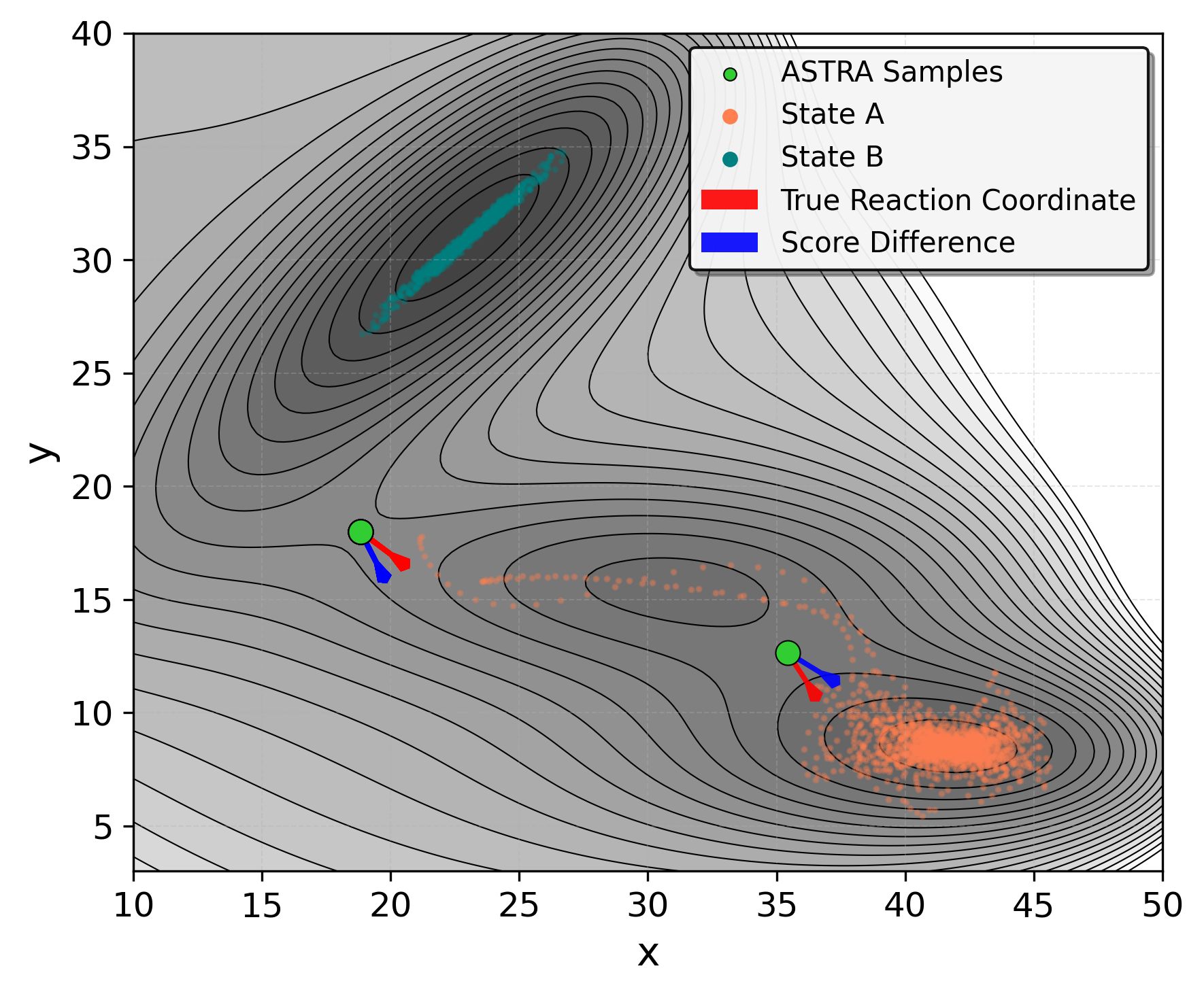}
        \caption{Müller-Brown potential.}
        \label{fig:abl:val_sda_mb}
    \end{subfigure}
    \hfill 
    \begin{subfigure}[b]{0.48\textwidth}
        \centering
        \includegraphics[width=\textwidth]{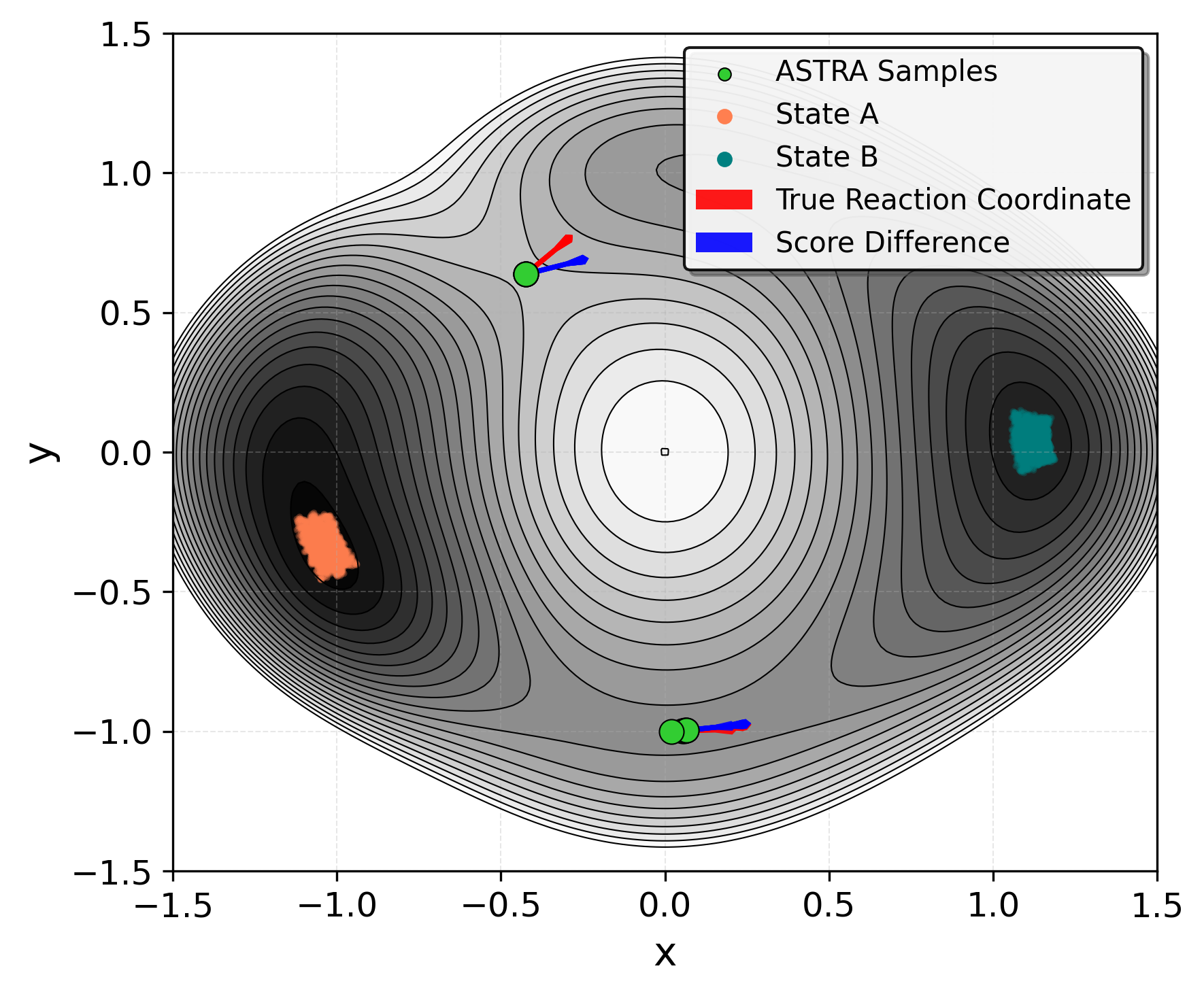}
        \caption{Double path potential.}
        \label{fig:abl:val_sda_dp}
    \end{subfigure}
    \caption{Validation of Score Difference Approximation. The score difference vector (red) closely aligns with the true reaction coordinate (blue) on the (a) Müller-Brown potential and (b) double path potential.}
    \label{fig:abl:val_sda} 
\end{figure}

To validate the claim that the score difference approximates the reaction coordinate, i.e., $s_{\text{A}}(\mathbf{x}, t) - s_{\text{B}}(\mathbf{x}, t) \approx \mathbf{r}$, we compare the two vectors in Figure~\ref{fig:abl:val_sda}. 
The visualization shows strong alignment between our approximation (blue arrows) and the true reaction coordinate (red arrows). 
This is quantified by the high average cosine similarity: 0.9275 on the Müller-Brown potential and 0.9584 on the double path potential.

\newpage
\section{ASTRA Sampling Algorithm}
\begin{algorithm}[h!]
\caption{Sampling with Score-Based Interpolation and Score-Aligned Ascent}
\label{alg:ASTRA_sampling}
\begin{algorithmic}[1]
    \Require
    Denoising model $\boldsymbol{\epsilon}_\theta(\mathbf{x}_t, t, c)$ with classifier-free guidance;
    Energy-based force field $\mathbf{F}(\mathbf{x})$;
    Total number of time steps $T$;
    Guidance pause time step $T_{\text{pause}}$;
    Number of optimization steps $N_{\text{opt}}$;
    Optimizer hyperparameters $\Theta_{\text{opt}}$.
    
    \Statex \textbf{Output:} Sampled molecular conformation $\mathbf{x}_0$.
    \Statex
    
    \Function{SBI\_Combine}{$\mathbf{x}_t, t$}
        \State Compute conditional noise estimates:
        $\boldsymbol{\epsilon}_t^A \gets \boldsymbol{\epsilon}_\theta(\mathbf{x}_t, t, A)$, $\boldsymbol{\epsilon}_t^B \gets \boldsymbol{\epsilon}_\theta(\mathbf{x}_t, t, B)$.
        \State Compute the interpolation factor:
        $\kappa_t \gets \text{SBI}(\boldsymbol{\epsilon}_t^A, \boldsymbol{\epsilon}_t^B, \mathbf{x}_t, t)$.
        \State Combine noise estimates: $\boldsymbol{\epsilon}_{\text{comb}} \gets \boldsymbol{\epsilon}_t^B + \kappa_t (\boldsymbol{\epsilon}_t^A - \boldsymbol{\epsilon}_t^B)$.
        \State \Return $\boldsymbol{\epsilon}_{\text{comb}}, \boldsymbol{\epsilon}_t^A, \boldsymbol{\epsilon}_t^B$
    \EndFunction
    \Statex

    \State Initialize positions $\mathbf{x}_T \sim \mathcal{N}(0, \mathbf{I})$.
    \Statex \textit{// Phase 1: Denoising with Score-Based Interpolation}
    \For{$t \gets T, T_{\text{pause}} + 1$}
        \State $\boldsymbol{\epsilon}_{\text{comb}}, \_, \_ \gets \text{SBI\_Combine}(\mathbf{x}_t, t)$
        \State Perform one reverse diffusion step to obtain $\mathbf{x}_{t-1}$ from $\mathbf{x}_t$ and $\boldsymbol{\epsilon}_{\text{comb}}$.
    \EndFor
    
    \Statex
    \Statex \textit{// Phase 2: Score-Aligned Ascent Optimization at latent time $T_{\text{pause}}$}
    \State Let $\mathbf{z} \gets \mathbf{x}_{T_{\text{pause}}}$. Initialize an optimizer $\mathcal{O}$ with $\Theta_{\text{opt}}$.
    \For{$k \gets 1, N_{\text{opt}}$}
        \State $\boldsymbol{\epsilon}_{\text{comb}}, \boldsymbol{\epsilon}_{T_{\text{pause}}}^A, \boldsymbol{\epsilon}_{T_{\text{pause}}}^B \gets \text{SBI\_Combine}(\mathbf{x}_{T_{\text{pause}}}, T_{\text{pause}})$
        \State Approximate the reaction coordinate: $\boldsymbol{\tau} \gets \boldsymbol{\epsilon}_{T_{\text{pause}}}^B - \boldsymbol{\epsilon}_{T_{\text{pause}}}^A$.
        \State Predict the clean sample: $\hat{\mathbf{x}}_0 \gets \text{predict\_x$_0$}(\mathbf{z}, T_{\text{pause}}, \boldsymbol{\epsilon}_{\text{comb}})$.
        \State Evaluate the force from the external field: $\mathbf{f} \gets \mathbf{F}(\hat{\mathbf{x}}_0)$.
        \State Compute the Score-Aligned Ascent force: $\mathbf{f}_{\text{SAA}} \gets \mathbf{f} - 2 \frac{\mathbf{f} \cdot \boldsymbol{\tau}}{\|\boldsymbol{\tau}\|^2_2} \boldsymbol{\tau}$.
        \State Update positions with gradient descent: $\mathbf{z} \gets \mathcal{O}(\mathbf{z}, \nabla_{\mathbf{z}}\mathcal{L})$, where $\nabla_{\mathbf{z}}\mathcal{L} = -\mathbf{f}_{\text{SAA}}$.
    \EndFor
    \State $\mathbf{x}_{T_{\text{pause}}} \gets \mathbf{z}$.
    
    \Statex
    \Statex \textit{// Phase 3: Resumed Denoising}
    \For{$t \gets T_{\text{pause}}, 1$}
        \State $\boldsymbol{\epsilon}_{\text{comb}}, \_, \_ \gets \text{SBI\_Combine}(\mathbf{x}_t, t)$
        \State Perform one reverse diffusion step to obtain $\mathbf{x}_{t-1}$ from $\mathbf{x}_t$ and $\boldsymbol{\epsilon}_{\text{comb}}$.
    \EndFor
    
    \Statex \Return $\mathbf{x}_0$.
\end{algorithmic}
\end{algorithm}

\newpage
\section{Extended analysis of ASTRA dynamics and robustness}\label{analysis_ASTRA}
\subsection{Modules characterization} \label{modules_characterization}
In this section we attempt to characterize and interpret the behavior of ASTRA. We base our discussions on the alanine dipeptide systems as its multiple and geometrically diverse transition pathways provide a challenging task. All ablation studies results can be found in Section~\ref{ablation_studies}. The ASTRA framework is composed of two fundamental modules, Score-Based Interpolation (SBI) for generating an initial guess of the transition ensemble and Score-Aligned Ascent (SAA) for its subsequent refinement. Therefore, we evaluate the efficacy and contribution of each individual components through ablation studies that are discussed using the committor and Transition State optimization metrics.

\subsubsection{Score-Based Interpolation}

We examine the effectiveness of the SBI module as an initialization strategy for SAA by replacing it with established transition state (TS) guess methods, and by comparing two distinct SBI-based approaches. Specifically, we benchmark our primary method—Isodensity Interpolation (II)—against alternative initialization techniques including Simple Averaging (SA), geodesic interpolation ~\cite{zhu2019geodesic}, and linear interpolation in Cartesian space.

From SBI-based approaches, ASTRA successfully recovers at least two out three TSs as shown in Figure \ref{fig:abl:ala2}. These results are in agreement with prior results obtained using our force field \cite{raja2025action}. While SAA shines in its convergence capabilities, SBI - as an informed initialization prior to SAA optimization - is the main feature allowing \textit{a priori} sampling of a diverse TS ensemble. In particular, by sampling across a broad region of the (approximate) iso-probability surface connecting the two metastable states—under their respective conditional generative models, II provides a diverse set of initial guesses that helps avoid premature convergence to a single dominant pathway. This allows to sample all three transition states.

Interestingly, this comprehensive sampling of the TS is only observed when employing a guidance scale of 1.0. This phenomenon is illustrated in Figure \ref{fig:abl:and_nosaa_guid} and Fig. \ref{fig:abl:sa_nosaa_guid}, where increasing the guidance scale from 1.0 to 3.0 progressively concentrates the TS guesses into the direct region separating the high-probability metastable states. This behavior is consistent with a stronger guidance driving the sampling toward more localized, higher-density regions: a reduction in directional diversity constrains the sampling trajectories to fewer dominant modes.
Notably, SA fails to generate initial guesses that converge to all three TSs under SAA, whereas II succeeds, albeit with the third TS (bottom-right region of the Ramachandran plot) being sampled less frequently (Figure \ref{fig:abl:and_nosaa_guid}). Although this third mode remains underrepresented, the fact that II recovers it at all highlights its potential to access transition channels missed by simpler interpolation strategies.
Looking ahead, further improving the diversity and breadth of PES coverage during initialization could enhance the robustness of the overall method, particularly in systems exhibiting complex, multi-pathway reaction dynamics.


To evaluate the advantage of SBI over traditional TS guess methods for initializing SAA, we compare it against linear interpolation and geodesic interpolation. In both cases, we construct a path of 25 intermediate structures and extract the five central images, which are then perturbed with latent noise at a PR = 0.05 level before being used as inputs to SAA (see Figure \ref{fig:abl:geo_abl}. To ensure fair comparison and assess the ability of each method to converge toward distinct TSs, we avoid random starting configurations and instead ensure that the interpolated paths pass sufficiently close to the true TS region. This allows us to sample all TSs.
Results presented in Table \ref{tab:geo_lin_neb} demonstrate that SAA initialized with both linear and geodesic interpolation can yield reasonable TS candidates, as supported by committor values and dimer convergence metrics. However, SBI consistently produces slightly more accurate TS guesses, as measured by lower RMSD and energy difference to the reference TS.
We highlight a key limitation of classical interpolation-based methods: they require prior structural knowledge of the TS region or the mechanism, to construct meaningful and diverse paths. In contrast, SBI operates without any such prior, generating diverse and physically plausible TS guesses directly from the conditional generative model. The broad coverage of the intermediate region between the two basins, facilitates the discovery of multiple valid transition states without prior access to the PES.

\subsubsection{Score-Aligned Ascent}\label{saa_analysis}
We proceed to analyzing the behavior of SAA by sweeping the hyperparameter space. The results are reported Table \ref{tab:hyperparameter_ablations} and \ref{tab:hyperparameter_ablations_committor}.

We start with the Pause Ratio (PR) that defines at which step is the reverse diffusion process halted to perform SAA before resuming the last few denoising steps. The PR is critical for the overall performance of the method. We define PR as $\frac{\text{Pause time step}}{\text{Total number of time steps}}$. The PR intrinsically manages the quality of the force field prediction, and the level of discrepancy between the SAA ascent direction and the structure it is optimized on, in terms of noise level. 
In fact, the force field should be evaluated on a clean structure and therefore we should use the posterior expectation as input. However the score difference defining the ascent direction should be evaluated at the latent level.
Consequently, whether the optimization is carried out in latent space or on the denoised structure introduces an inherent trade-off: in either case, a discrepancy arises between the domain where the ascent direction or the forces are computed and where they are applied. In fact, we observe in Figure \ref{fig:abl:noised_opt} nearly identical performances whether optimization is performed directly in latent space or on the denoised structure. In our implementation, we choose to perform the optimization on the denoised structure and subsequently project it back to the latent space corresponding to the PR level, using the same predicted score used to compute the posterior expectation. Then, the next score evaluation can be computed, and so on. 

In addition, we find that pausing the reverse diffusion at a nonzero latent noise level plays a critical role in correcting for the partially non-physical nature of the ascent direction in SAA, which can otherwise drive samples away from the true data manifold. This is a key feature of the method. Specifically, the difference between conditional scores yields a highly informative ascent direction, but one that is more reliable in latent space - consistent with prior observations that denoising score matching models tend to struggle at very low noise levels \cite{de2024target}. 
The strength of the method is that the outcome of this latent ascent can be effectively mapped back to the data manifold by simply resuming the reverse SDE from the pause time step to completion. In fact, we observe in Table F2 that computing SAA at $\tau=0$  is not numerically stable. Increasing the pausing ratio generally increases the number of samples that have physical structures. However, when pausing too far from the SDE endpoint, the distribution of committer values spreads out of the TS region range. 
We attribute this degradation partially to the aforementioned discrepancy between the structure on which the forces are computed and the structure on which the ascent direction is computed. We also hypothesize that the II scores may actively drive samples out of the narrow TS region at high pause ratios. Therefore, we find that there exists an optimal pause ratio around 0.05 that achieves the best balance between these constraints.  

When increasing the Optimization Step (OS), we observe a clear convergence, across all metrics,  demonstrating the stability and reliability of SAA. We note that the trends of both the committor analysis, and dimer optimization are consistent with one another. Our optimization algorithm reaches a plateau around 750 OS achieving about a 85\% convergence rate, 20 steps needed for convergence and guess TS structures as close as a 0.01\AA~and a -0.5$\text{kcal.mol}^{-1}$ energy difference away from the optimized ground truth TS.
Changing the number of optimization steps, we also identify a characteristic of SAA. In fact, we observe that SAA can drive samples out of the data manifold. As OS increases, the number of stable samples decreases before reaching a plateau of about 75\% of physical structures. As non-physical structures can be easily filtered out by running a short MD simulation from them, we report the \% of Committor value between 0.4 and 0.6, with respect to the number of stable (filtered) structures. We may in part interpret harsher SAA conditions as a filter here. As the distribution narrows around the TS regions, non promising TS guesses from SBI are driven too far away from the manifold for them to be corrected by the subsequent denoising. They can then be easily filtered out, achieving in the end an excellent quality of the TS guesses, of almost all structures.

The Step Size (SS) presents an optimum value of around 0.01 {\AA} corresponding to a standard tradeoff between a slow convergence, and an overshooting leading to non-physical structures. We emphasize that the filtered ASTRA samples still remain close to the true TS in harsh SS conditions, although there are logically very few of them.

\subsection{Baselines} \label{Baselines}

\subsubsection{Alanine dipeptide}
Our main baseline for alanine dipeptide is NEB, a well-established method for optimizing an MEP and locating the saddle point along it. Table~\ref{tab:geo_lin_neb} compares SAA against NEB starting from the same initial structures or paths obtained by linear and geodesic interpolation. 
For SAA, we retain the five middle points of each interpolated path as independent inputs. 
For NEB, we use the entire interpolated path (built from energy-minimized endpoints, as required to obtain a meaningful highest-energy image). 
The endpoint structures used to construct the interpolations are the same as in Section~\ref{saa_analysis}, and the resulting NEB-optimized paths are shown in Figure~\ref{fig:abl:geo_lin_opt_neb}. 

When its inputs are well behaved, NEB places the highest-energy image very close to a true saddle point. NEB-optimized images that subsequently converge under Dimer lie within $\sim 0.07$~Å RMSD of the reference TS. However, NEB's reliance on the input path quality is consequential. From minimized geodesic interpolation, NEB recovers both main TSs of alanine dipeptide. From minimized linear interpolation, by contrast, NEB converges to only one of the two main TSs (Figure~\ref{fig:abl:geo_lin_opt_neb}), as the linear path passes through high-energy, sterically clashing intermediates (Figure~\ref{fig:abl:ala2_basline}) that pull the band toward a single basin of attraction, and the optimization cannot escape this topology. SAA, by noising its input back into latent space and projecting it onto the data manifold via the remaining denoising steps, can correct such non-physical intermediates and recovers both TSs from the same linearly interpolated paths.  


We further observe that NEB is highly sensitive to whether the endpoints have been energy-minimized. Starting from non-minimized geodesic interpolation, almost all NEB images become unstable during optimization (only 5 of 56 stable; 0\% Dimer convergence; Table~\ref{tab:geo_lin_neb}), whereas SAA still recovers high-quality TS guesses (85\% Dimer convergence) under the same conditions.   

Together, these observations indicate that SAA is more robust than NEB both to quality of the initial interpolation (linear vs. geodesic) and to the quality of the endpoints (minimized vs. non-minimized). We emphasize that this is a comparison of the optimization step alone. The broader advantage of ASTRA lies one level up, in removing the need for any interpolation between user-chosen endpoints. ASTRA does not require minimized endpoint structures or pre-specified path interpolation, which allows broader coverage of the TS ensemble than is straightforward to obtain with NEB (Figure~\ref{fig:abl:geo_abl}). For NEB, achieving similar diversity typically requires running multiple calculations from different conformer pairs of the endpoints. ASTRA generates this diversity natively from the metastable-state diffusion model.


\subsubsection{Electrocyclic reaction}\label{sec:dasa_baselines}
We took multiple established TS generation methods as baselines for the DASA system to ensure that ASTRA covers all relevant transition states.
We ran a CREST simulation~\cite{pracht2020automated, pracht2024crest} with the default settings for both the reactant and product.
This resulted in 20 reactant and 162 product conformers.
We then carried out a Newton Trajectory 2 scan~\cite{unsleber2020exploration} for each reactant and product conformer with the associations and dissociations defined by the two oxygen-hydrogen bonds and the carbon-carbon bond within the formed 5-membered ring.
Additionally, we carried out double-ended searches.
We ran one double-ended search per conformer, where we selected the conformer of the other side to be the one with the lowest RMSD.
The double-ended search was then carried out with both the Direct Max Flux (DMF) method~\cite{koda2024locating} and with a B-Spline-based optimization~\cite{vaucher2018minimum}.
The initial trajectory was determined by geodesic interpolation~\cite{zhu2019geodesic} for DMF and with linear interpolation in Cartesian space for the B-Spline-based optimization.
All generated guess structures were subjected to the identical Dimer optimization algorithm with a maximum of 300 optimization steps.

\section{Ablation Studies}\label{ablation_studies}
\subsection{Effect of Score-Based Interpolation and Score-Aligned Ascent}
\label{abl:ib_saa}
\begin{figure}[htb!]
    \centering
    
    \begin{subfigure}{0.48\textwidth}
        \centering
        \includegraphics[width=\linewidth]{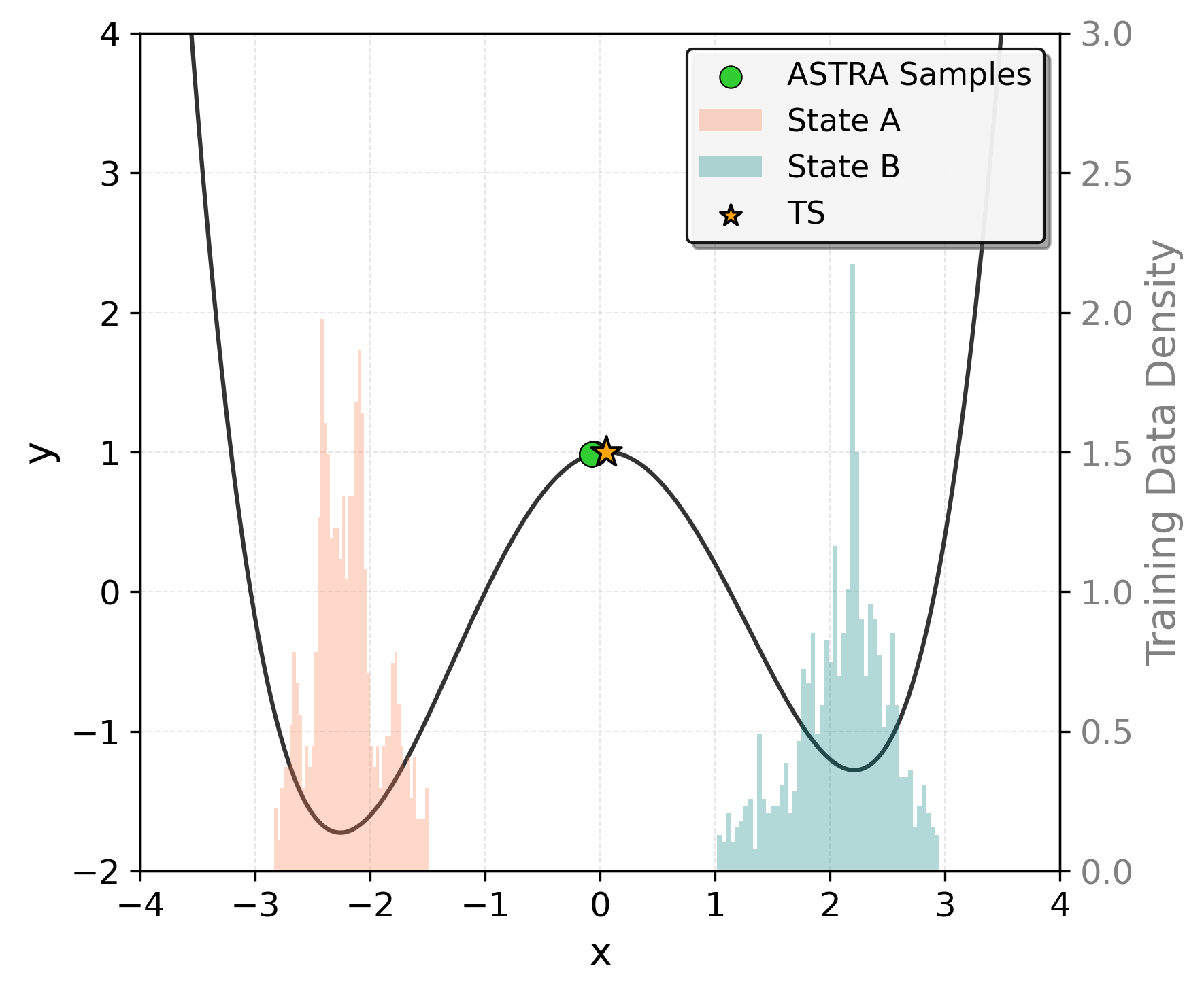}
        \caption{II.}
        \label{fig:abl:1d_ib}
    \end{subfigure}
    \hfill
    \begin{subfigure}{0.48\textwidth}
        \centering
        \includegraphics[width=\linewidth]{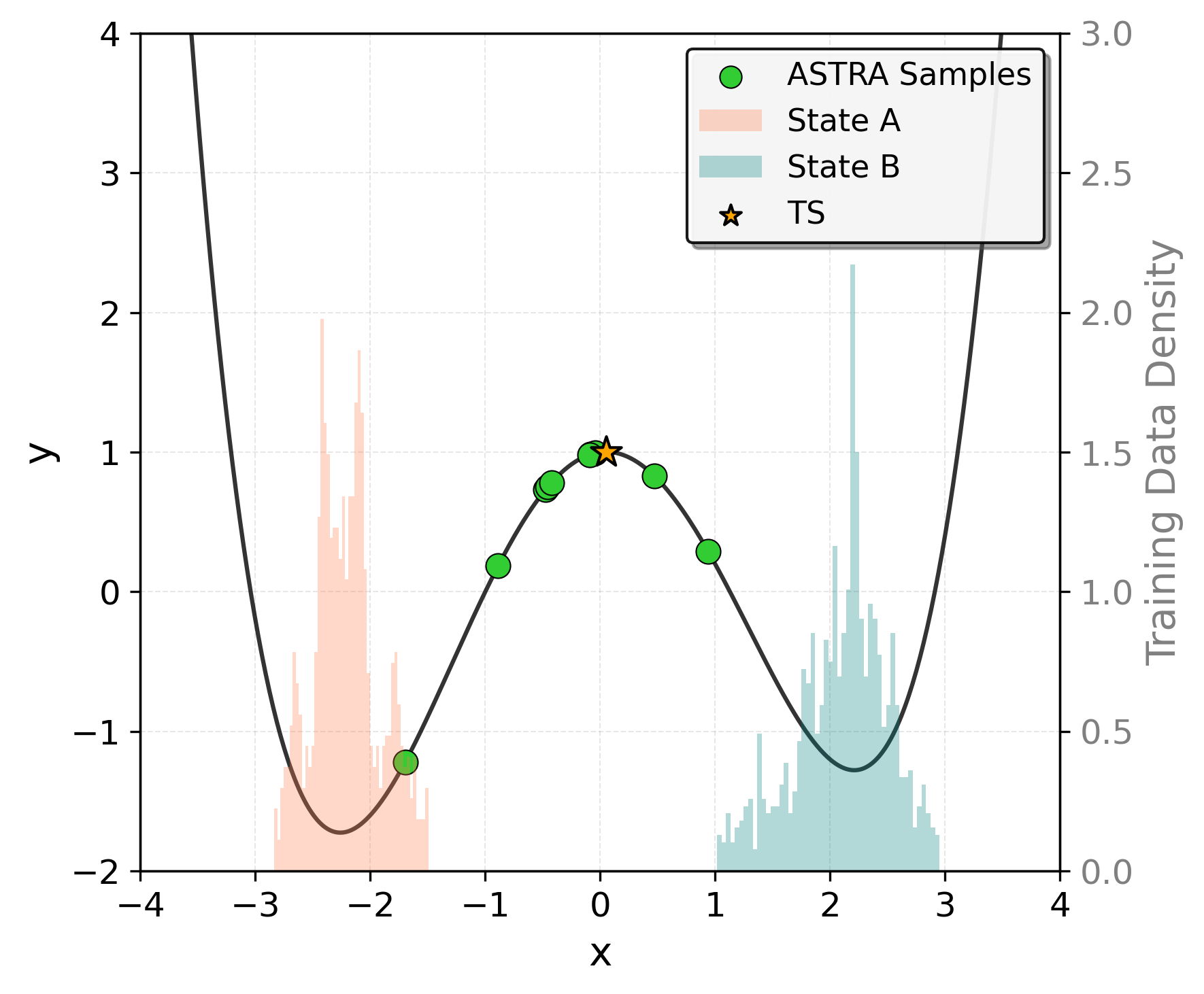}
        \caption{SA.}
        \label{fig:abl:1d_sa}
    \end{subfigure}
    
    \vspace{0.5cm} 

    \begin{subfigure}{0.48\textwidth}
        \centering
        \includegraphics[width=\linewidth]{figures/ablations/1d_ib+saa.png}
        \caption{II + SAA.}
        \label{fig:abl:1d_ib+saa}
    \end{subfigure}
    \hfill
    \begin{subfigure}{0.48\textwidth}
        \centering
        \includegraphics[width=\linewidth]{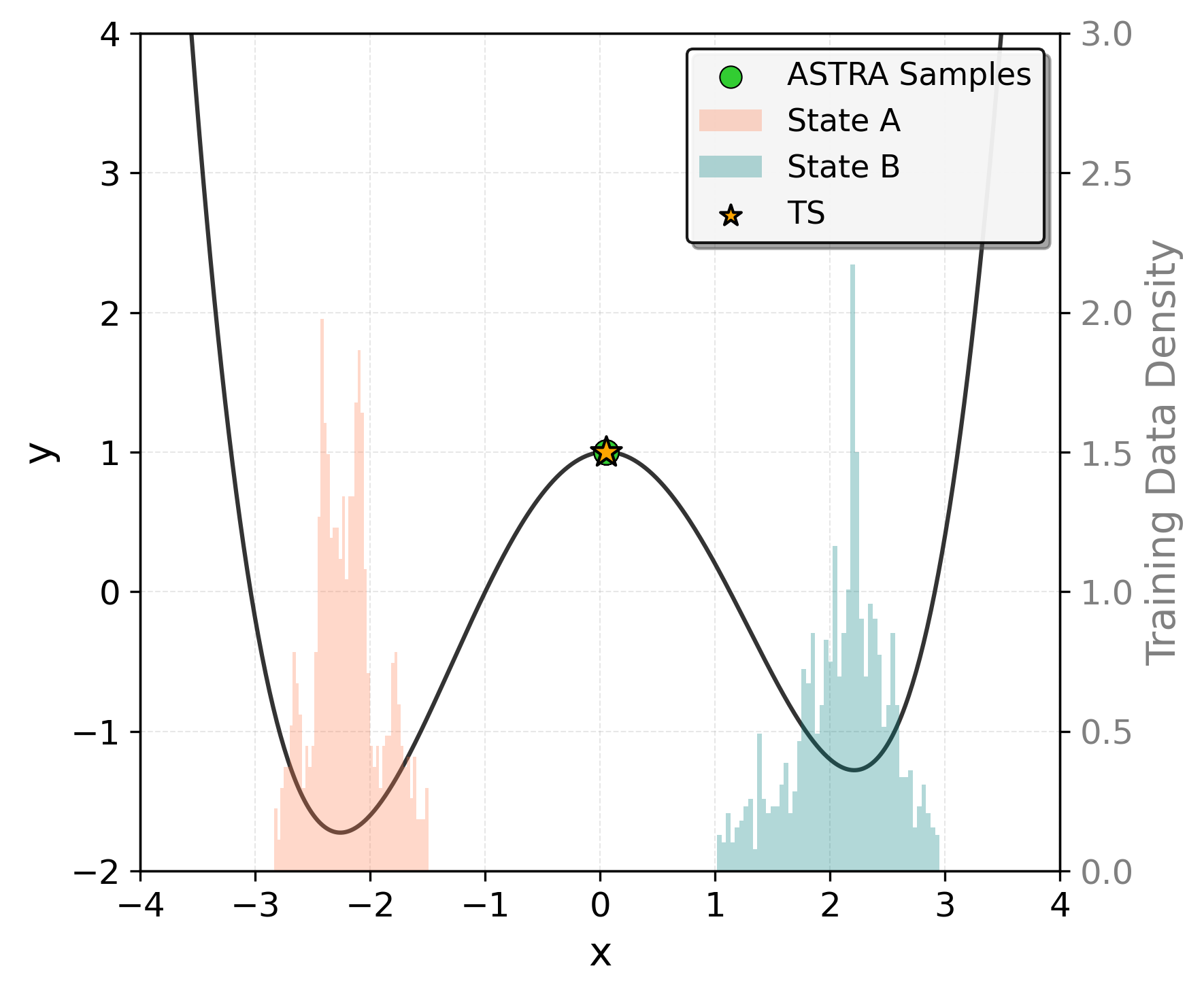}
        \caption{SA + SAA.}
        \label{fig:abl:1d_sa+saa}
    \end{subfigure}
    
    \caption{Impact of Score-Based Interpolation and Score-Aligned Ascent on double well potential.}
    \label{fig:abl:1d}
\end{figure}

We investigate the effects of two interpolation methods, \textbf{Isodensity Interpolation (II)} and \textbf{Simple Averaging (SA)}, both with and without the application of \textbf{Score-Aligned Ascent (SAA)}, on the double well potential, as shown in Figure~\ref{fig:abl:1d}. 
When used alone, SA fails to capture the true transition state, instead generating samples scattered between the two basins. 
In contrast, II produces samples closer to the true transition state. 
The subsequent application of SAA further refines these outcomes. 
For both interpolation methods, SAA guides the diffusion process toward the transition state (TS) region, correcting deviations and ensuring the final configurations closely align with the true TS. 
Notably, the combination of II and SAA yields the most accurate and physically meaningful transition path.

\begin{figure}[htb!]
    \centering
    
    \begin{subfigure}{0.48\textwidth}
        \centering
        \includegraphics[width=\linewidth]{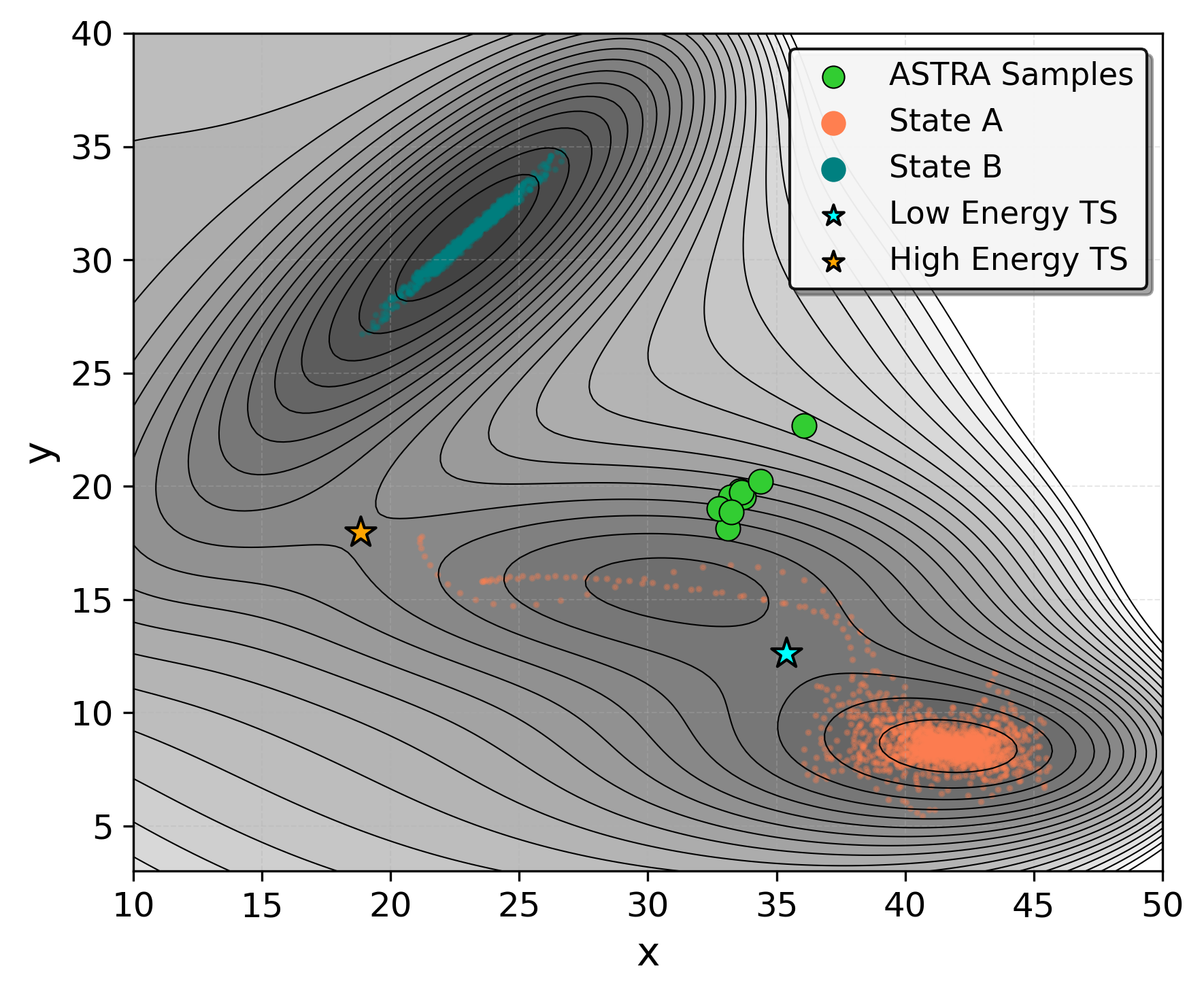}
        \caption{II.}
        \label{fig:abl:mb_ib}
    \end{subfigure}
    \hfill
    \begin{subfigure}{0.48\textwidth}
        \centering
        \includegraphics[width=\linewidth]{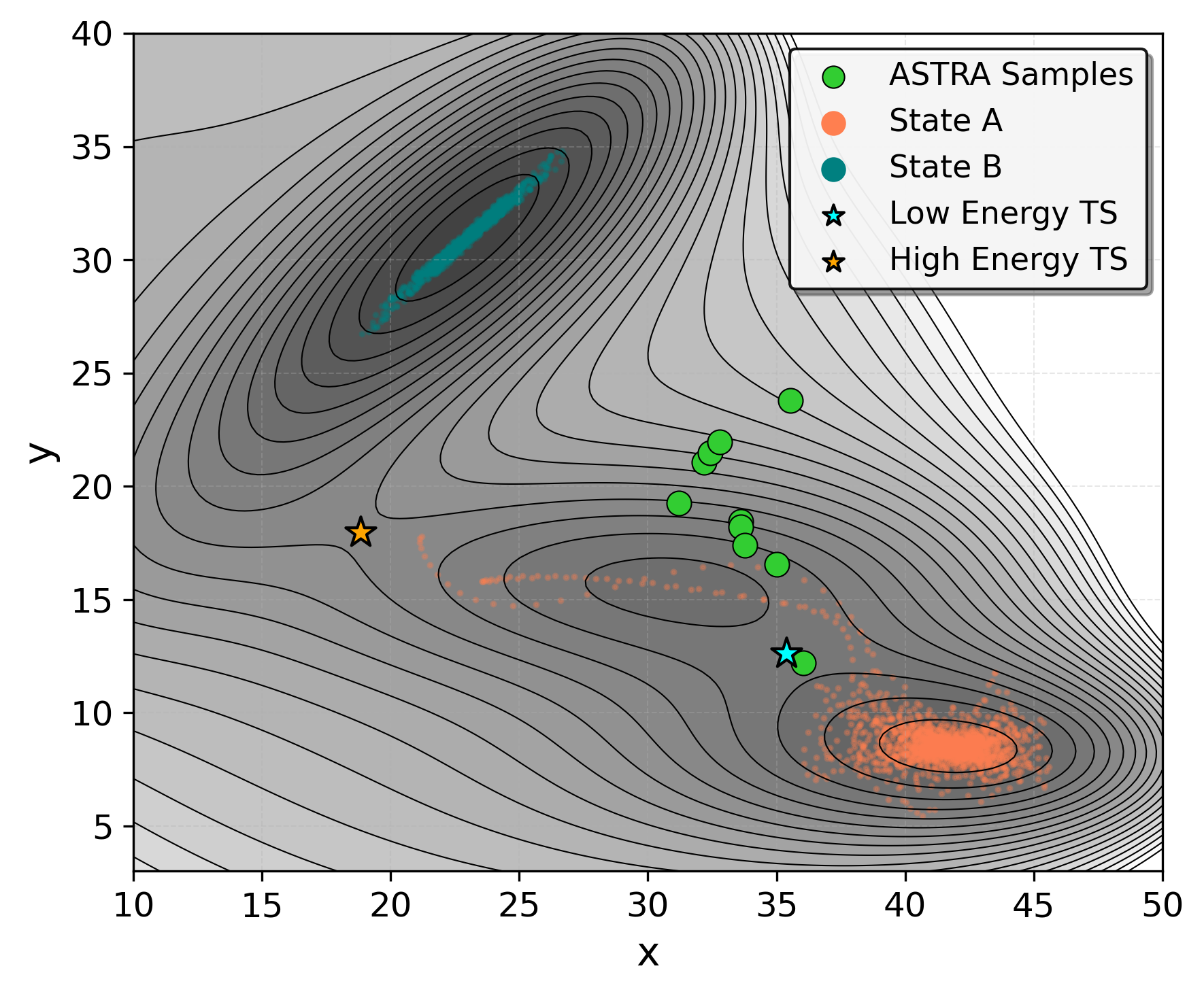}
        \caption{SA.}
        \label{fig:abl:mb_sa}
    \end{subfigure}
    
    \vspace{0.5cm} 

    \begin{subfigure}{0.48\textwidth}
        \centering
        \includegraphics[width=\linewidth]{figures/ablations/mb_ib+saa.png}
        \caption{II + SAA.}
        \label{fig:abl:mb_ib+saa}
    \end{subfigure}
    \hfill
    \begin{subfigure}{0.48\textwidth}
        \centering
        \includegraphics[width=\linewidth]{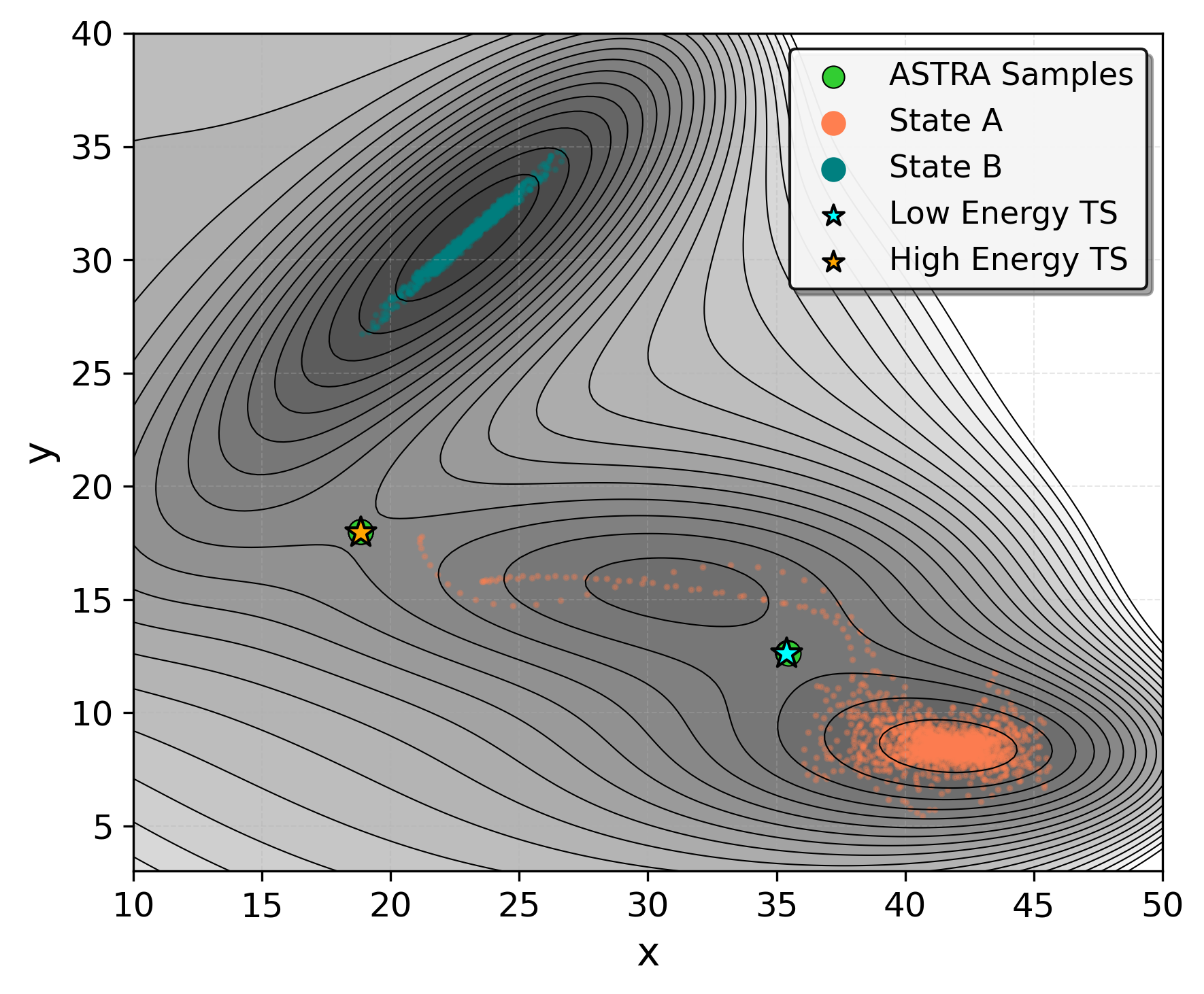}
        \caption{SA + SAA.}
        \label{fig:abl:mb_sa+saa}
    \end{subfigure}
    
    \caption{Impact of Score-Based Interpolation and Score-Aligned Ascent on Müller-Brown potential.}
    \label{fig:abl:mb}
\end{figure}

\begin{figure}[htb!]
    \centering
    
    \begin{subfigure}{0.48\textwidth}
        \centering
        \includegraphics[width=\linewidth]{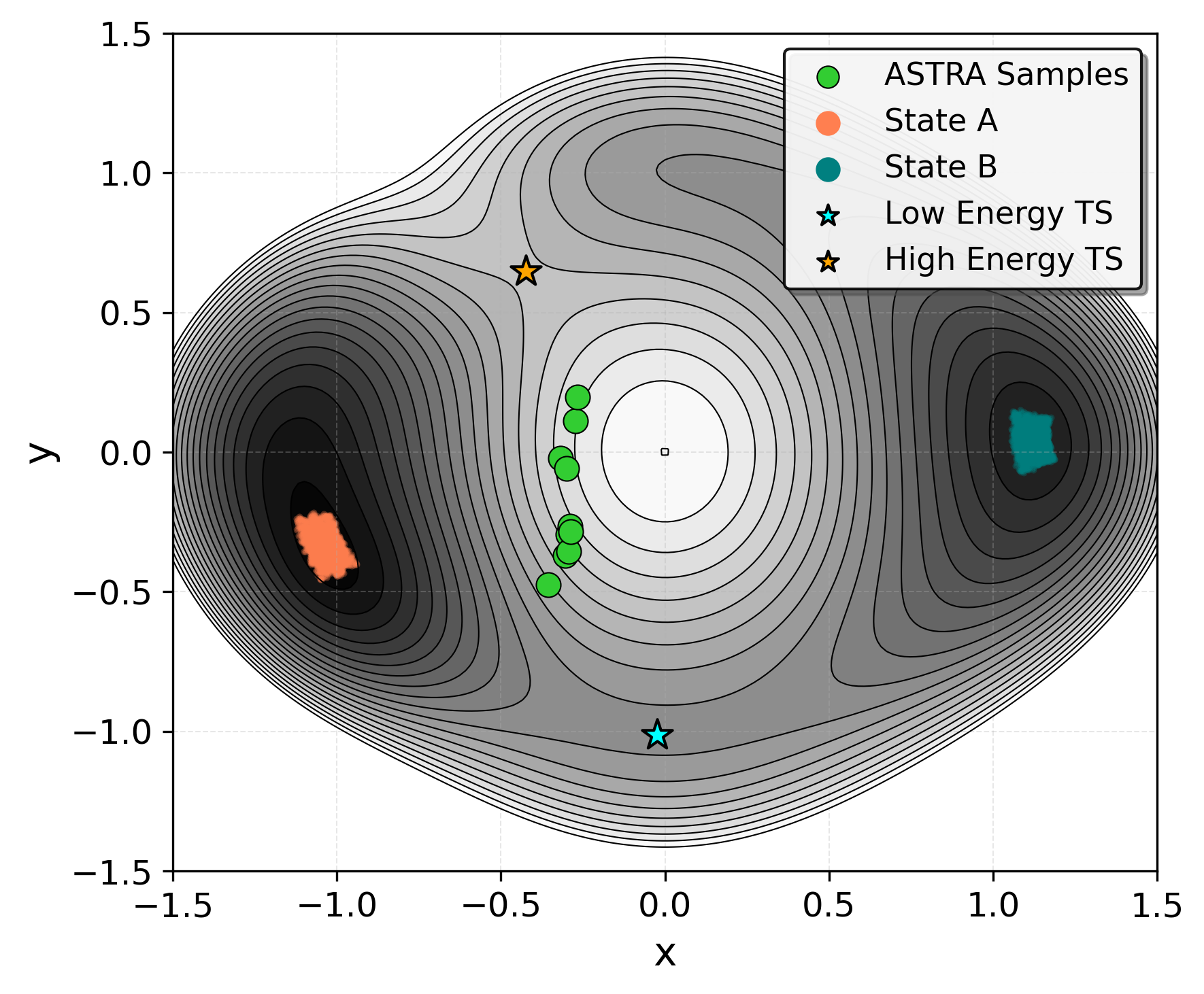}
        \caption{II.}
        \label{fig:abl:dp_ib}
    \end{subfigure}
    \hfill
    \begin{subfigure}{0.48\textwidth}
        \centering
        \includegraphics[width=\linewidth]{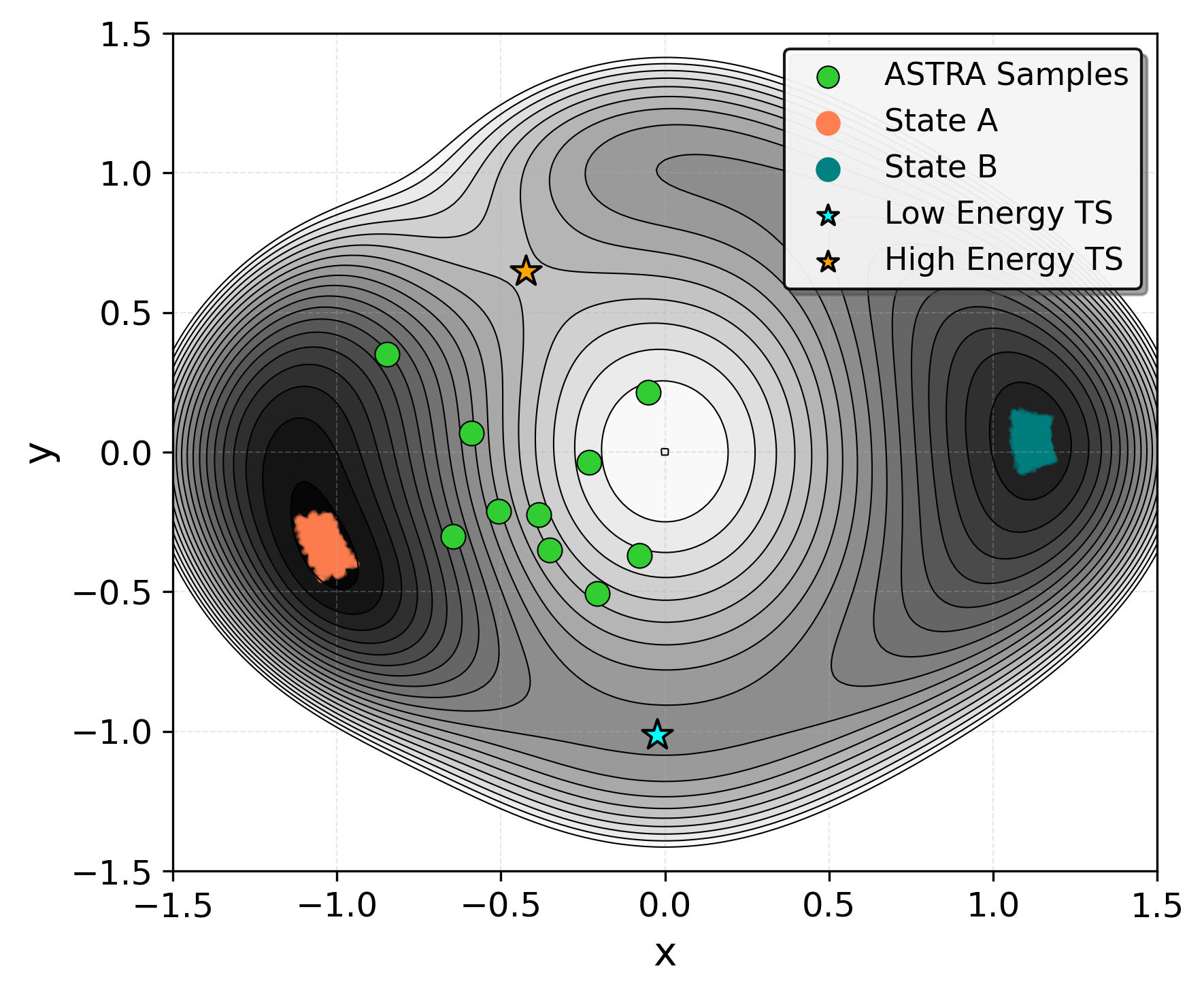}
        \caption{SA.}
        \label{fig:abl:dp_sa}
    \end{subfigure}
    
    \vspace{0.5cm} 

    \begin{subfigure}{0.48\textwidth}
        \centering
        \includegraphics[width=\linewidth]{figures/ablations/dp_ib+saa.png}
        \caption{II + SAA.}
        \label{fig:abl:dp_ib+saa}
    \end{subfigure}
    \hfill
    \begin{subfigure}{0.48\textwidth}
        \centering
        \includegraphics[width=\linewidth]{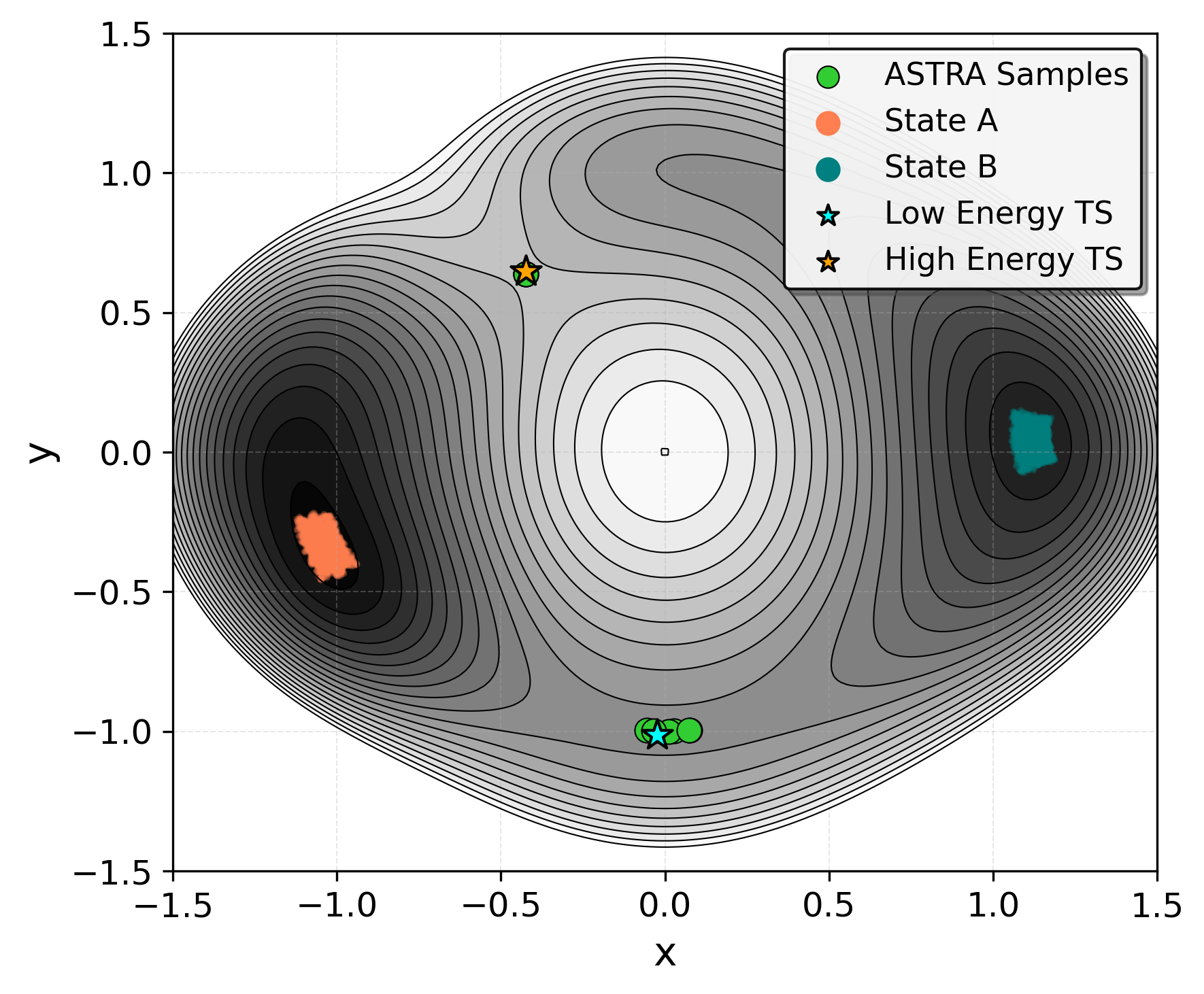}
        \caption{SA + SAA.}
        \label{fig:abl:dp_sa+saa}
    \end{subfigure}
    
    \caption{Impact of Score-Based Interpolation and Score-Aligned Ascent on double path potential.}
    \label{fig:abl:dp}
\end{figure}

\newpage
We extended this ablation study to the two-dimensional potential energy surfaces of the Müller-Brown and double path potential. 
On both systems, as shown in Figures~\ref{fig:abl:mb} and~\ref{fig:abl:dp}, II draws a dividing line between the two states, while SA is less effective, producing a scattered distribution of samples, several of which are located in non-physical high-energy regions. 
Neither interpolation method alone is sufficient for precise TS localization. 
The application of SAA proves crucial, refining the samples by collapsing the broad distributions onto the precise locations of the low- and high-energy transition states. 
As quantified by the $L_2$ distance statistics in Table~\ref{tab:l2dist_PES}, both interpolation methods alone yield substantial errors, whereas subsequently applying SAA dramatically reduces the distance to the nearest transition state, confirming its critical role.

\newpage

\begin{table}[!htb]
\centering
\caption{L2 Distance statistics for analytical potentials from the closer transition state. The best results are \textbf{bolded}, and the second-best results are \textit{underlined}.}
\label{tab:l2dist_PES}
\begin{tabular}{lcccc}
\toprule
\textbf{Potential} & \textbf{II} & \textbf{II+SAA} & \textbf{SA} & \textbf{SA+SAA} \\
\midrule
\textbf{Double Well} & 0.1174 $\pm$ 0.0101 & \textbf{0.0002 $\pm$ 0.0000} & 0.5870 $\pm$ 0.5000 & \textbf{0.0002 $\pm$ 0.0000} \\
\textbf{Müller-Brown} & 7.3632 $\pm$ 1.0707 & \textbf{0.0376 $\pm$ 0.0044} & 6.8723 $\pm$ 3.1354 & \underline{0.0449 $\pm$ 0.0107} \\
\textbf{Double Path} & 0.6810 $\pm$ 0.1008 & \underline{0.0511 $\pm$ 0.0377} & 0.6998 $\pm$ 0.1511 & \textbf{0.0384 $\pm$ 0.0358} \\
\bottomrule
\end{tabular}
\end{table}

\begin{figure}[htb!]
    \centering
    
    \begin{subfigure}{0.48\textwidth}
        \centering
        \includegraphics[width=\linewidth]{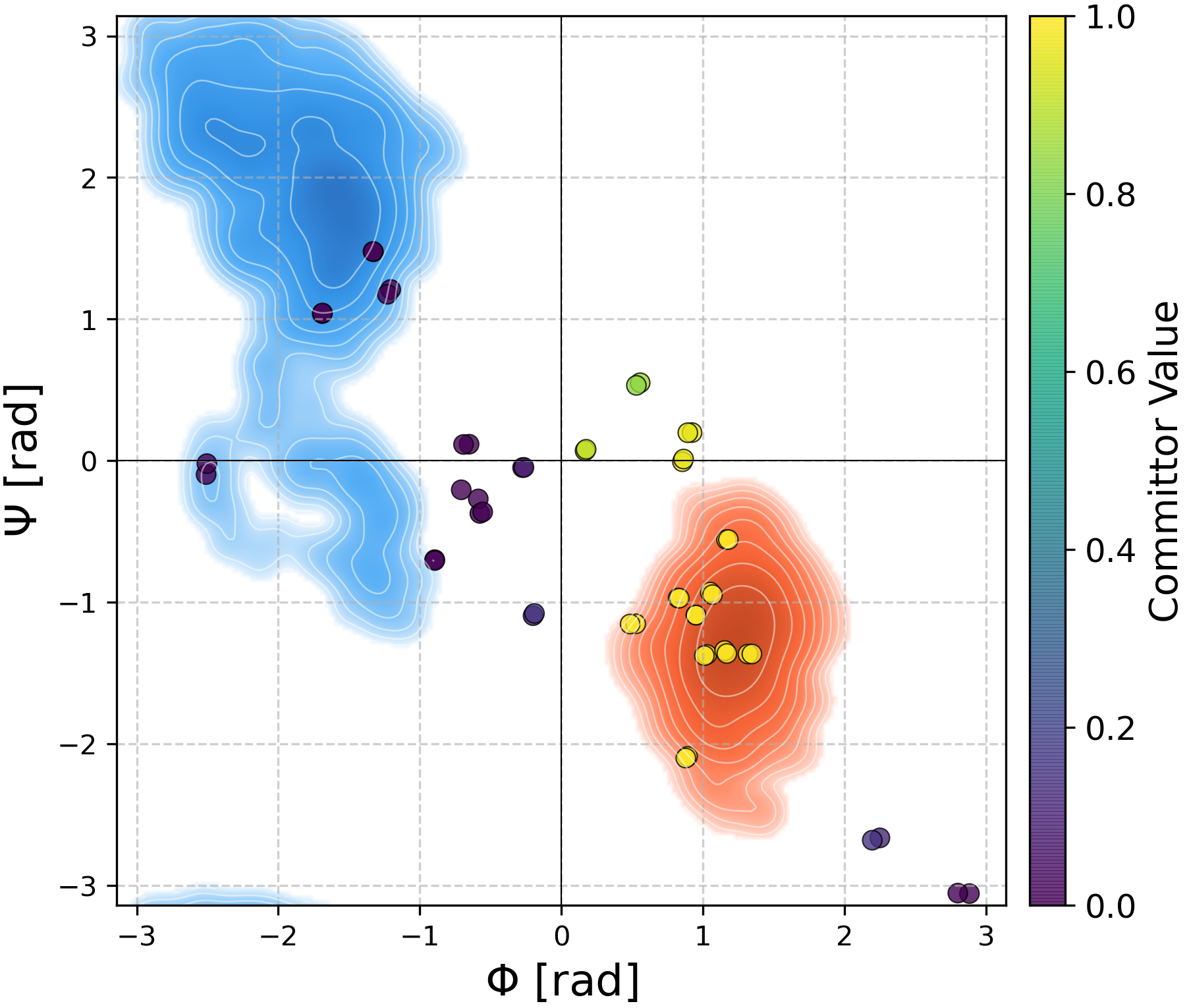}
        \caption{II.}
        \label{fig:abl:ala2_ib}
    \end{subfigure}
    \hfill
    \begin{subfigure}{0.48\textwidth}
        \centering
        \includegraphics[width=\linewidth]{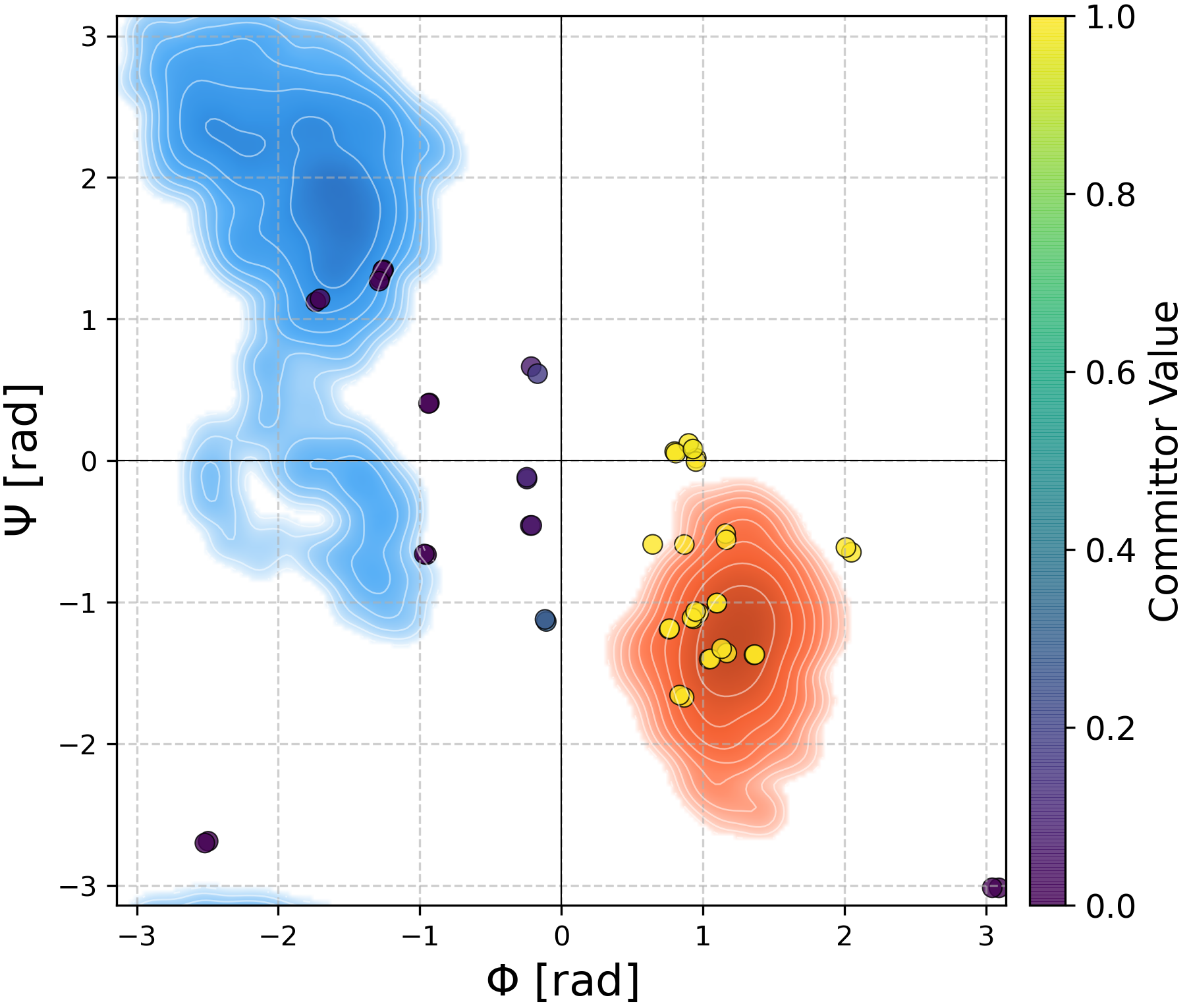}
        \caption{SA.}
        \label{fig:abl:ala2_sa}
    \end{subfigure}
    
    \vspace{0.5cm} 

    \begin{subfigure}{0.48\textwidth}
        \centering
        \includegraphics[width=\linewidth]{figures/ablations/ala2_ib+saa.png}
        \caption{II + SAA.}
        \label{fig:abl:ala2_ib+saa}
    \end{subfigure}
    \hfill
    \begin{subfigure}{0.48\textwidth}
        \centering
        \includegraphics[width=\linewidth]{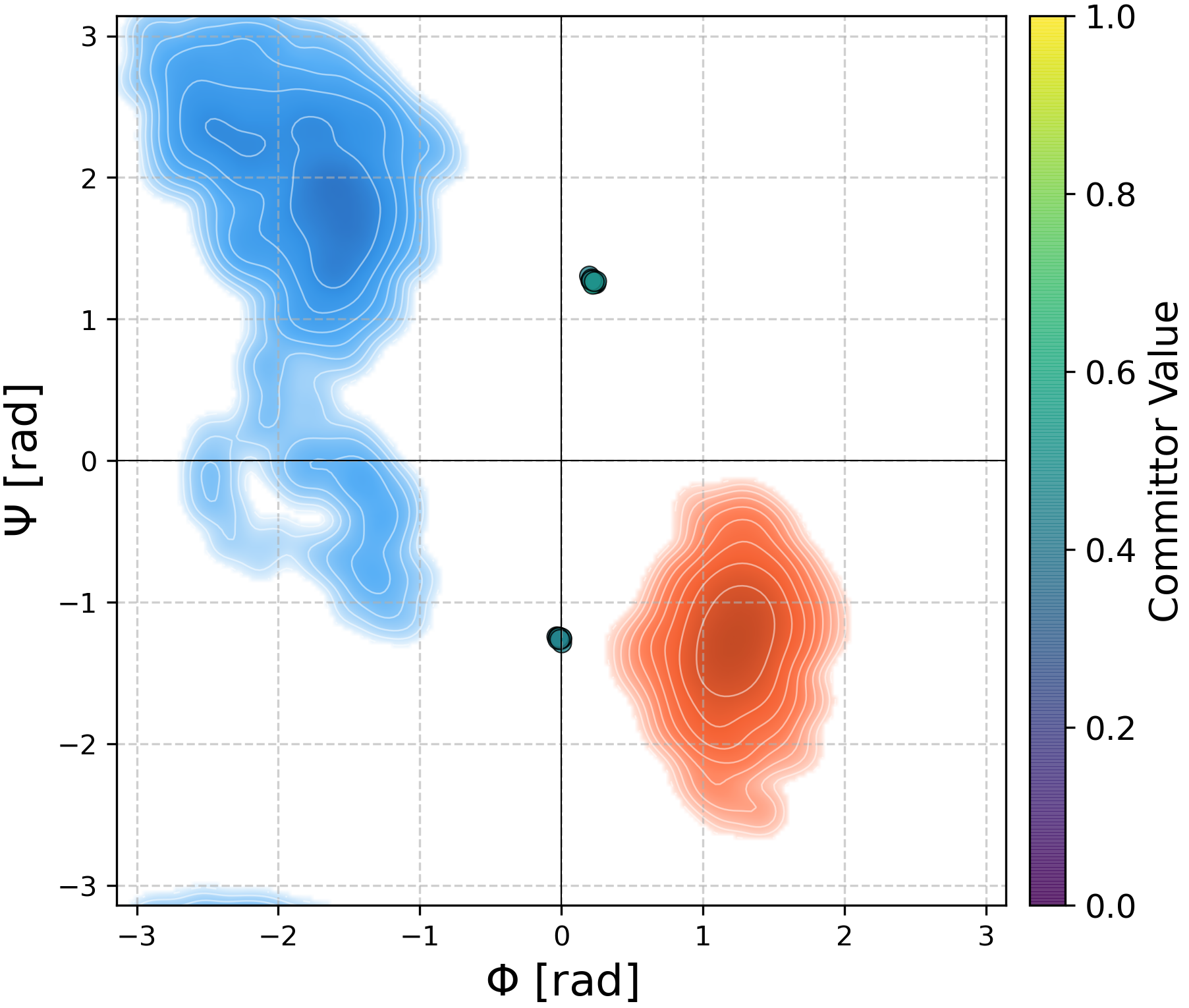}
        \caption{SA + SAA.}
        \label{fig:abl:ala2_sa+saa}
    \end{subfigure}
    
    \caption{Impact of Score-Based Interpolation and Score-Aligned Ascent on alanine dipeptide.}
    \label{fig:abl:ala2}
\end{figure}

\newpage
We further extend our ablation study to alanine dipeptide to assess the method's performance in a high-dimensional system. 
The results, presented in Figure~\ref{fig:abl:ala2}, highlight the critical role of SAA.
When used alone, both II and SA incorrectly identify the transition state region by generating samples located between the two basins. 
Upon introducing SAA, the performance of both methods improves notably.
SAA successfully guides the generated samples to probable TS regions, causing them to converge precisely onto the known, physically meaningful transition states on the potential energy surface.

\begin{figure}[htb!]
    \centering
    
    \begin{subfigure}{0.48\textwidth}
        \centering
        \includegraphics[width=\linewidth]{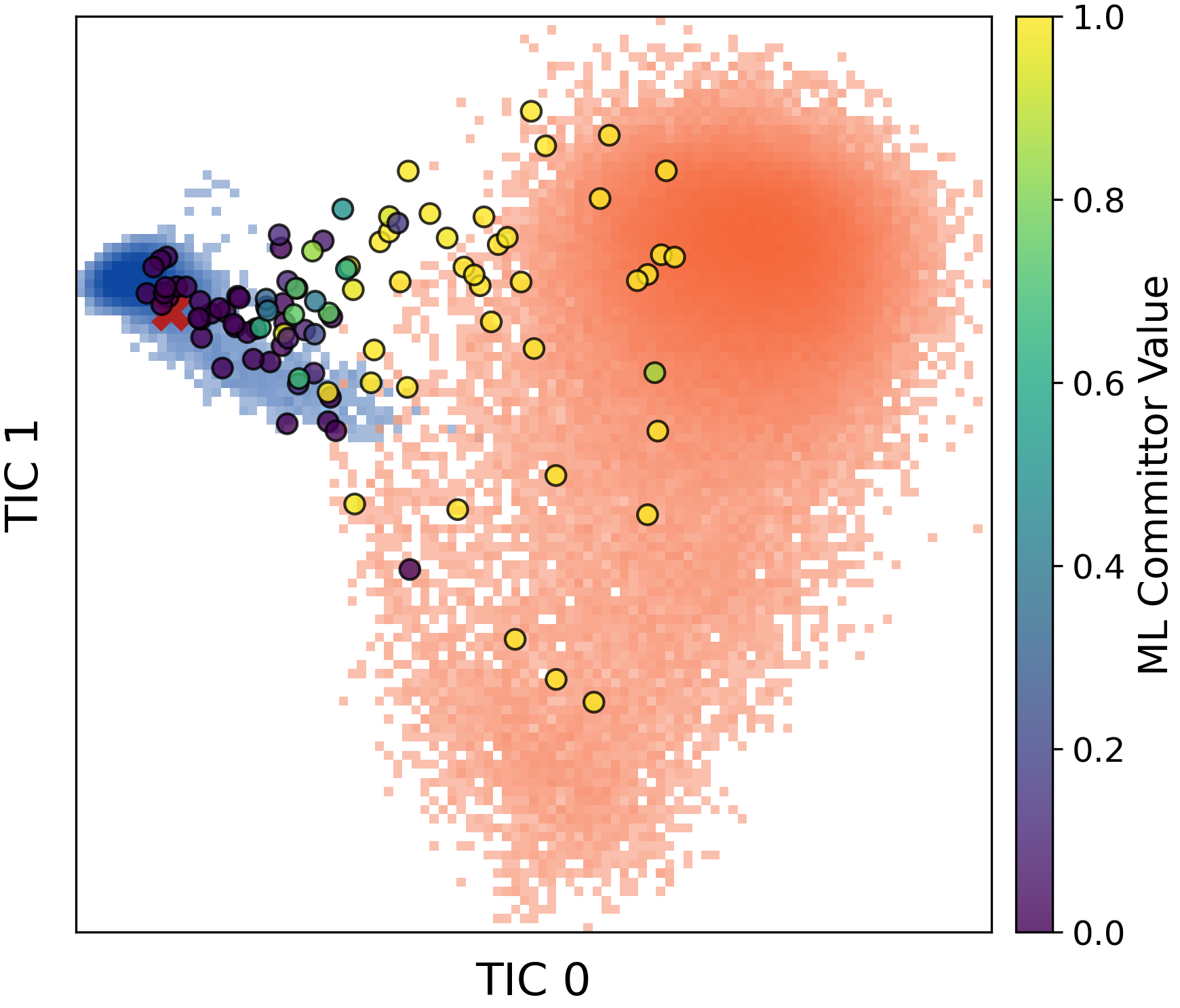}
        \caption{II.}
        \label{fig:abl:chig_ib}
    \end{subfigure}
    \hfill
    \begin{subfigure}{0.48\textwidth}
        \centering
        \includegraphics[width=\linewidth]{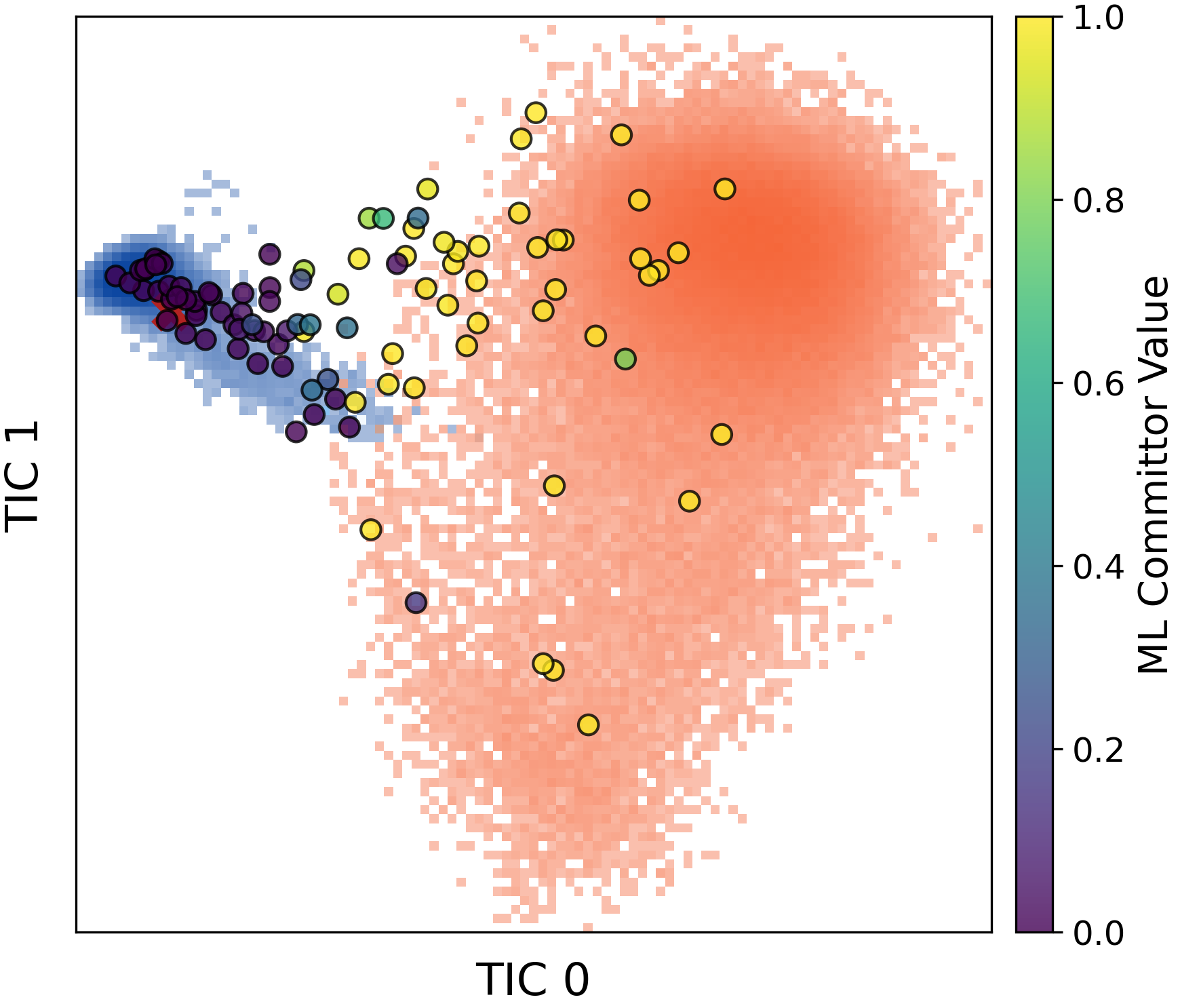}
        \caption{SA.}
        \label{fig:abl:chig_sa}
    \end{subfigure}
    
    \vspace{0.5cm} 

    \begin{subfigure}{0.48\textwidth}
        \centering
        \includegraphics[width=\linewidth]{figures/ablations/chig_ib+saa.png}
        \caption{II + SAA.}
        \label{fig:abl:chig_ib+saa}
    \end{subfigure}
    \hfill
    \begin{subfigure}{0.48\textwidth}
        \centering
        \includegraphics[width=\linewidth]{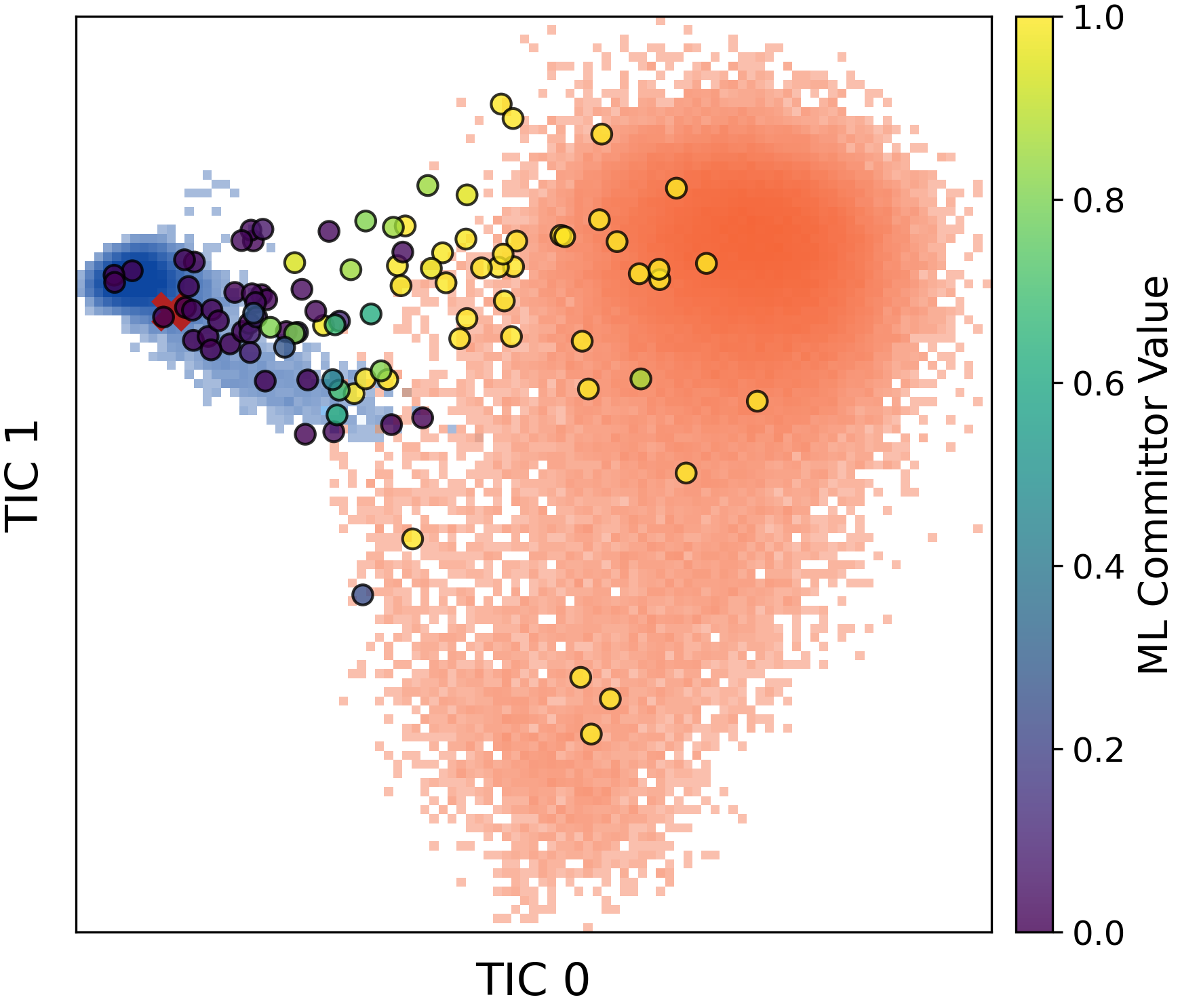}
        \caption{SA + SAA.}
        \label{fig:abl:chig_sa+saa}
    \end{subfigure}
    
    \caption{Impact of Score-Based Interpolation and Score-Aligned Ascent on Chignolin.}
    \label{fig:abl:chig}
\end{figure}

\newpage
A similar trend is observed for the folding of chignolin, depicted in Figure~\ref{fig:abl:chig}. 
While both interpolation methods can generate a coarse path between the folded and unfolded states, the resulting samples are diffuse and fail to define the transition pathway clearly.
The addition of SAA is essential for refining this pathway, guiding the scattered points toward the transition state region. 
This result demonstrates that the combination of II  and SAA is a robust and effective method for identifying transition states in complex biomolecular systems.

\begin{figure}[htb!]
    \centering
    
    \begin{subfigure}{0.48\textwidth}
        \centering
        \includegraphics[width=\linewidth]{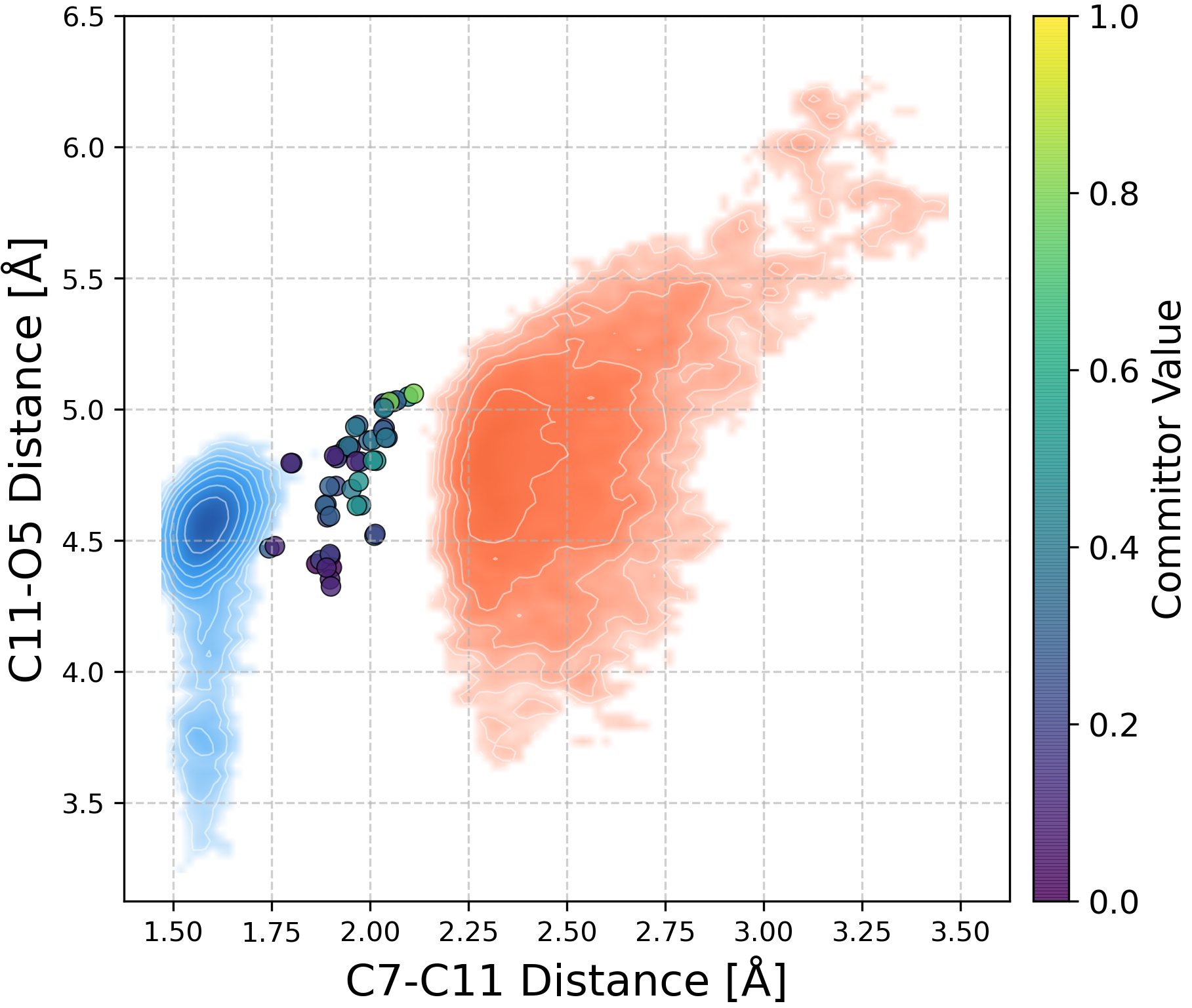}
        \caption{II.}
        \label{fig:abl:dasa_ib}
    \end{subfigure}
    \hfill
    \begin{subfigure}{0.48\textwidth}
        \centering
        \includegraphics[width=\linewidth]{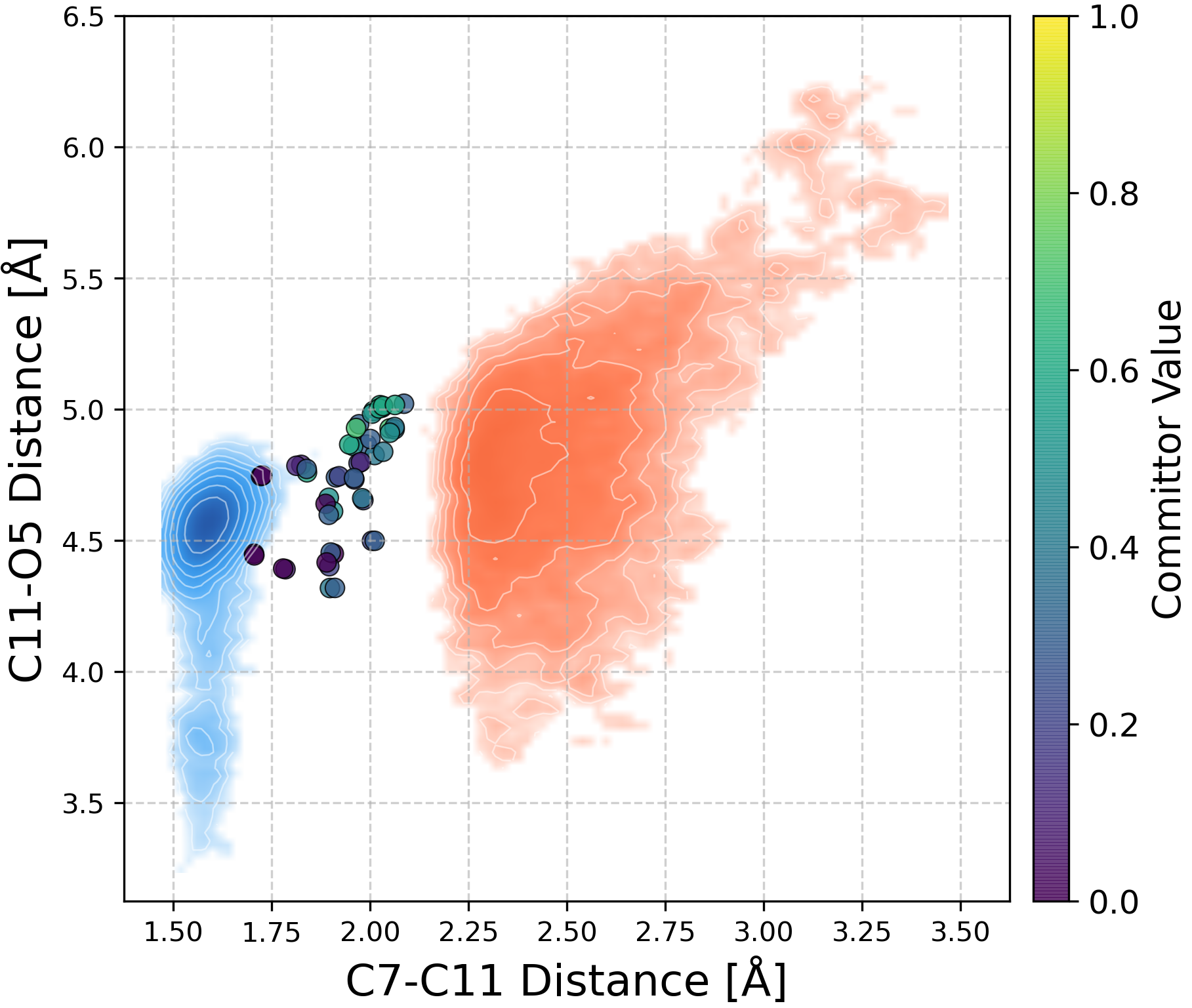}
        \caption{SA.}
        \label{fig:abl:dasa_sa}
    \end{subfigure}
    
    \vspace{0.5cm} 

    \begin{subfigure}{0.48\textwidth}
        \centering
        \includegraphics[width=\linewidth]{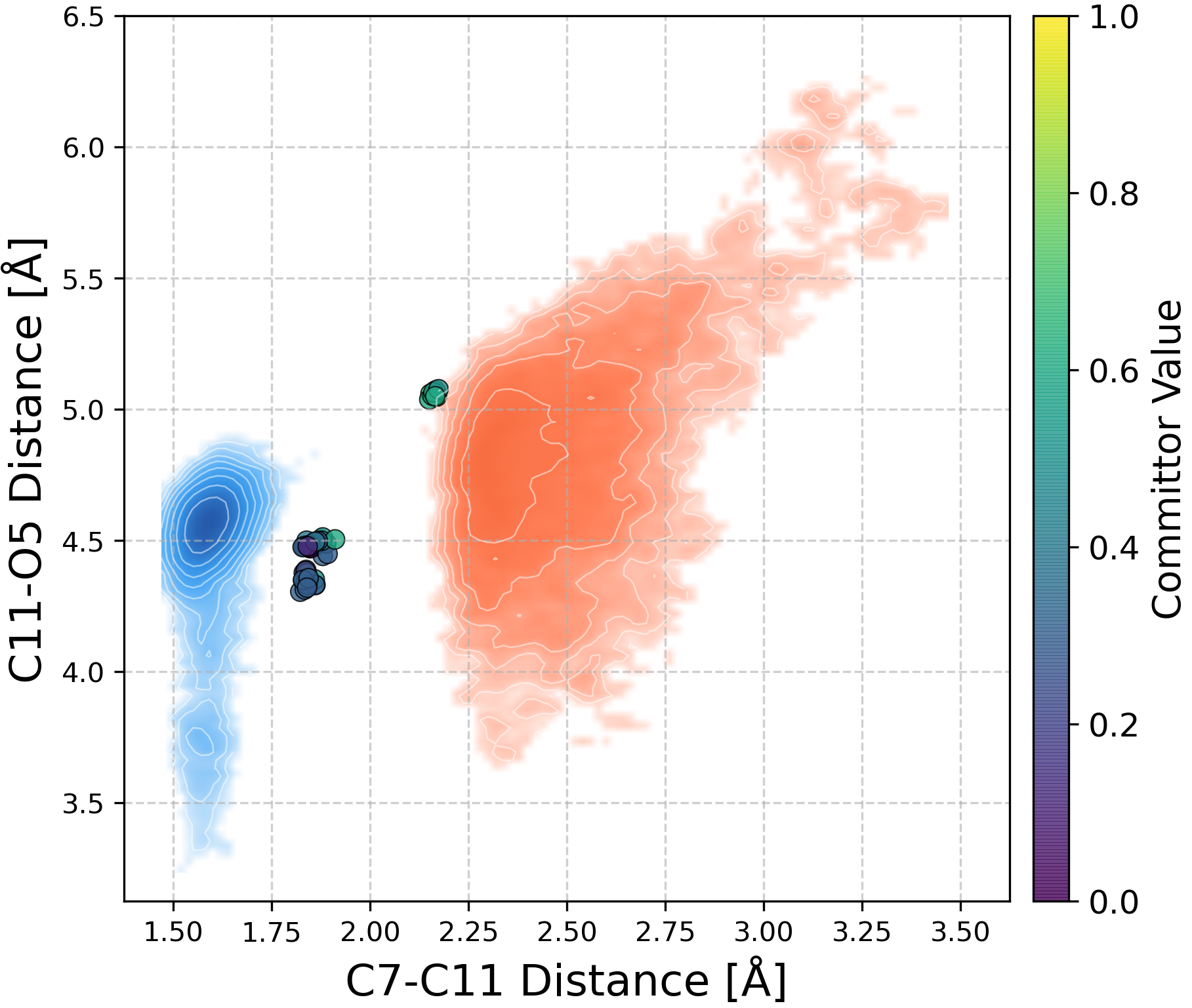}
        \caption{II + SAA.}
        \label{fig:abl:dasa_ib+saa}
    \end{subfigure}
    \hfill
    \begin{subfigure}{0.48\textwidth}
        \centering
        \includegraphics[width=\linewidth]{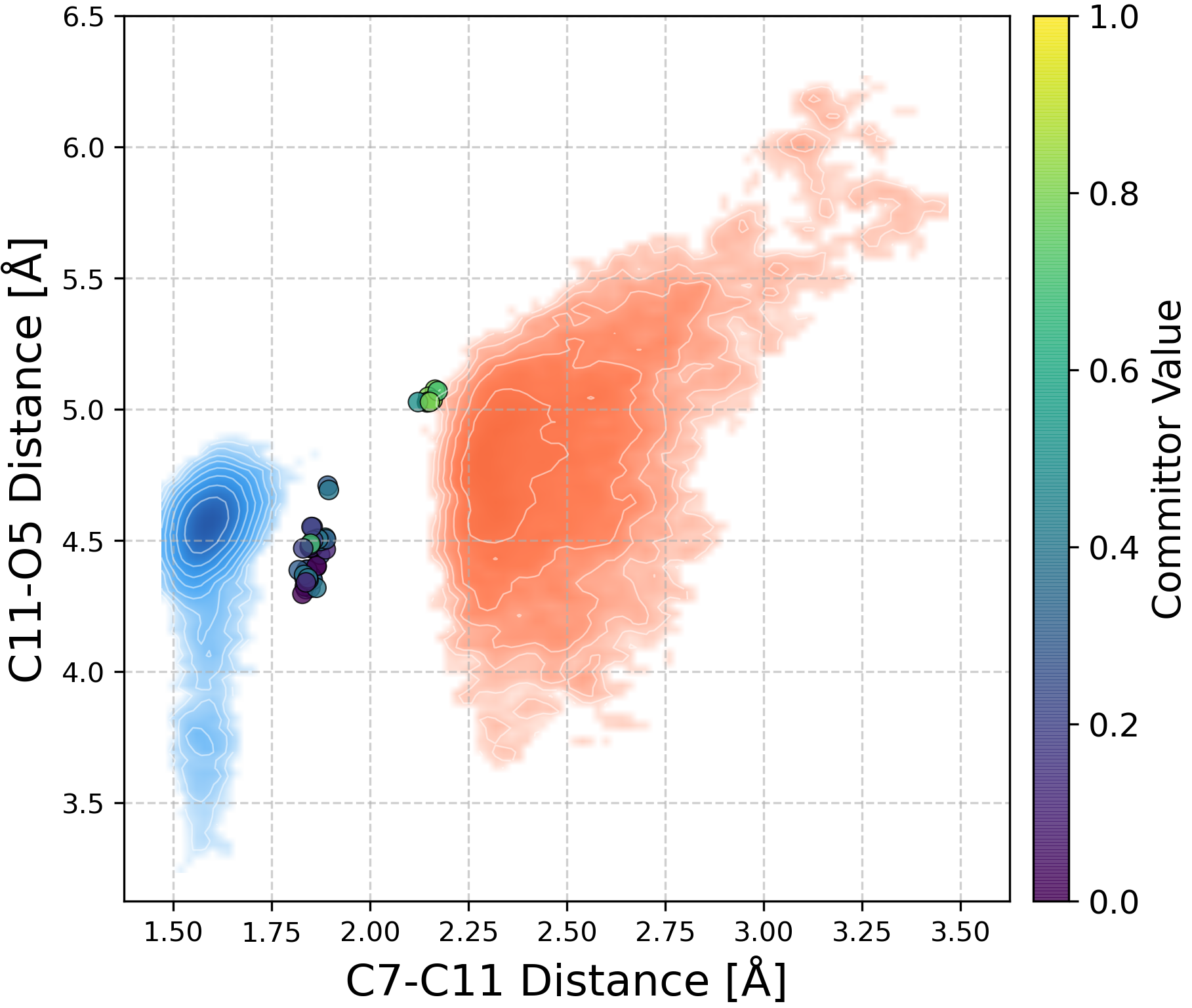}
        \caption{SA + SAA.}
        \label{fig:abl:dasa_sa+saa}
    \end{subfigure}
    
    \caption{Impact of Score-Based Interpolation and Score-Aligned Ascent on DASA reaction.}
    \label{fig:abl:dasa}
\end{figure}

\newpage
For the electrocyclical reaction, as illustrated in Figure~\ref{fig:abl:dasa}, both II and SA fail to precisely locate the TS of the reaction. 
Instead, the resulting samples are broadly distributed in the region between the two energy minima corresponding to the open (colored) and closed (colorless) isomers. 
The application of SAA is critical, as it refines this scattered distribution, driving the system to the more probable TS structures. 
This result underscores the method's ability to navigate the complex PES of a chemical reaction.

    
    

    

\newpage
\subsection{Committor Analysis of Chemical Systems}
\label{abl:committor}
\begin{figure}[htb!]
    \centering
    
    \begin{subfigure}{0.48\textwidth}
        \centering
        \includegraphics[width=\linewidth]{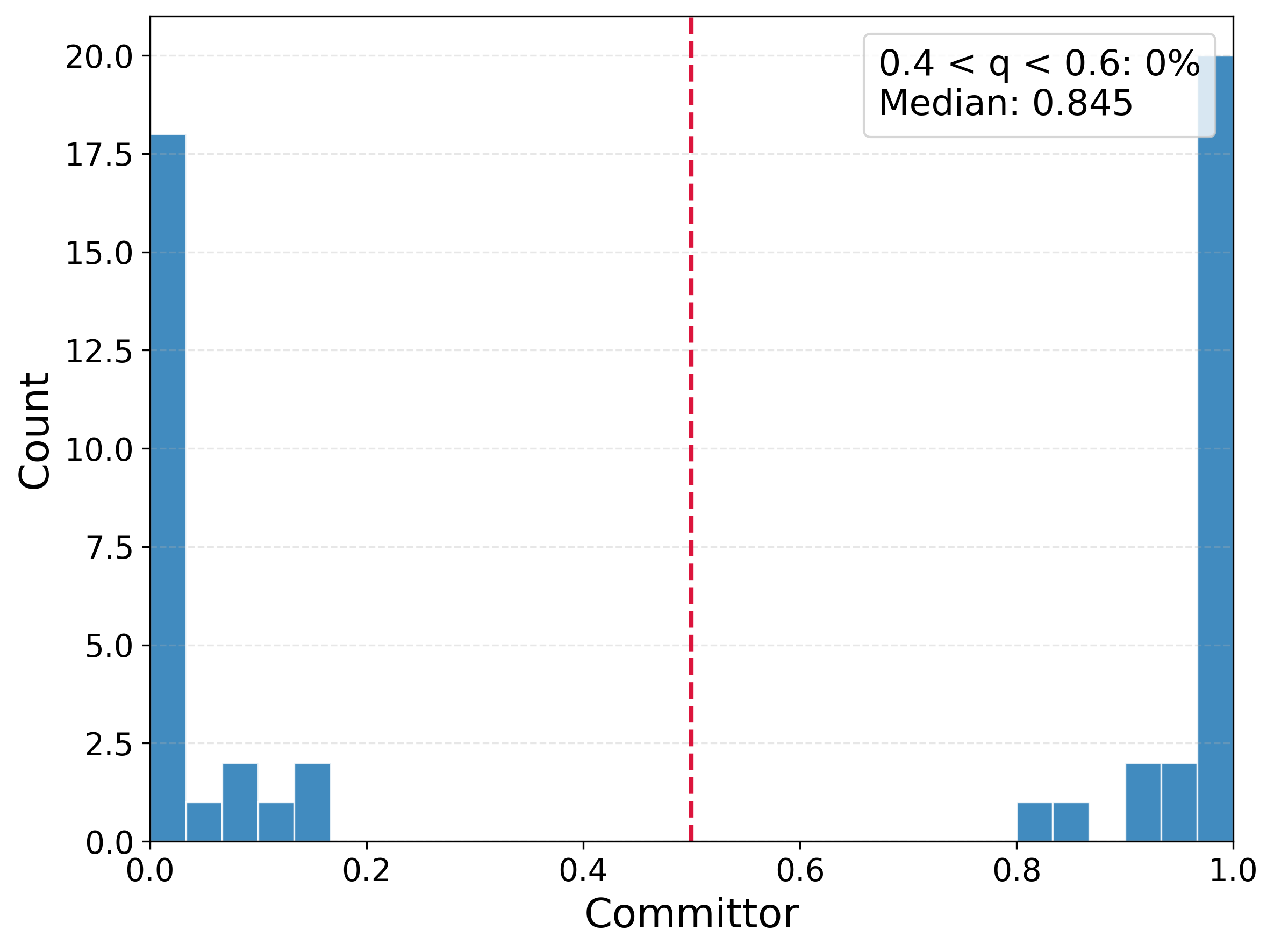}
        \caption{II.}
        \label{fig:abl:com_ala2_ib}
    \end{subfigure}
    \hfill
    \begin{subfigure}{0.48\textwidth}
        \centering
        \includegraphics[width=\linewidth]{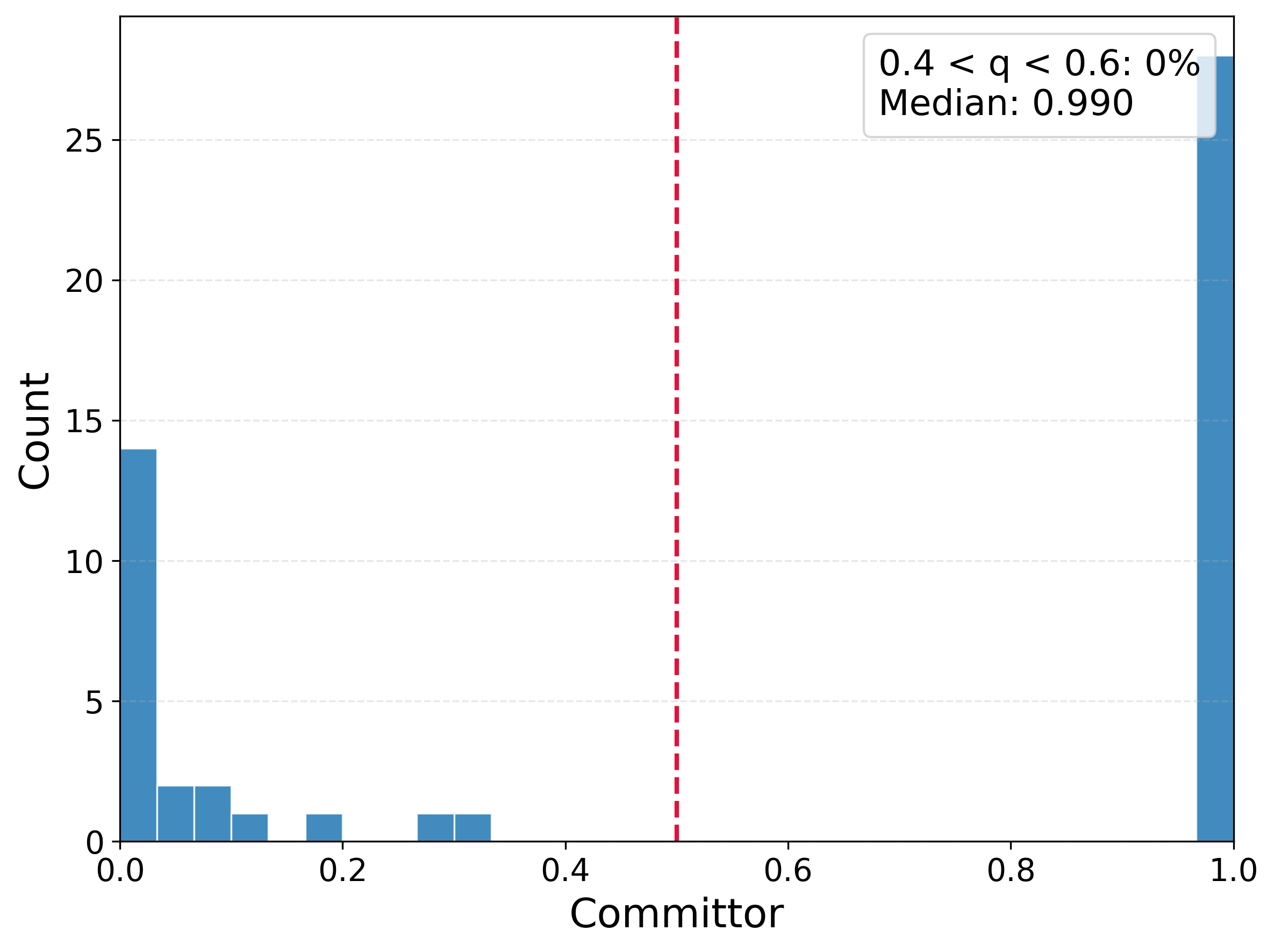}
        \caption{SA.}
        \label{fig:abl:com_ala2_sa}
    \end{subfigure}
    
    \vspace{0.5cm} 

    \begin{subfigure}{0.48\textwidth}
        \centering
        \includegraphics[width=\linewidth]{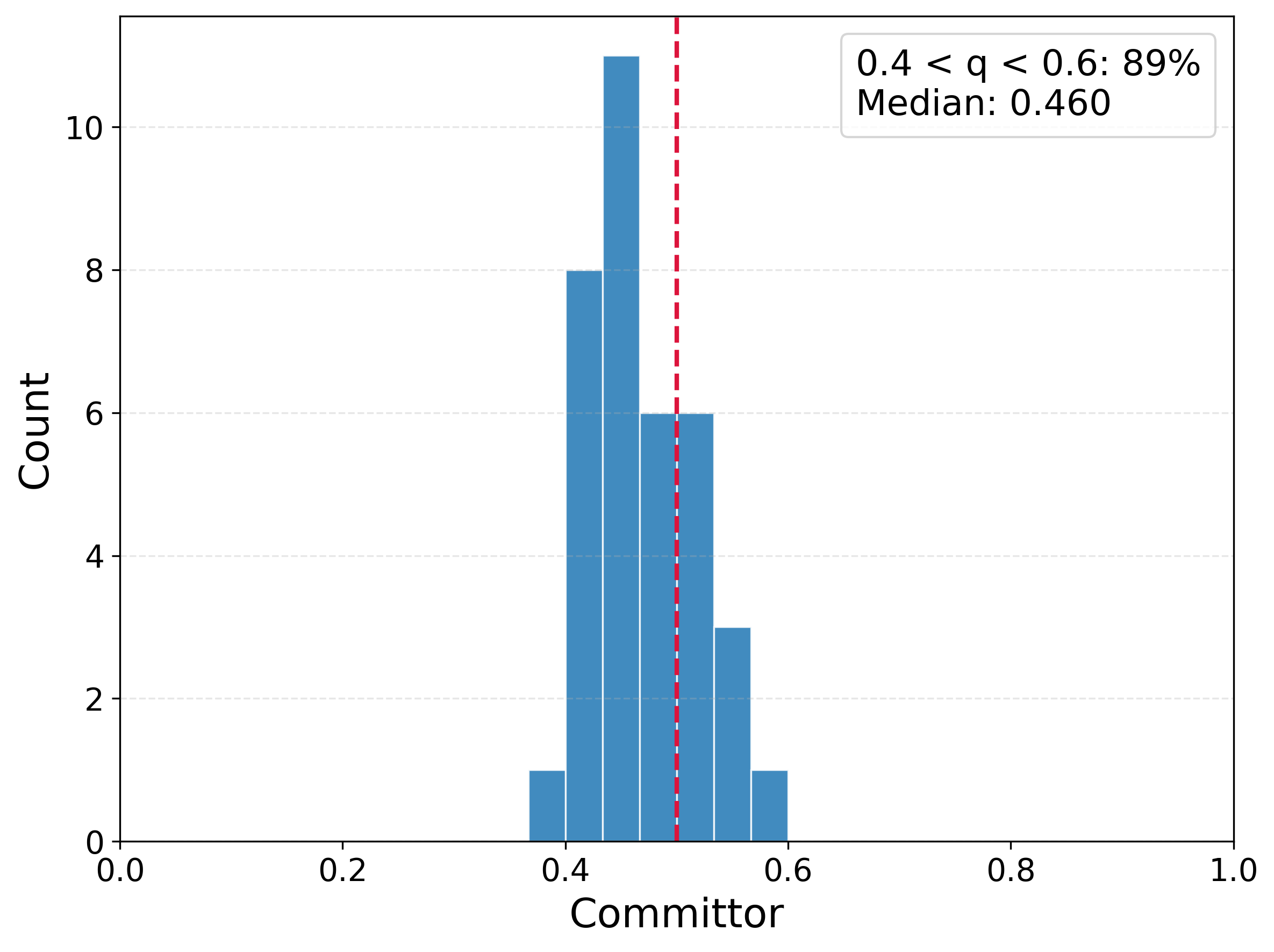}
        \caption{II + SAA.}
        \label{fig:abl:com_ala2_ib+saa}
    \end{subfigure}
    \hfill
    \begin{subfigure}{0.48\textwidth}
        \centering
        \includegraphics[width=\linewidth]{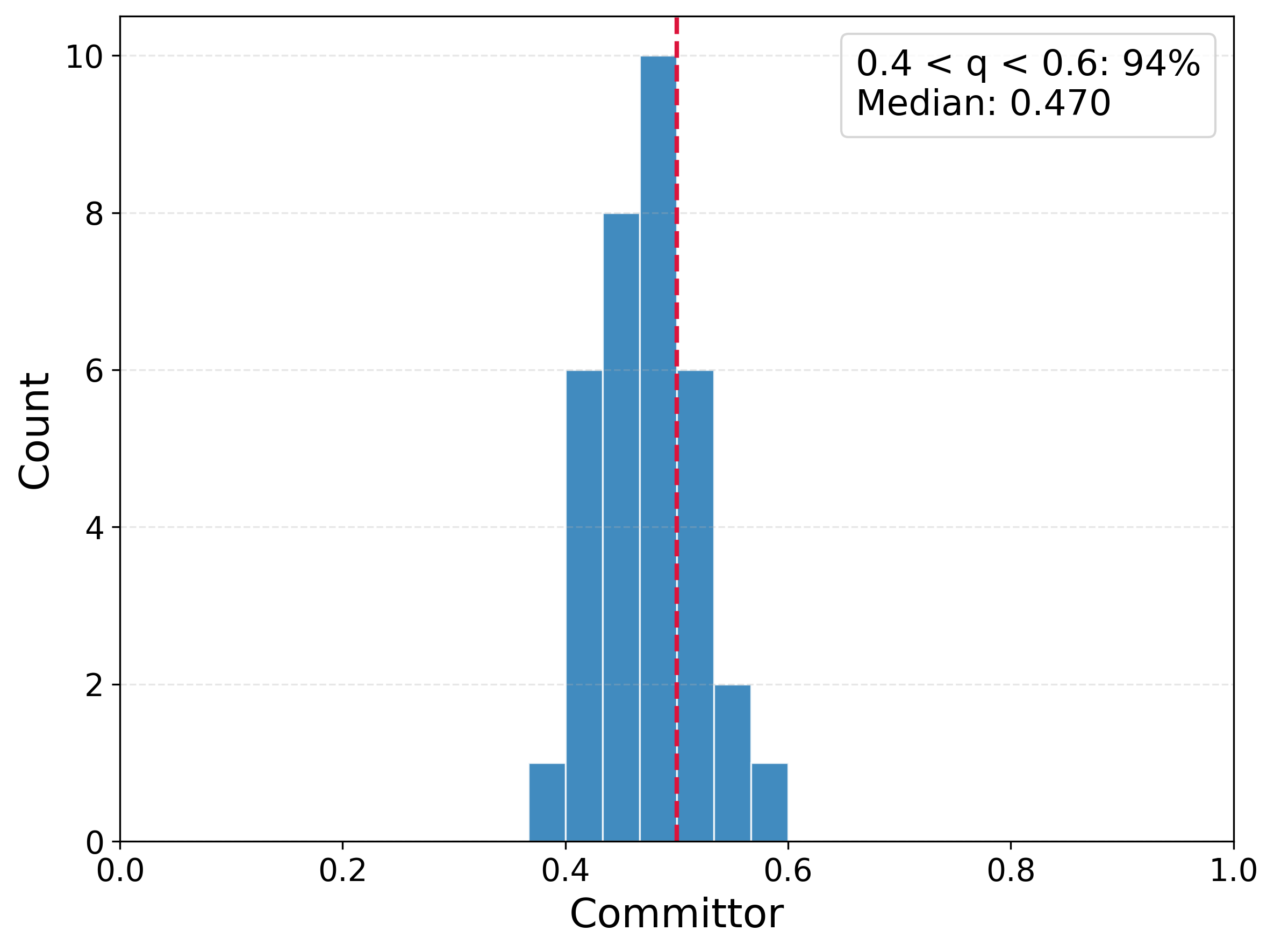}
        \caption{SA + SAA.}
        \label{fig:abl:com_ala2_sa+saa}
    \end{subfigure}
    
    \caption{Histogram of calculated committors on alanine dipeptide.}
    \label{fig:abl:com_ala2}
\end{figure}

In Figure~\ref{fig:abl:com_ala2}, we report histograms of the distributions of committor values calculated based on the samples in Figure~\ref{fig:abl:ala2}.
We observe that given the partial coverage of the metastable states by our MD simulations at 300~K, simply using Score-Based Interpolation allows sampling an intermediate region but not the TSs themselves. The SAA algorithm successfully directs the samples towards the true TSs as demonstrated by a peaked distribution of the committor values around 0.5. SuperDiff AND surpasses SA in our experiments.

\newpage
We also report ML committor values computed with the trained models from \cite{kang2024computing} to compare the effects of SBI and SAA algorithm. We observe that SAA consistently increases the number of samples close to the $q=0.5$ region. We note that the ML committor model draws much sharper isocommittor surfaces compared to our MD evaluation for alanine dipeptide.
\label{abl:committor_chig}
\begin{figure}[htb!]
    \centering
    
    \begin{subfigure}{0.48\textwidth}
        \centering
        \includegraphics[width=\linewidth]{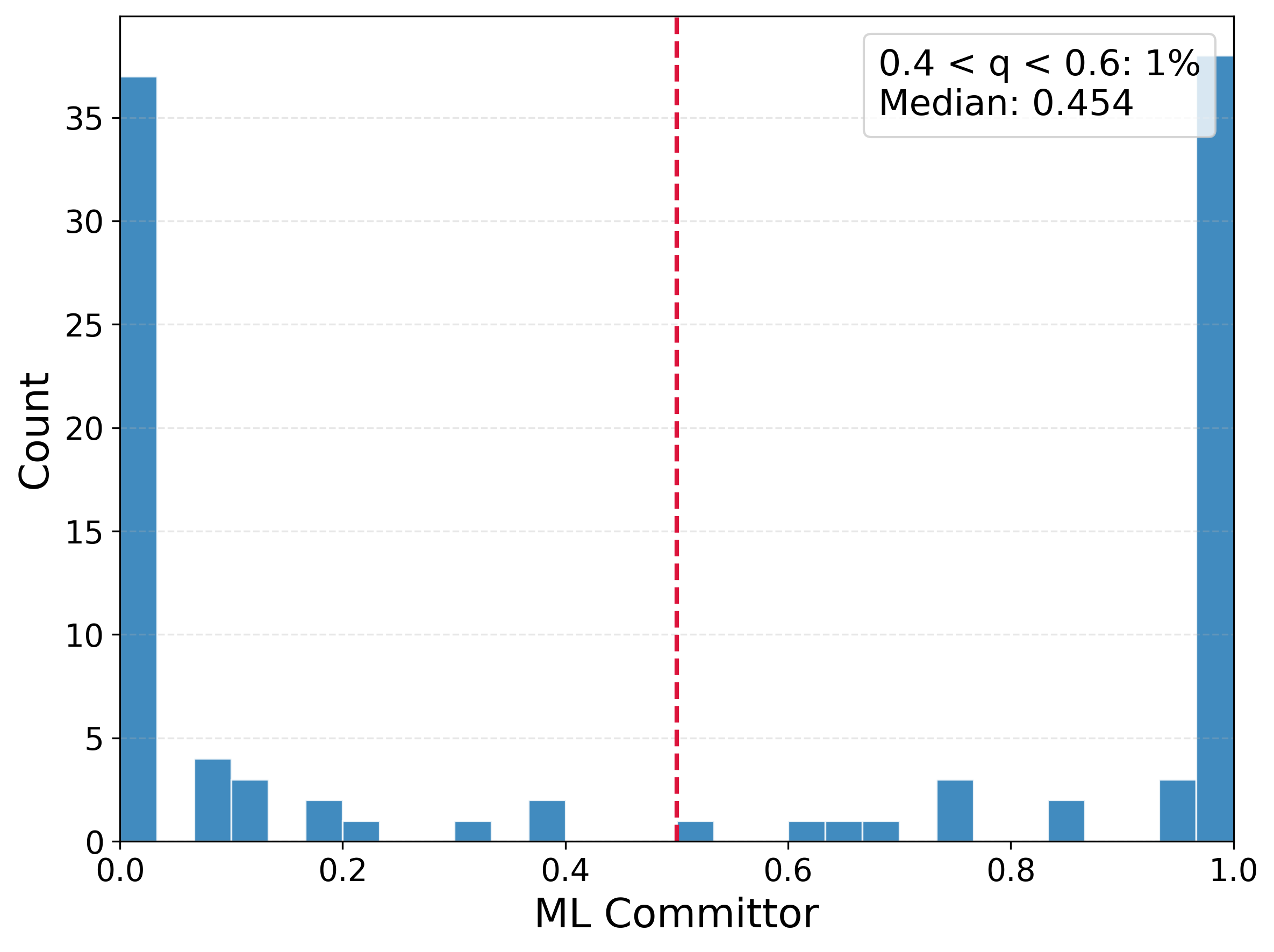}
        \caption{II.}
        \label{fig:abl:com_chig_ib}
    \end{subfigure}
    \hfill
    \begin{subfigure}{0.48\textwidth}
        \centering
        \includegraphics[width=\linewidth]{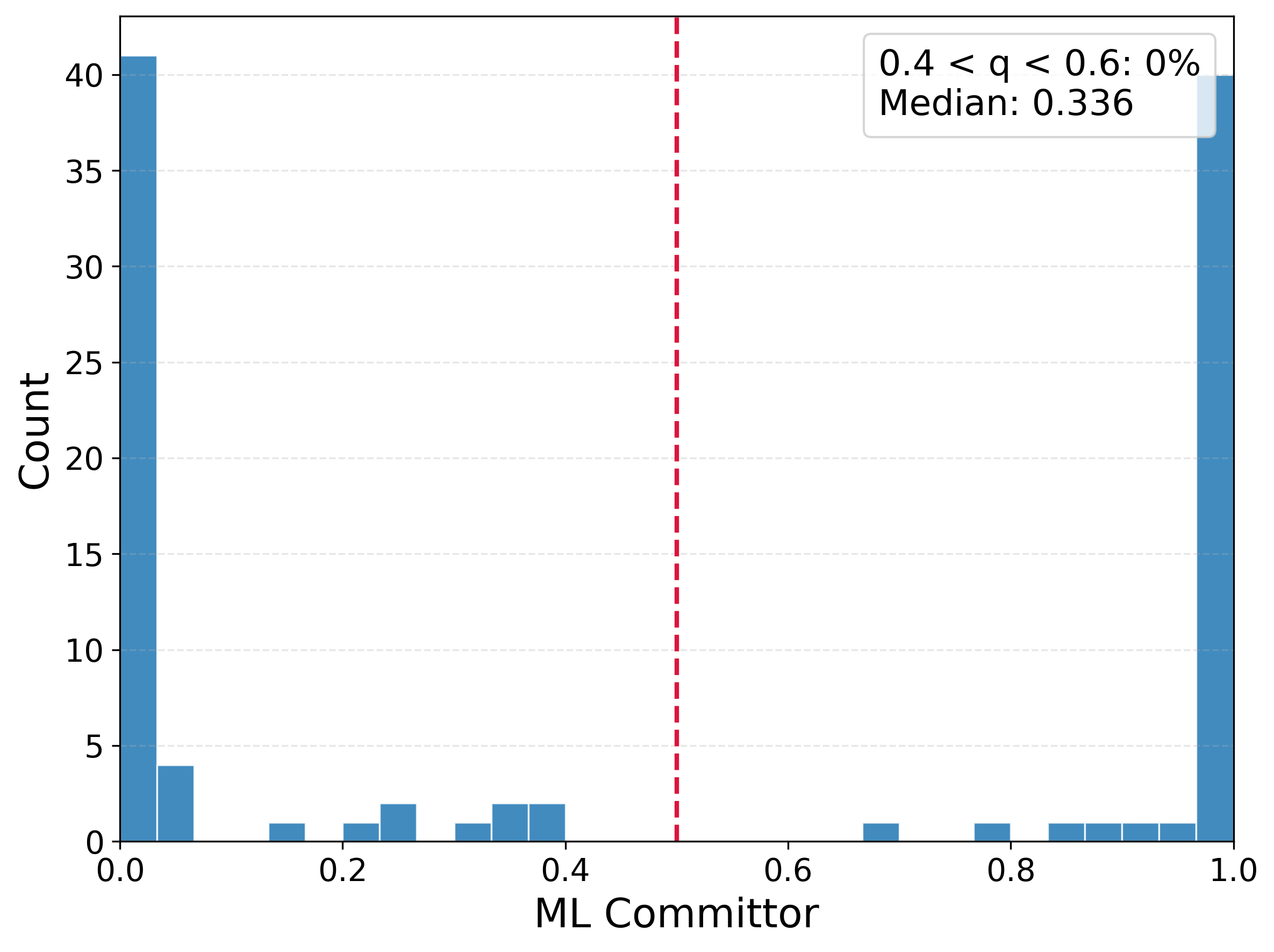}
        \caption{SA.}
        \label{fig:abl:com_chig_sa}
    \end{subfigure}
    
    \vspace{0.5cm} 

    \begin{subfigure}{0.48\textwidth}
        \centering
        \includegraphics[width=\linewidth]{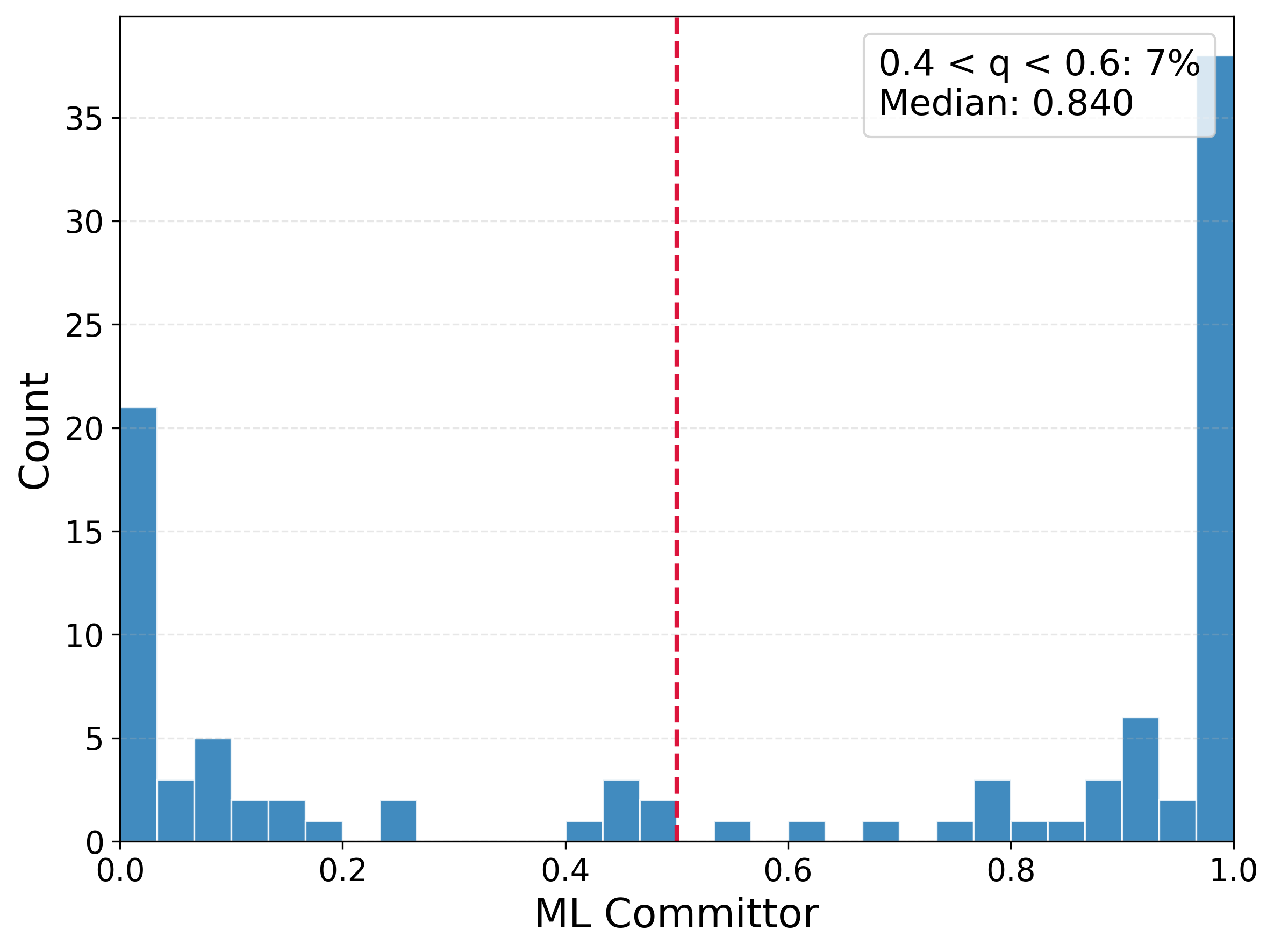}
        \caption{II + SAA.}
        \label{fig:abl:com_chig_ib+saa}
    \end{subfigure}
    \hfill
    \begin{subfigure}{0.48\textwidth}
        \centering
        \includegraphics[width=\linewidth]{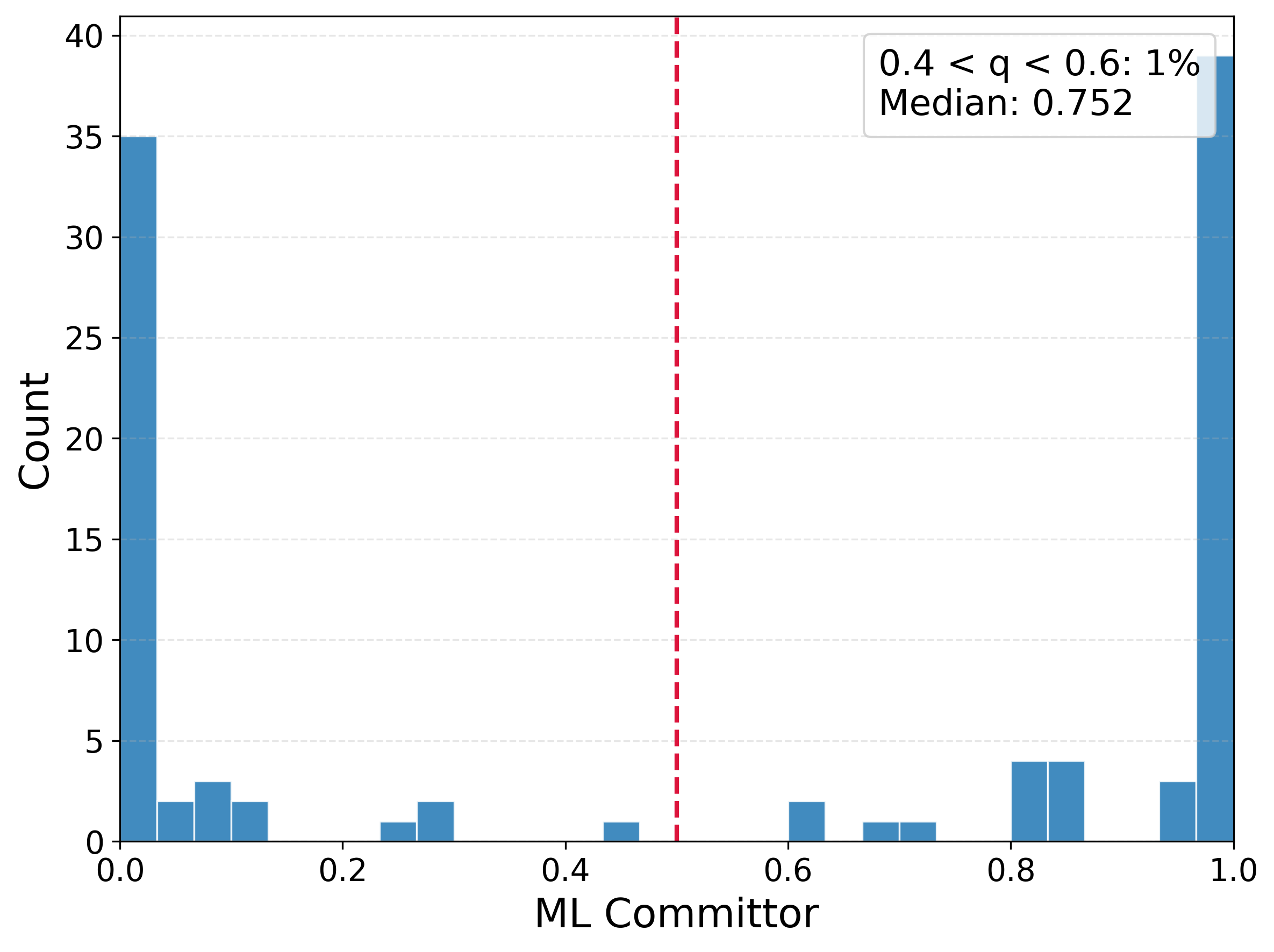}
        \caption{SA + SAA.}
        \label{fig:abl:com_chig_sa+saa}
    \end{subfigure}
    
    \caption{Histogram of ML predicted committors on Chignolin.}
    \label{fig:abl:com_chig}
\end{figure}

\newpage

\label{abl:committor_dasa}
\begin{figure}[htb!]
    \centering
    
    \begin{subfigure}{0.48\textwidth}
        \centering
        \includegraphics[width=\linewidth]{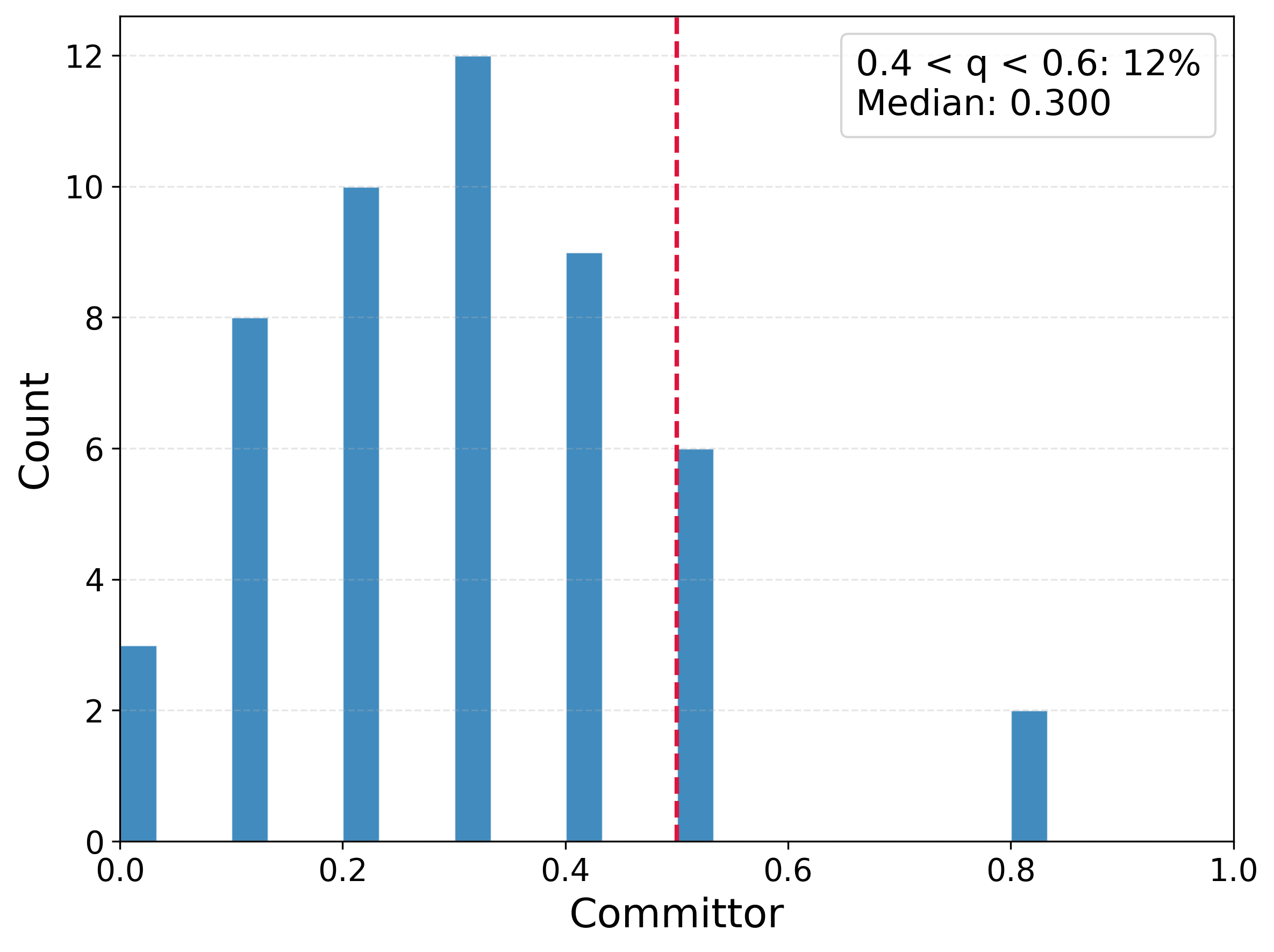}
        \caption{II.}
        \label{fig:abl:com_dasa_ib}
    \end{subfigure}
    \hfill
    \begin{subfigure}{0.48\textwidth}
        \centering
        \includegraphics[width=\linewidth]{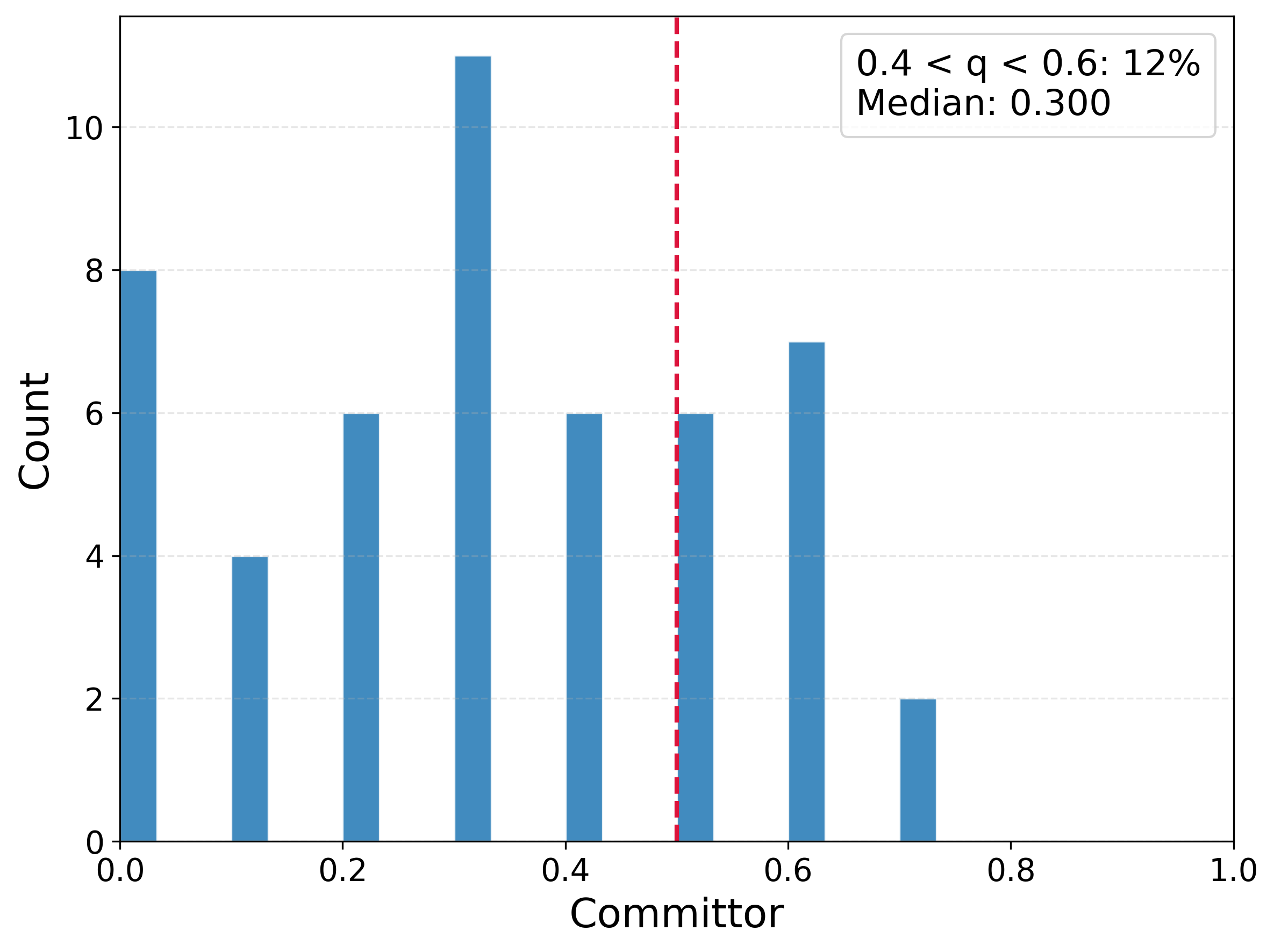}
        \caption{SA.}
        \label{fig:abl:com_dasa_sa}
    \end{subfigure}
    
    \vspace{0.5cm} 

    \begin{subfigure}{0.48\textwidth}
        \centering
        \includegraphics[width=\linewidth]{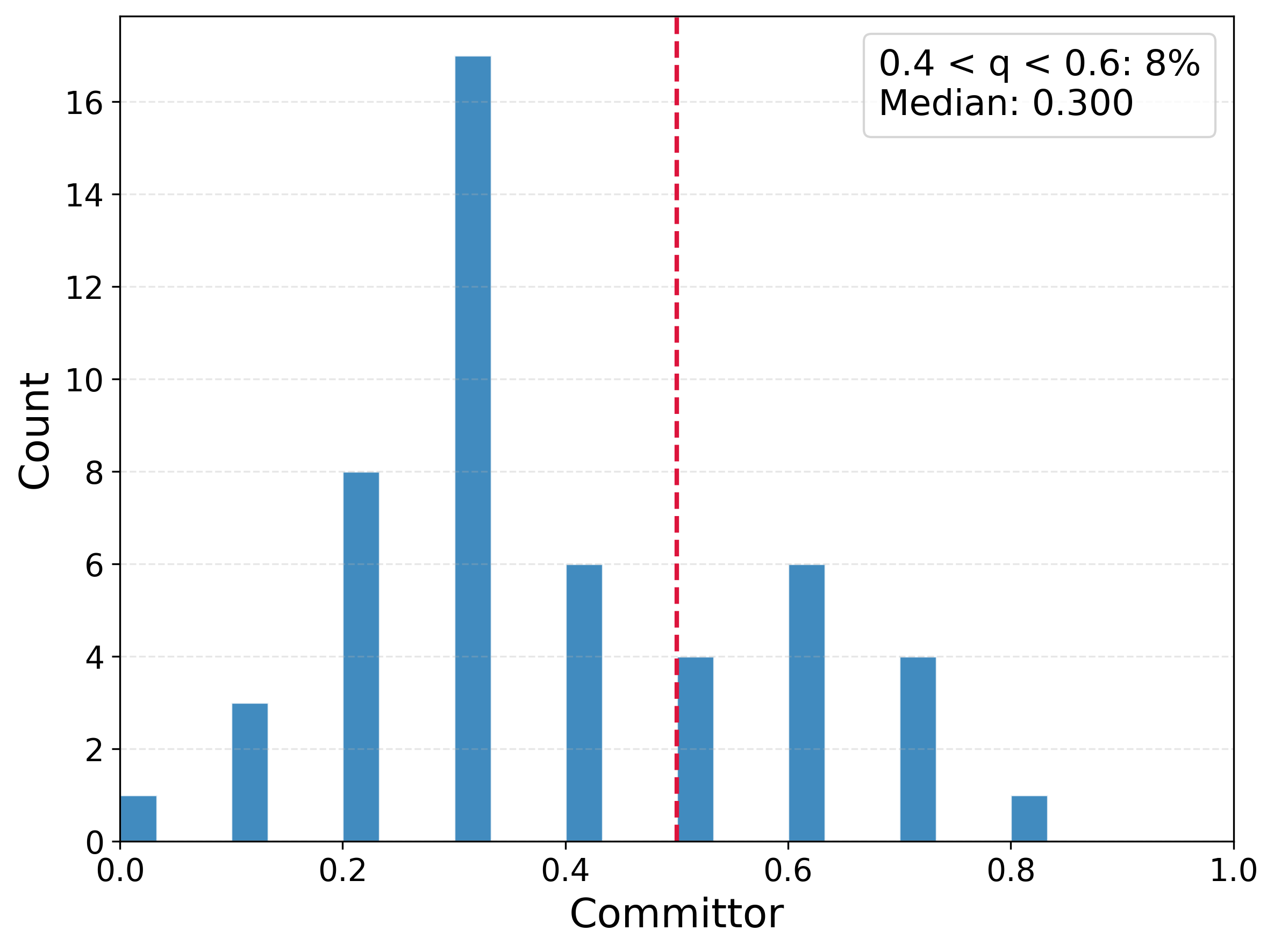}
        \caption{II + SAA.}
        \label{fig:abl:com_dasa_ib+saa}
    \end{subfigure}
    \hfill
    \begin{subfigure}{0.48\textwidth}
        \centering
        \includegraphics[width=\linewidth]{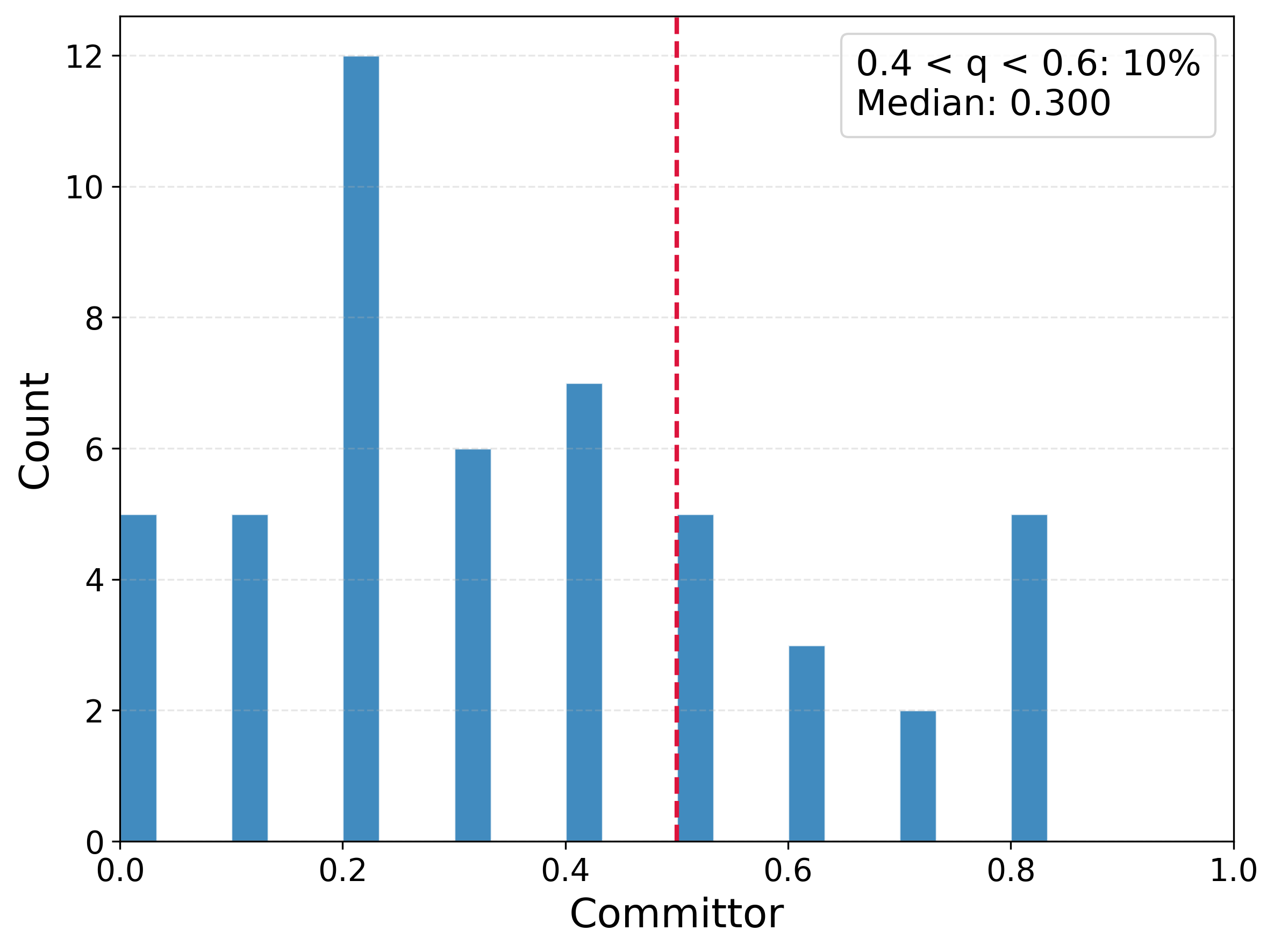}
        \caption{SA + SAA.}
        \label{fig:abl:com_dasa_sa+saa}
    \end{subfigure}
    
    \caption{Histogram of ML predicted committors on DASA reaction.}
    \label{fig:abl:com_dasa}
\end{figure}

\clearpage
\subsection{Sensitivity to Hyperparameters}
We present a detailed hyperparameter sweep for ASTRA, and a comparison of SAA against NEB. These results are the basis for the discussion in Sections \ref{modules_characterization} and \ref{Baselines}.

\begin{table}[!htb]
\centering
\caption{Hyperparameter Ablation Study evaluated by running the Dimer method on ASTRA generated samples. The main result configuration is shown first. Each subsequent row ablates a single hyperparameter, with the modified value in \textbf{bold}. No value means that the SAA was unstable and the force field eventually returns NaN. We report median values for all metrics. Higher energy difference is better as all differences are negative.}
\label{tab:hyperparameter_ablations}
\sisetup{
    detect-weight=true, 
    detect-family=true,
    table-align-text-post=false 
}
\footnotesize
\begin{tabular}{lcccccccc}
\toprule
\multirow{2}{*}{\textbf{Method}} & 
\multirow{2}{*}{\textbf{GS}} & 
\multirow{2}{*}{\textbf{PR}} & 
\multirow{2}{*}{\textbf{OS}} & 
\multirow{2}{*}{\textbf{SS}} & 
\multirow{2}{*}{\textbf{Conv. Rate$\uparrow$}} & 
\multirow{2}{*}{\textbf{Conv. Step$\downarrow$}} & 
\textbf{Pos. RMSD$\downarrow$} &           
\textbf{Energy Diff.$\uparrow$} \\      
\cmidrule(lr){8-8} \cmidrule(lr){9-9}
& & & & & & & (\AA) & (kcal$\cdot$mol$^{-1}$) \\
\midrule
Main & 1.0 & 0.05 & 1000 & 0.0100 & 0.84 & 14 & 0.010 & -0.50 \\
\midrule
\multicolumn{9}{l}{\textit{Ablation on the Guidance Scale (GS)}} \\
GS = 2.0 & \textbf{2.0} & 0.05 & 1000 & 0.0100 & 0.86 & 34 & 0.011 & -0.51 \\
GS = 3.0 & \textbf{3.0} & 0.05 & 1000 & 0.0100 & 0.87 & 11 & 0.011 & -0.46 \\
\midrule
\multicolumn{9}{l}{\textit{Ablation on the Pause Ratio (PR)}} \\
PR = 0.00 & 1.0 & \textbf{0.00} & 1000 & 0.0100 & {--} & {--} & {--} & {--} \\
PR = 0.25 & 1.0 & \textbf{0.25} & 1000 & 0.0100 & {0.78} & 70 & 0.05 & -0.12 \\
PR = 0.75 & 1.0 & \textbf{0.75} & 1000 & 0.0100 & {0.41} & 129 & 0.018 & -1.1 \\
PR = 0.10 & 1.0 & \textbf{0.10} & 1000 & 0.0100 & 0.13 & 294 & 0.06 & -1.2 \\
\midrule
\multicolumn{9}{l}{\textit{Ablation on the Optimization Steps (OS)}} \\
Steps = 100 & 1.0 & 0.05 & \textbf{100} & 0.0100 & 0.27 & 193 & 0.38 & -1.8 \\
Steps = 250 & 1.0 & 0.05 & \textbf{250} & 0.0100 & 0.55 & 161 & 0.17 & -0.83 \\
Steps = 500 & 1.0 & 0.05 & \textbf{500} & 0.0100 & 0.79 & 84 & 0.03 & -0.53 \\
Steps = 750 & 1.0 & 0.05 & \textbf{750} & 0.0100 & 0.84 & 17 & 0.01 & -0.50 \\
Steps = 1250 & 1.0 & 0.05 & \textbf{1250} & 0.0100 & 0.84 & 20 & 0.01 & -0.49 \\
Steps = 1500 & 1.0 & 0.05 & \textbf{1500} & 0.0100 & 0.88 & 14 & 0.01 & -0.48 \\
\midrule
\multicolumn{9}{l}{\textit{Ablation on the Step Size (SS)}} \\
SS = 0.0005 & 1.0 & 0.05 & 1000 & \textbf{0.0005} & 0.25 & 209 & 0.47 & -2.8\\
SS = 0.001 & 1.0 & 0.05 & 1000 & \textbf{0.0010} & 0.31 & {--} & 0.36 & -1.6 \\ 
SS = 0.005 & 1.0 & 0.05 & 1000 & \textbf{0.005} & 0.80 & 18 & 0.018 & -0.51\\
SS = 0.02 & 1.0 & 0.05 & 1000 & \textbf{0.02} & 0.81 & 76 & 0.018 & -0.49\\
SS = 0.05 & 1.0 & 0.05 & 1000 & \textbf{0.05} & 0.50 & 108 & 0.07 & -0.54\\
\bottomrule
\end{tabular}
\end{table}

\newpage
\begin{table}[!htb]
\centering
\caption{Hyperparameter Ablation Study evaluated by computing the MD committor. The main result configuration is shown first. Each subsequent row ablates a single hyperparameter, with the modified value in \textbf{bold}. No value means that the SAA was unstable and the force field eventually returns NaN. The number of stable samples corresponds to the number of structure for which running MD for 2 ps starting, from these configurations, is stable. The metrics were evaluated by sampling 100 configurations from ASTRA.}
\label{tab:hyperparameter_ablations_committor}
\sisetup{
    detect-weight=true, 
    detect-family=true,
    table-align-text-post=false 
}
\footnotesize
\begin{tabular}{lccccccc}
\toprule
\multirow{2}{*}{\textbf{Method}} & 
\multirow{2}{*}{\textbf{GS}} & 
\multirow{2}{*}{\textbf{PR}} & 
\multirow{2}{*}{\textbf{OS}} & 
\multirow{2}{*}{\textbf{SS}} & 
\textbf{\% of Comm.$\uparrow$} & 
\textbf{Num. stable samples} & \\
\cmidrule(lr){6-6} \cmidrule(lr){7-7}
& & & & & in [0.4,0.6] & (out of 100) \\
\midrule
Main & 1.0 & 0.05 & 1000 & 0.0100 & 85 & 76 \\
\midrule
\multicolumn{7}{l}{\textit{Ablation on the Guidance Scale (GS)}} \\
GS = 2.0 & \textbf{2.0} & 0.05 & 1000 & 0.0100 & 81 & 88 \\
GS = 3.0 & \textbf{3.0} & 0.05 & 1000 & 0.0100 & 80 & 78 \\
\midrule
\multicolumn{7}{l}{\textit{Ablation on the Pause Ratio (PR)}} \\
PR = 0.00 & 1.0 & \textbf{0.00} & 1000 & 0.0100 & {--} & {--} \\
PR = 0.25 & 1.0 & \textbf{0.25} & 1000 & 0.0100 & 42 & 44 \\
PR = 0.75 & 1.0 & \textbf{0.75} & 1000 & 0.0100 & 61 & 79 \\
PR = 0.10 & 1.0 & \textbf{0.10} & 1000 & 0.0100 & 21 & 98 \\
\midrule
\multicolumn{7}{l}{\textit{Ablation on the Optimization Steps (OS)}} \\
Steps = 100 & 1.0 & 0.05 & \textbf{100} & 0.0100 & 19 & 96 \\
Steps = 250 & 1.0 & 0.05 & \textbf{250} & 0.0100 & 40 & 82 \\
Steps = 500 & 1.0 & 0.05 & \textbf{500} & 0.0100 & 75 & 76 \\
Steps = 750 & 1.0 & 0.05 & \textbf{750} & 0.0100 & 88 & 75 \\
Steps = 1250 & 1.0 & 0.05 & \textbf{1250} & 0.0100 & 86 & 76 \\
Steps = 1500 & 1.0 & 0.05 & \textbf{1500} & 0.0100 & 84 & 74 \\
\midrule
\multicolumn{7}{l}{\textit{Ablation on the Step Size (SS)}} \\
SS = 0.0005 & 1.0 & 0.05 & 1000 & \textbf{0.0005} & 12 & 100 \\
SS = 0.001 & 1.0 & 0.05 & 1000 & \textbf{0.0010} & 15 & 100 \\ 
SS = 0.005 & 1.0 & 0.05 & 1000 & \textbf{0.005} & 80 & 78 \\
SS = 0.02 & 1.0 & 0.05 & 1000 & \textbf{0.02} & 68 & 77 \\
SS = 0.05 & 1.0 & 0.05 & 1000 & \textbf{0.05} & 11 & 14 \\
\bottomrule
\end{tabular}
\end{table}

\clearpage
\subsubsection{Quality of samples from classic interpolation approaches} \label{quality linear geodesic}
\begin{figure}[htb!]
    \centering
    
    \begin{subfigure}{0.48\textwidth}
        \centering
        \includegraphics[width=\linewidth]{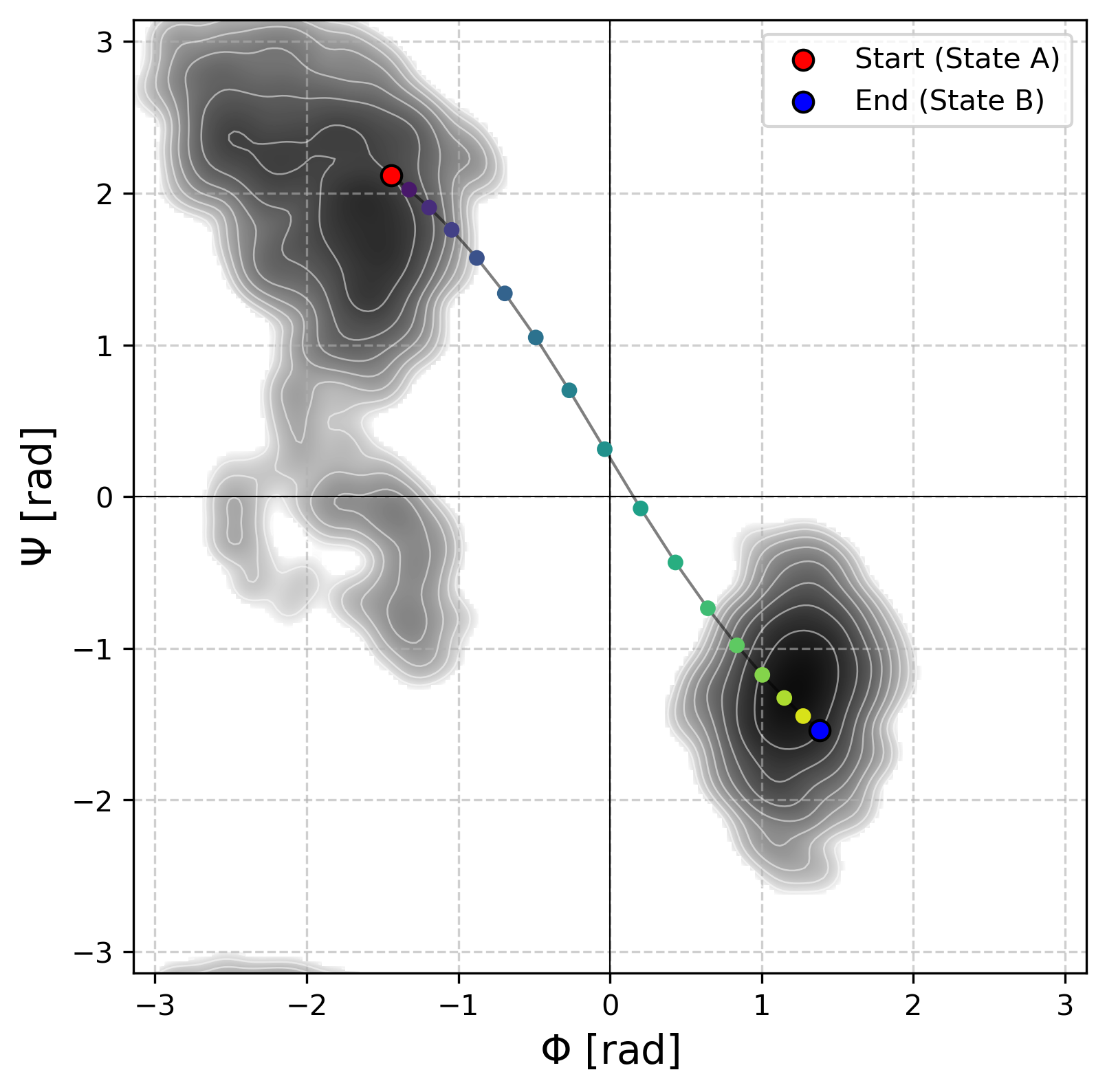}
        \caption{Linear interpolation.}
        \label{fig:abl:ala2_lin}
    \end{subfigure}
    \hfill
    \begin{subfigure}{0.48\textwidth}
        \centering
        \includegraphics[width=\linewidth]{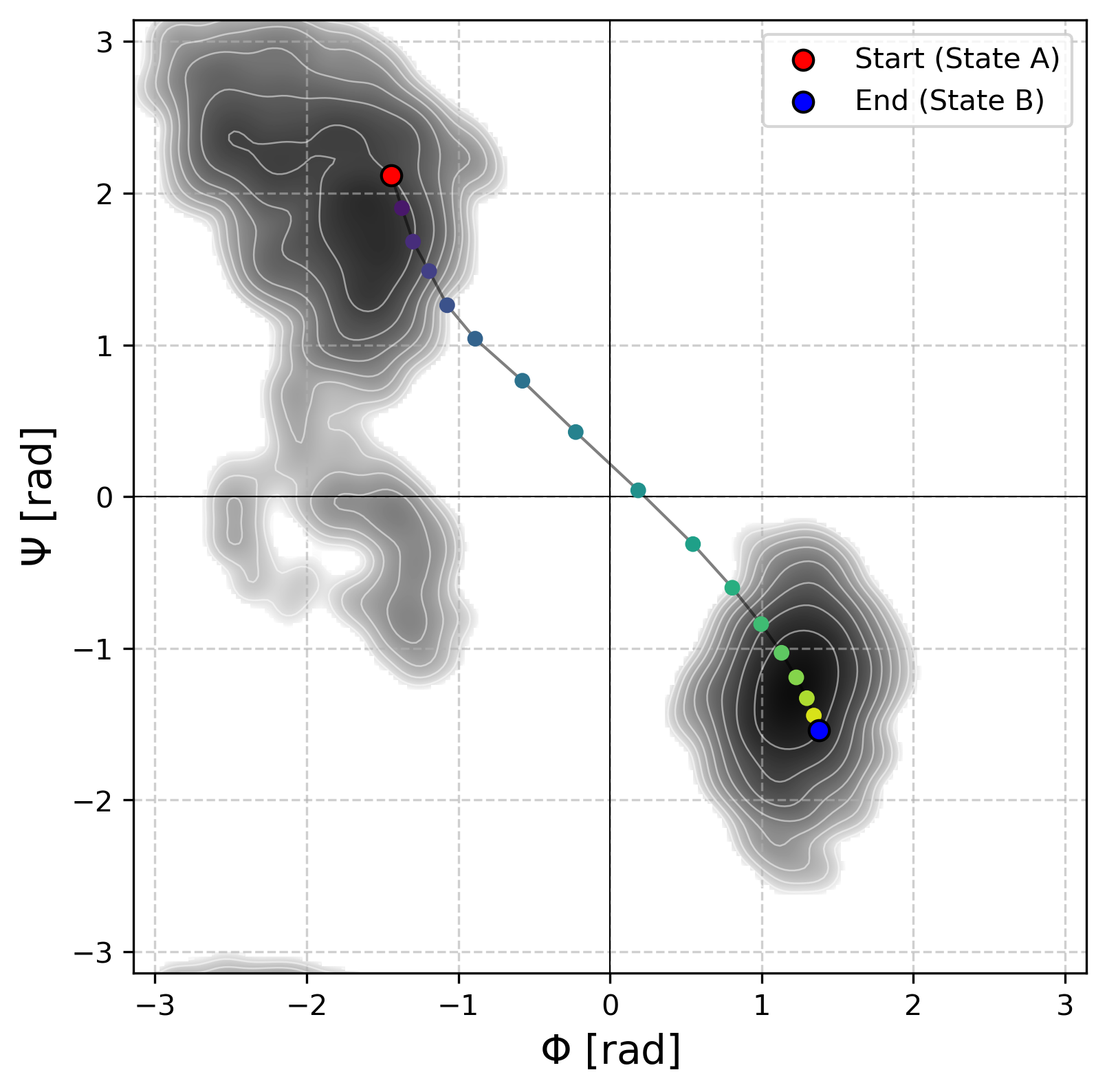}
        \caption{Geodesic interpolation.}
        \label{fig:abl:ala2_geo}
    \end{subfigure}
    
    \vspace{0.5cm} 

    \begin{subfigure}{0.48\textwidth}
        \centering
        \includegraphics[width=\linewidth]{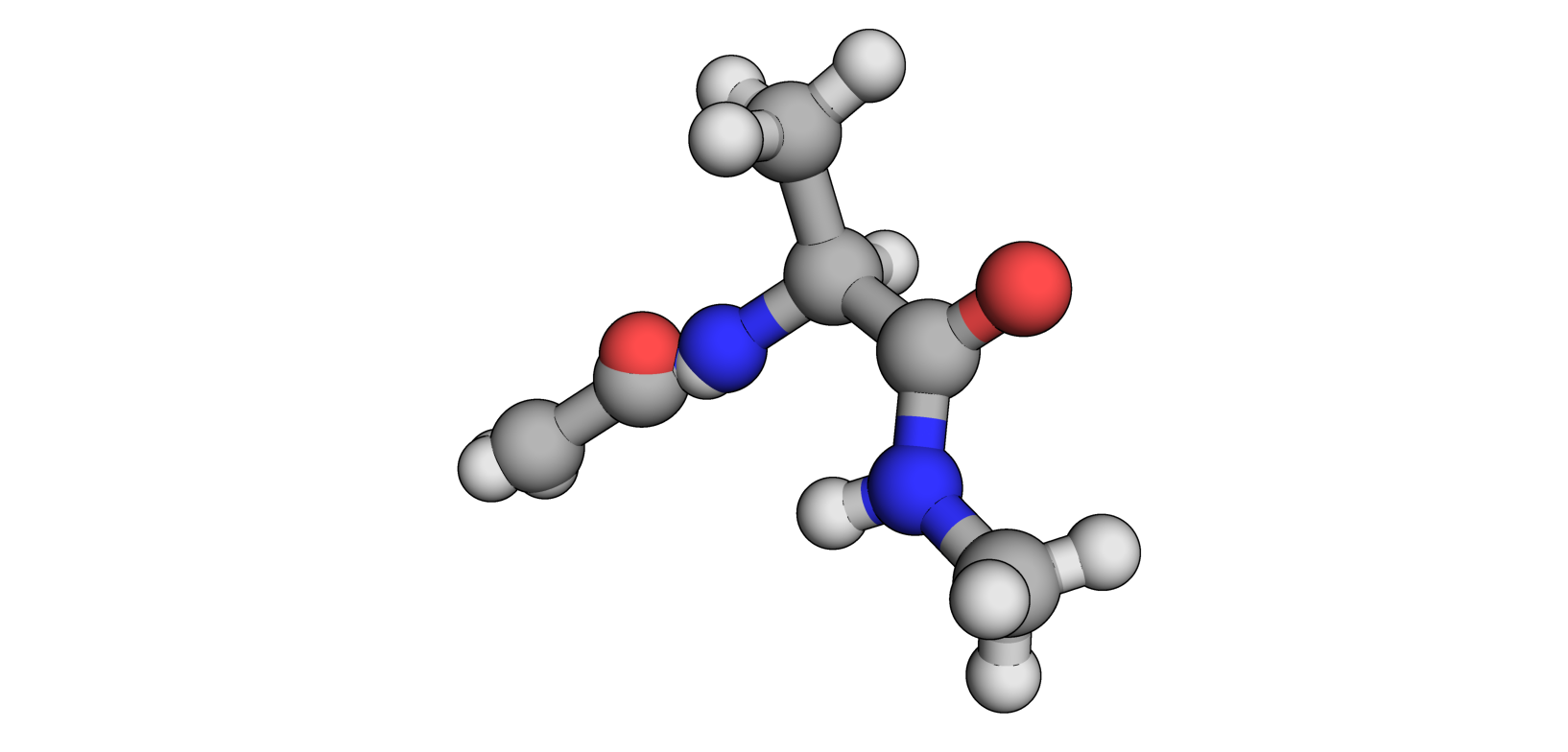}
        \caption{Structure of linear interpolation.}
        \label{fig:abl:ala2_lin_structure}
    \end{subfigure}
    \hfill
    \begin{subfigure}{0.48\textwidth}
        \centering
        \includegraphics[width=\linewidth]{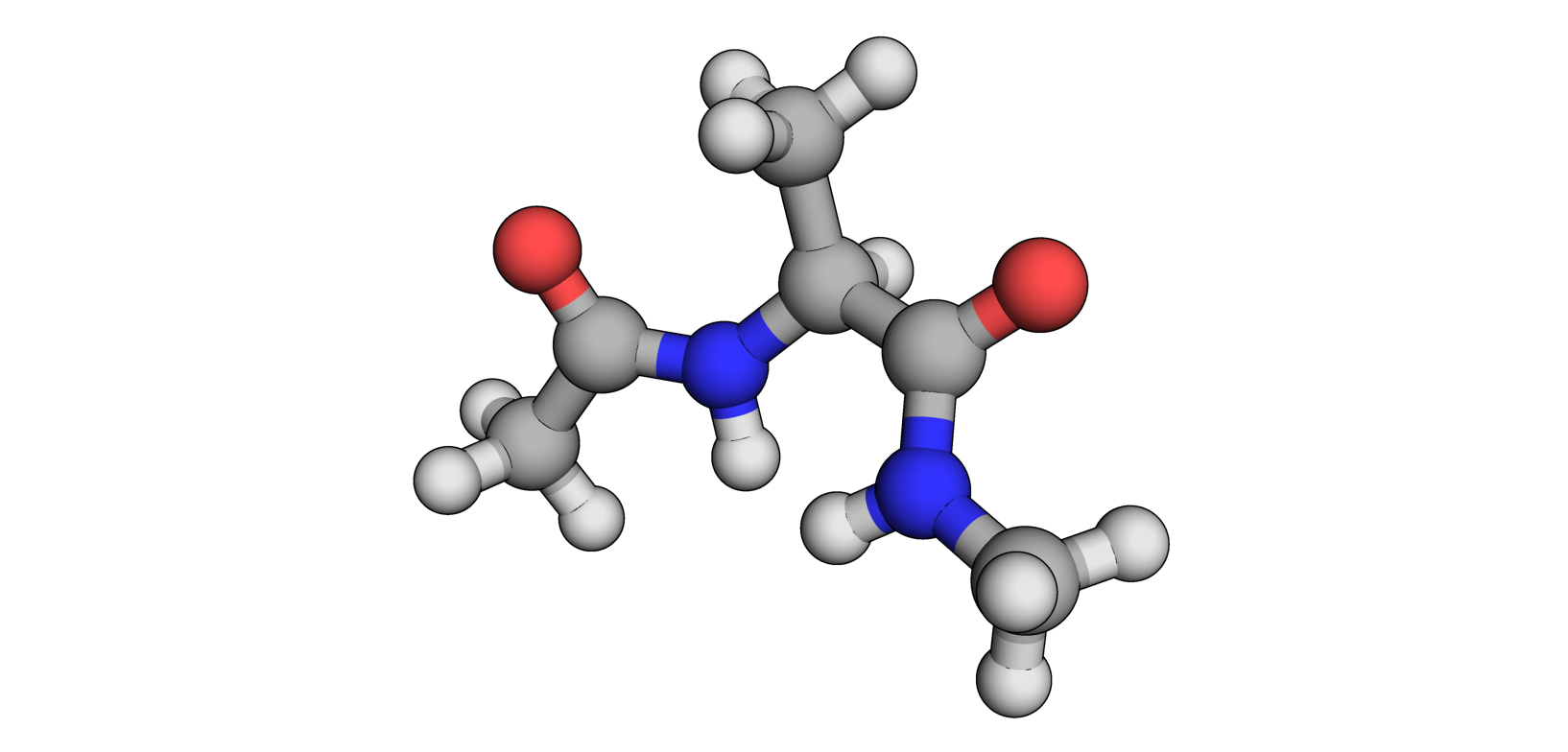}
        \caption{Structure of geodesic interpolation.}
        \label{fig:abl:ala2_geo_structure}
    \end{subfigure}
    
    \caption{Comparison of linear and geodesic interpolation for alanine dipeptide. The top row visualizes the interpolated pathways on the potential energy surface (PES), while the bottom row shows representative intermediate structures.}
    \label{fig:abl:ala2_basline}
\end{figure}

\begin{figure}[htb!]
    \centering
    
    \begin{subfigure}{0.48\textwidth}
        \centering
        \includegraphics[width=\linewidth]{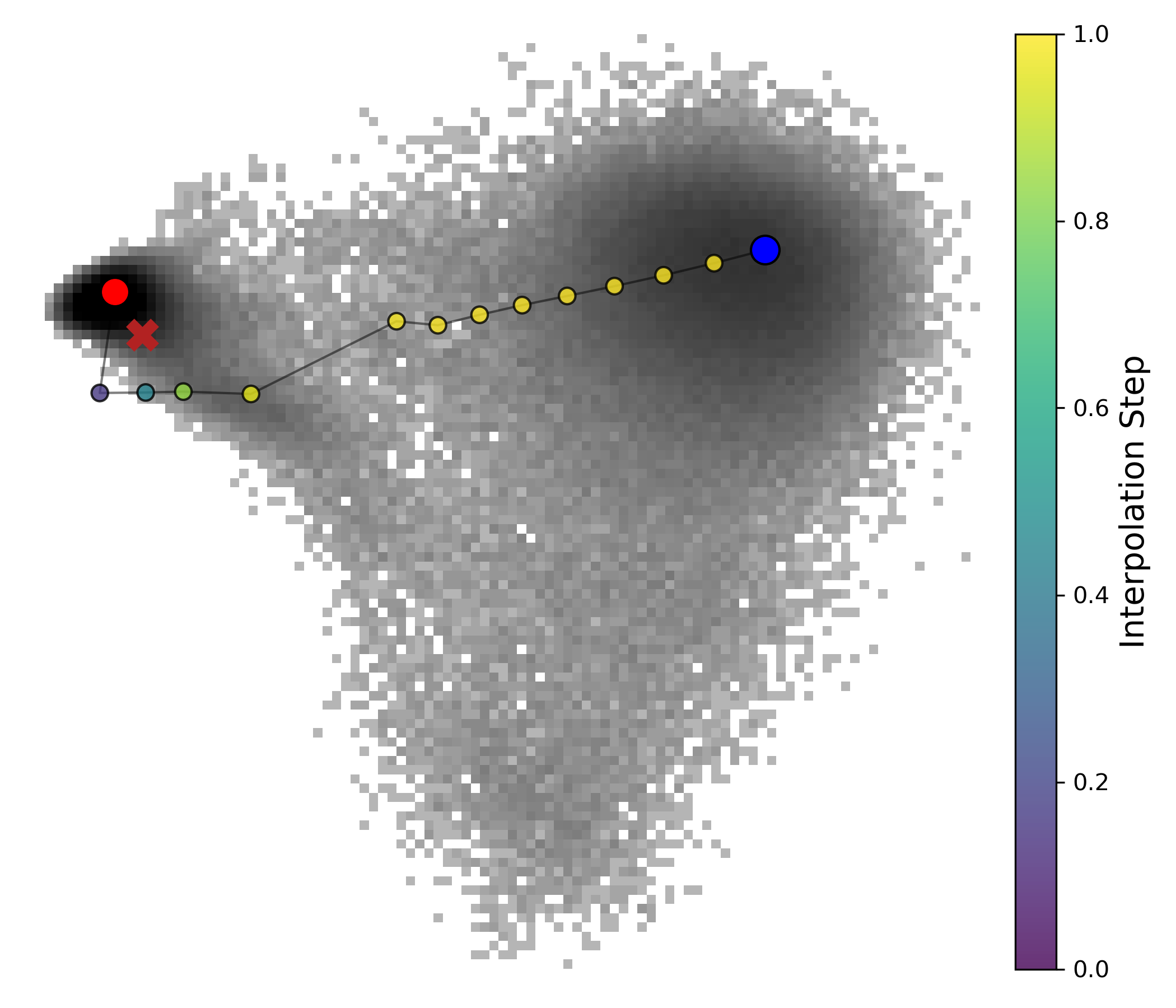}
        \caption{Linear interpolation.}
        \label{fig:abl:chig_lin}
    \end{subfigure}
    \hfill
    \begin{subfigure}{0.48\textwidth}
        \centering
        \includegraphics[width=\linewidth]{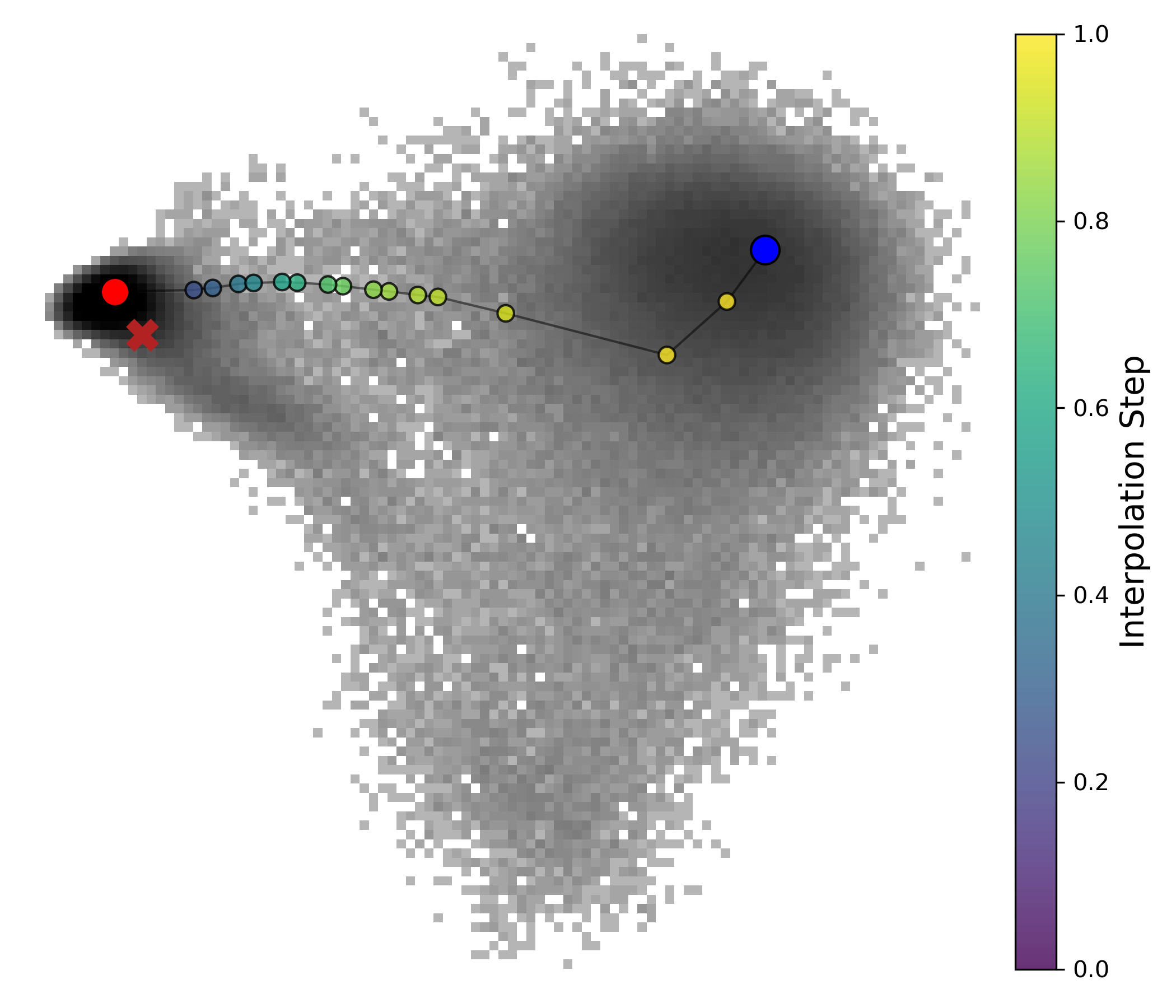}
        \caption{Geodesic interpolation.}
        \label{fig:abl:chig_geo}
    \end{subfigure}
    
    \vspace{0.5cm} 

    \begin{subfigure}{0.48\textwidth}
        \centering
        \includegraphics[width=\linewidth]{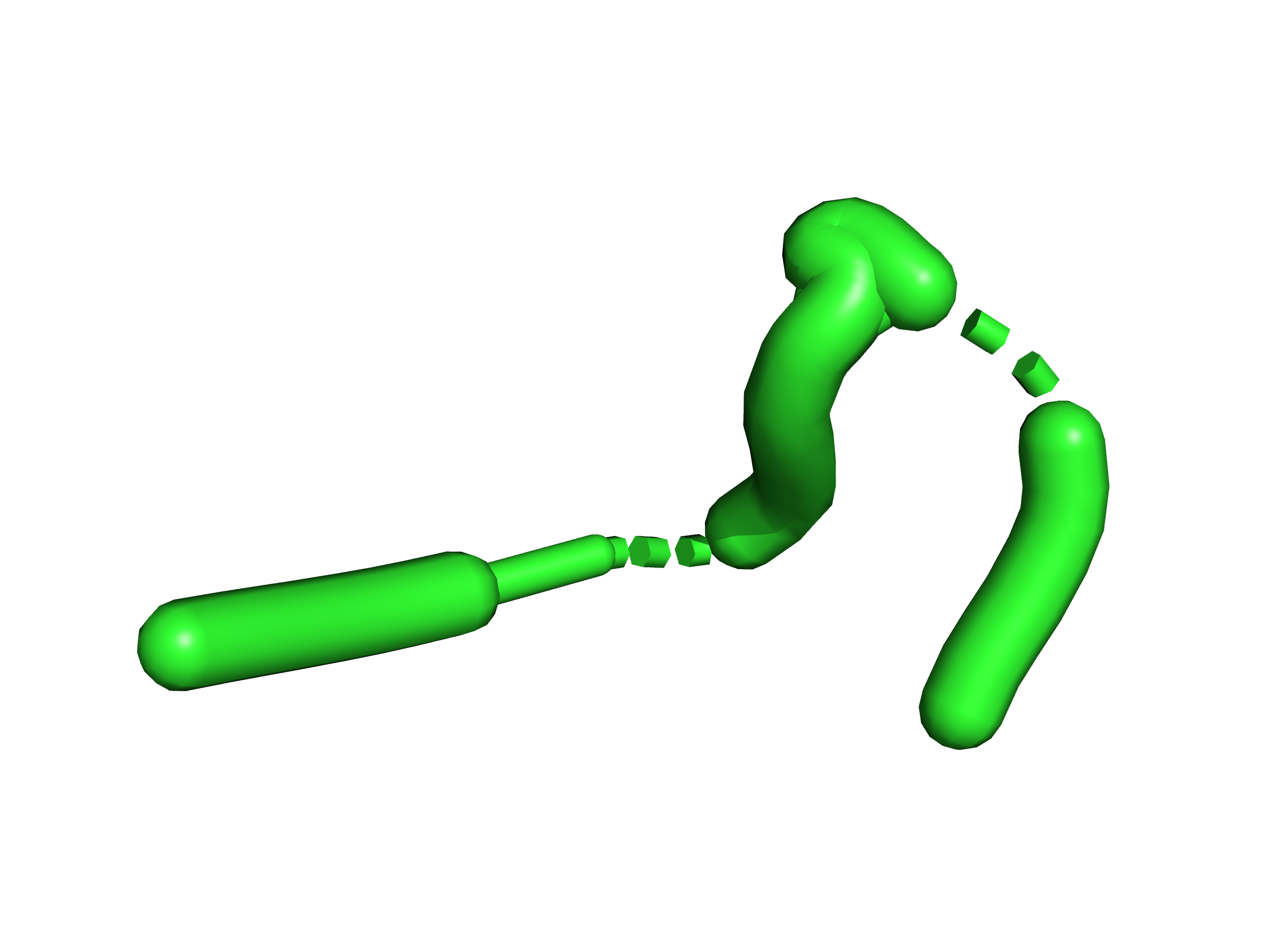}
        \caption{Structure of linear interpolation.}
        \label{fig:abl:chig_lin_structure}
    \end{subfigure}
    \hfill
    \begin{subfigure}{0.48\textwidth}
        \centering
        \includegraphics[width=\linewidth]{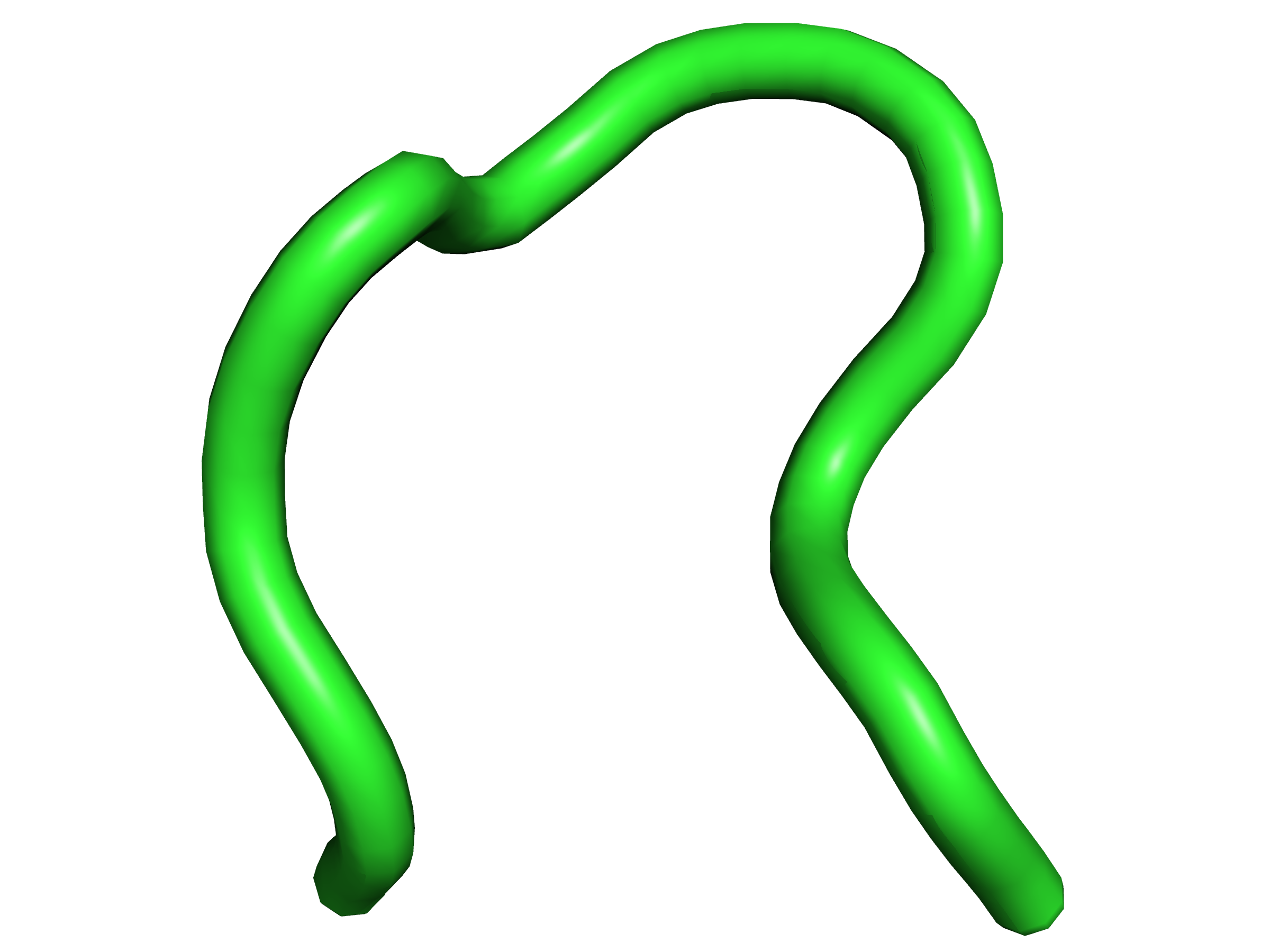}
        \caption{Structure of geodesic interpolation.}
        \label{fig:abl:chig_geo_structure}
    \end{subfigure}
    
    \caption{Comparison of linear and geodesic interpolation for the folding of chignolin. The top row displays the pathways on the PES, and the bottom row shows intermediate molecular structures.}
    \label{fig:abl:chig_basline}
\end{figure}

We evaluate two baseline methods for generating conformational pathways: linear interpolation, performed after applying the Kabsch algorithm~\cite{lawrence2019purely}, and geodesic interpolation based on internal coordinates, as proposed in Ref.~\citenum{zhu2019geodesic} (Figures~\ref{fig:abl:ala2_basline} and \ref{fig:abl:chig_basline}).
Specifically for the alanine dipeptide, both methods fail to produce physically realistic pathways, as they traverse significant energy barriers on the PES.
Linear interpolation, in particular, is prone to generating intermediate conformations with severe steric clashes, leading to geometrically invalid structures. 
While the geodesic method avoids such direct structural inconsistencies, the pathway it defines remains energetically prohibitive and is, therefore, an unviable representation of the transition. Therefore, further optimization is needed, using NEB for example. For such flexible systems with structurally very different pathways, these classical interpolation methods might struggle to provide a broad and valid set of initial guesses for optimization algorithms.

\clearpage

\subsection{Noise level for SAA optimization}
\begin{figure}[!htb]
    \centering
    \begin{subfigure}[b]{0.48\textwidth}
        \centering
        \includegraphics[width=\textwidth]{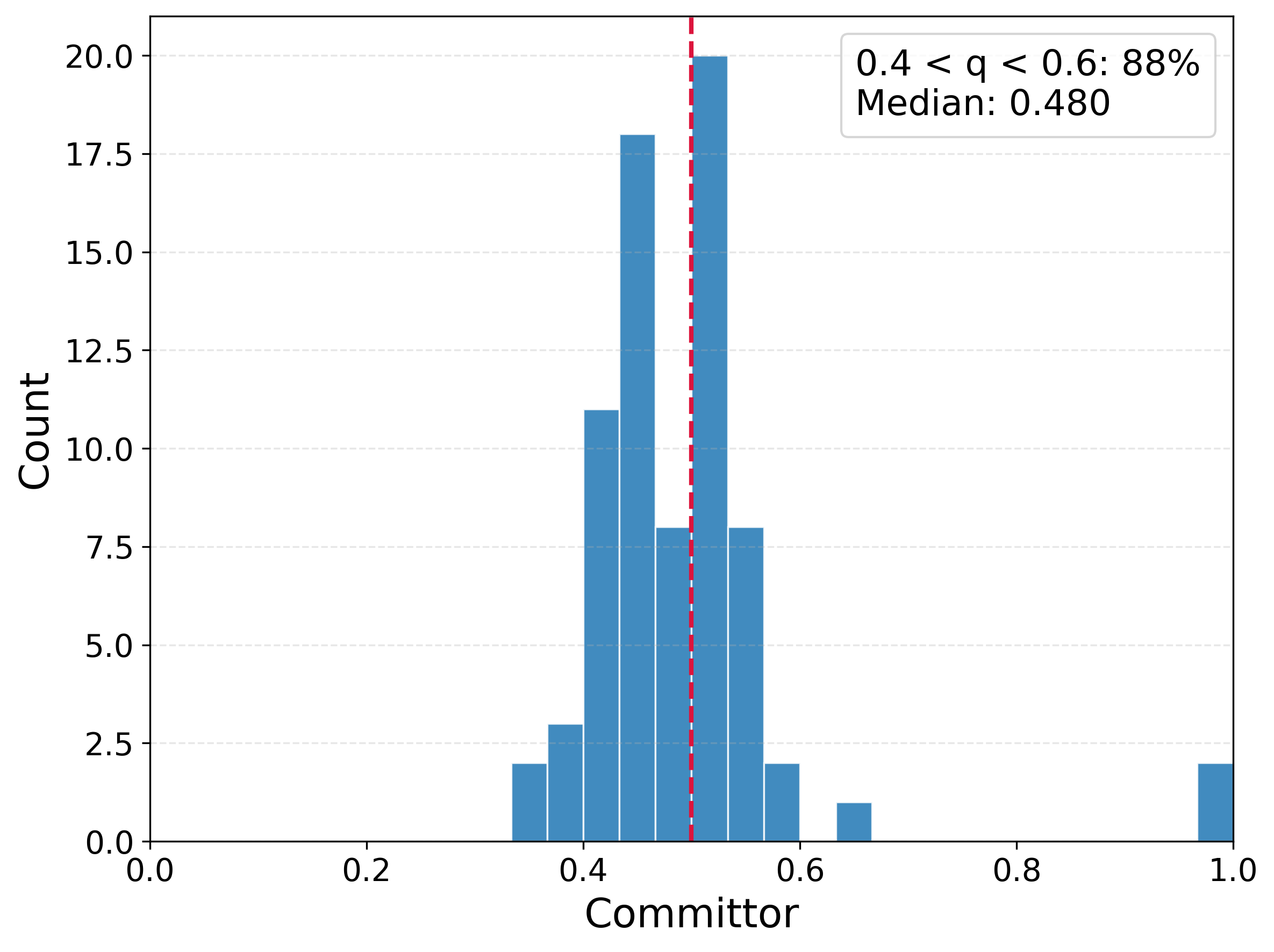}
        \caption{Latent-level optimization.}
        \label{fig:abl:noise_level_opt}
    \end{subfigure}
    \hfill 
    \begin{subfigure}[b]{0.48\textwidth}
        \centering
        \includegraphics[width=\textwidth]{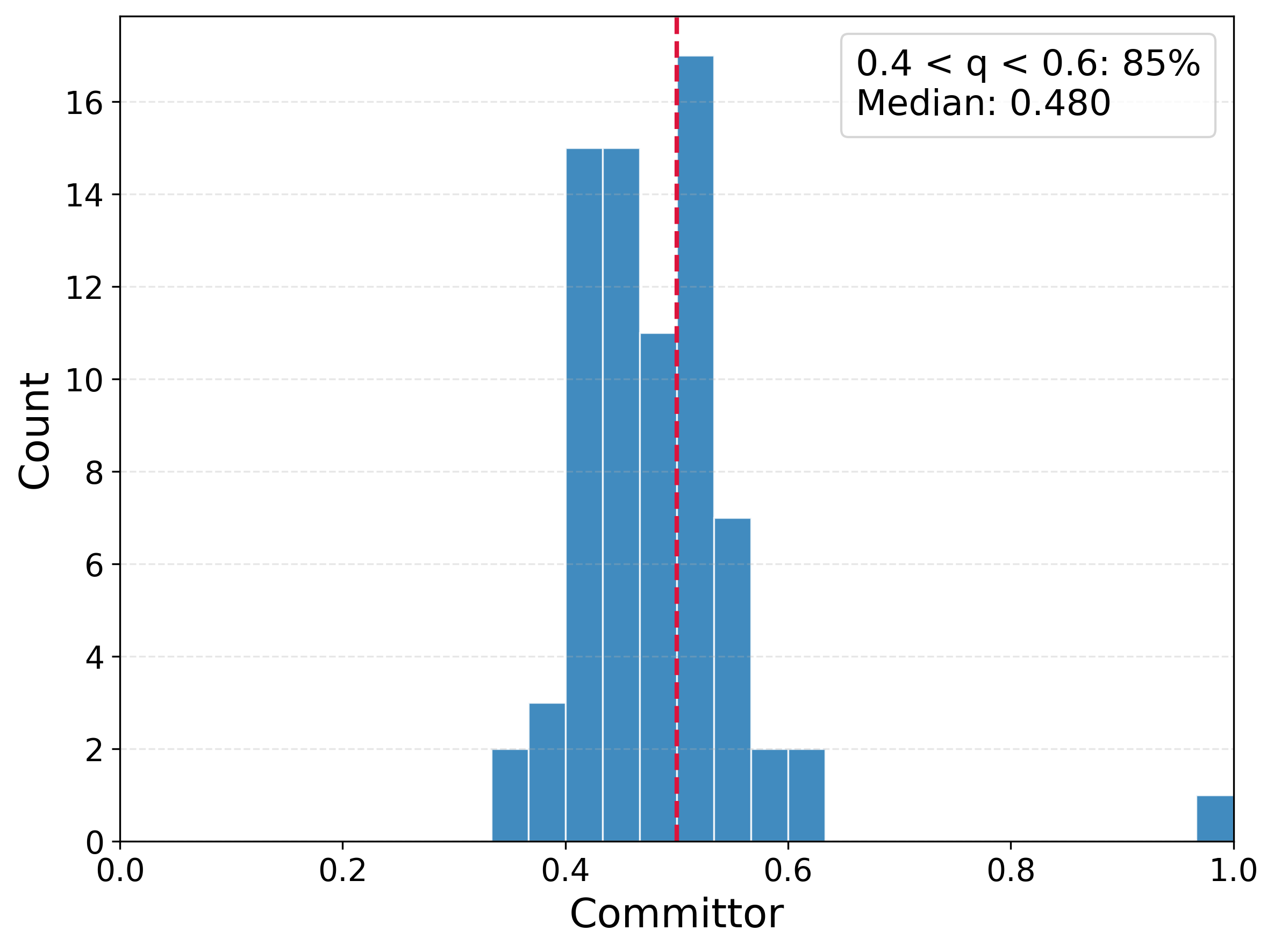}
        \caption{Clean structure-level optimization.}
        \label{fig:abl:clean_opt}
    \end{subfigure}
    \caption{Committor histograms comparison of a latent-level optimization and a clean structure-level optimization. This is for PR=0.05.}
    \label{fig:abl:noised_opt} 
\end{figure}
In Figure \ref{fig:abl:noised_opt}, we provide the comparison between optimizing at the latent level or at the clean structure level. The results show that the performance is almost identical. This stems from the fact that the force predictions are computed at the clean structure level while the SAA direction is computed at the latent level inducing a discrepancy in both cases. 

\newpage
\subsection{SAA and NEB from standard interpolation methods}

\begin{figure}[!htb]
    \centering
    \begin{subfigure}[b]{0.48\textwidth}
        \centering
        \includegraphics[width=\textwidth]{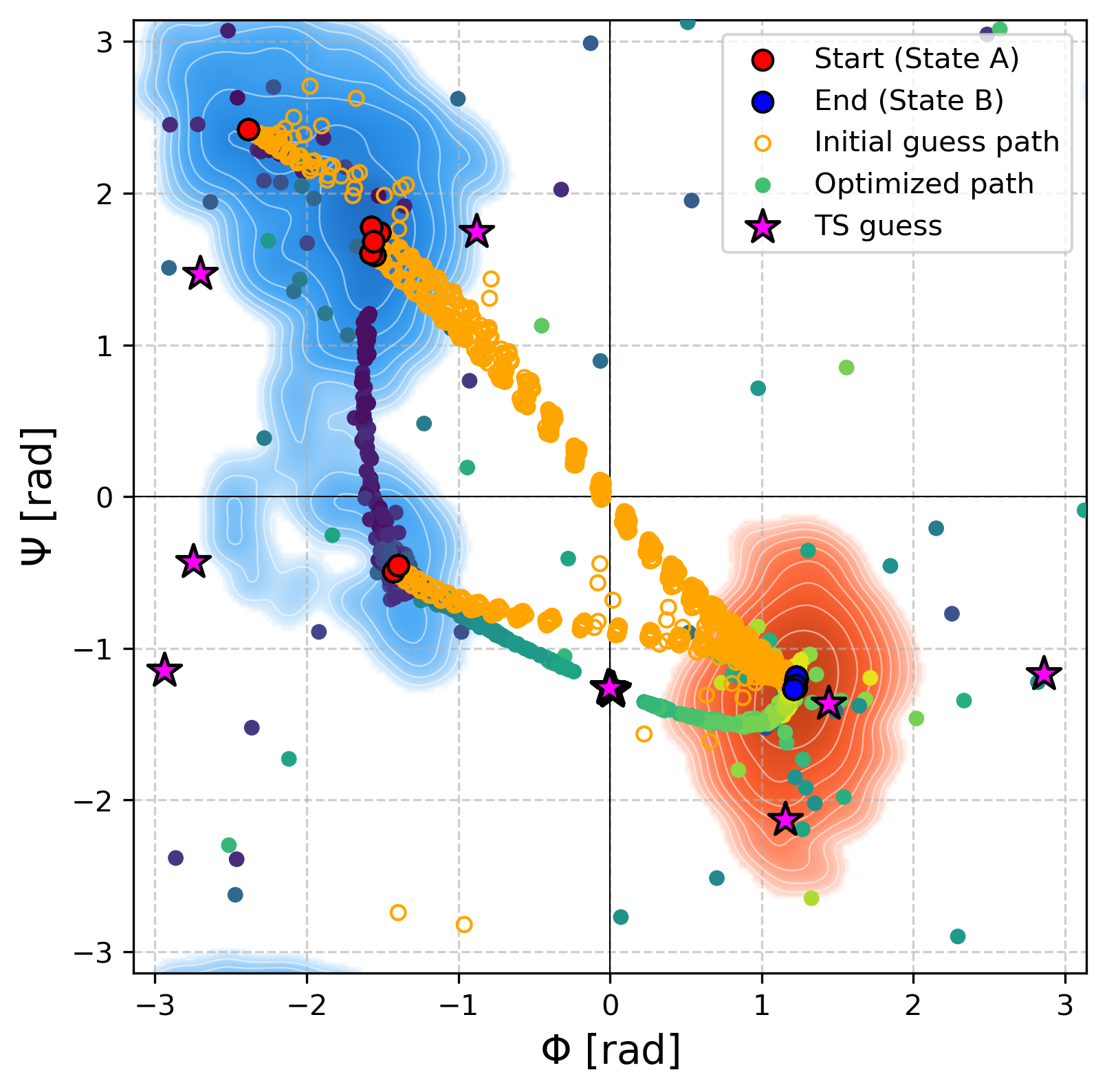}
        \caption{Linear Interpolation.}
        \label{fig:abl:lin_min_neb}
    \end{subfigure}
    \hfill 
    \begin{subfigure}[b]{0.48\textwidth}
        \centering
        \includegraphics[width=\textwidth]{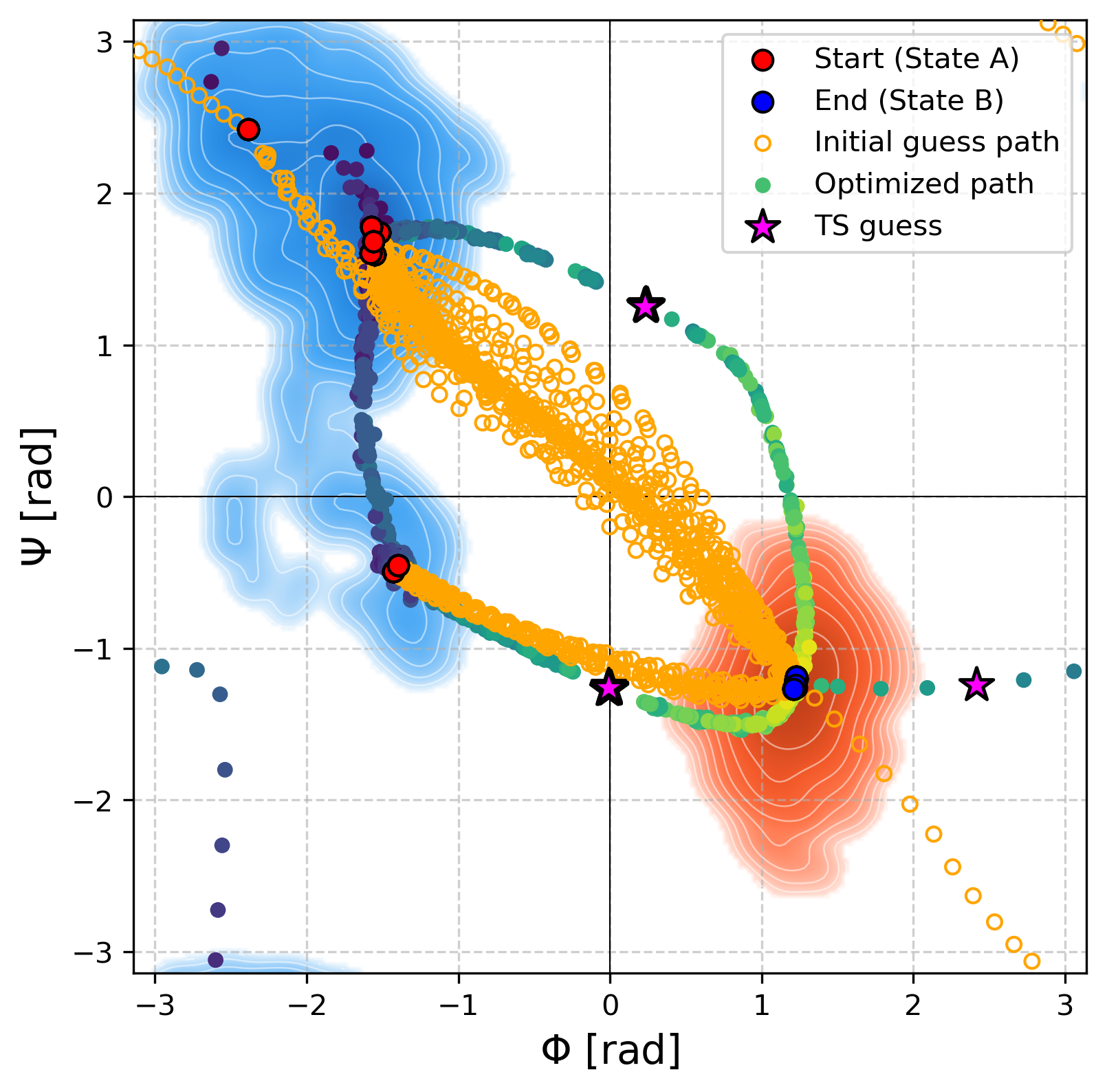}
        \caption{Geodesic Interpolation.}
        \label{fig:abl:geo_min_neb}
    \end{subfigure}
    \caption{NEB transition state guesses and minimum energy paths computed from linear and geodesic interpolation guess paths. The yellow empty circles correspond to the initial guesses computed by interpolating between the start (red) and end (blue) configurations using the given interpolation technique. The optimized paths are colored with the viridis color scheme.}
    \label{fig:abl:geo_lin_opt_neb} 
\end{figure}

\begin{figure}[!htb]
    \centering
    \begin{subfigure}[b]{0.48\textwidth}
        \centering
        \includegraphics[width=\textwidth]{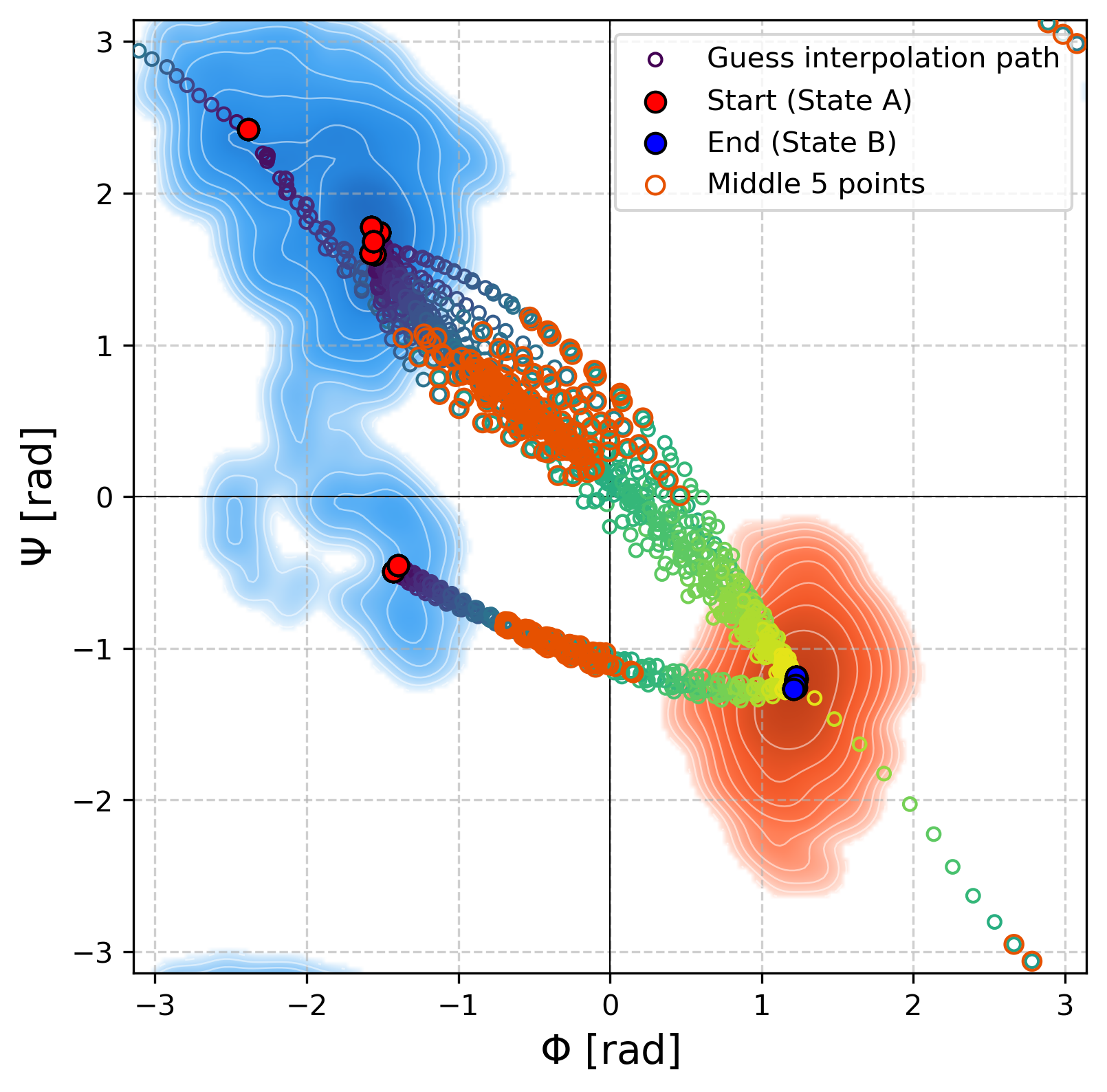}
        \caption{Minimized.}
        \label{fig:abl:geo_min}
    \end{subfigure}
    \hfill 
    \begin{subfigure}[b]{0.48\textwidth}
        \centering
        \includegraphics[width=\textwidth]{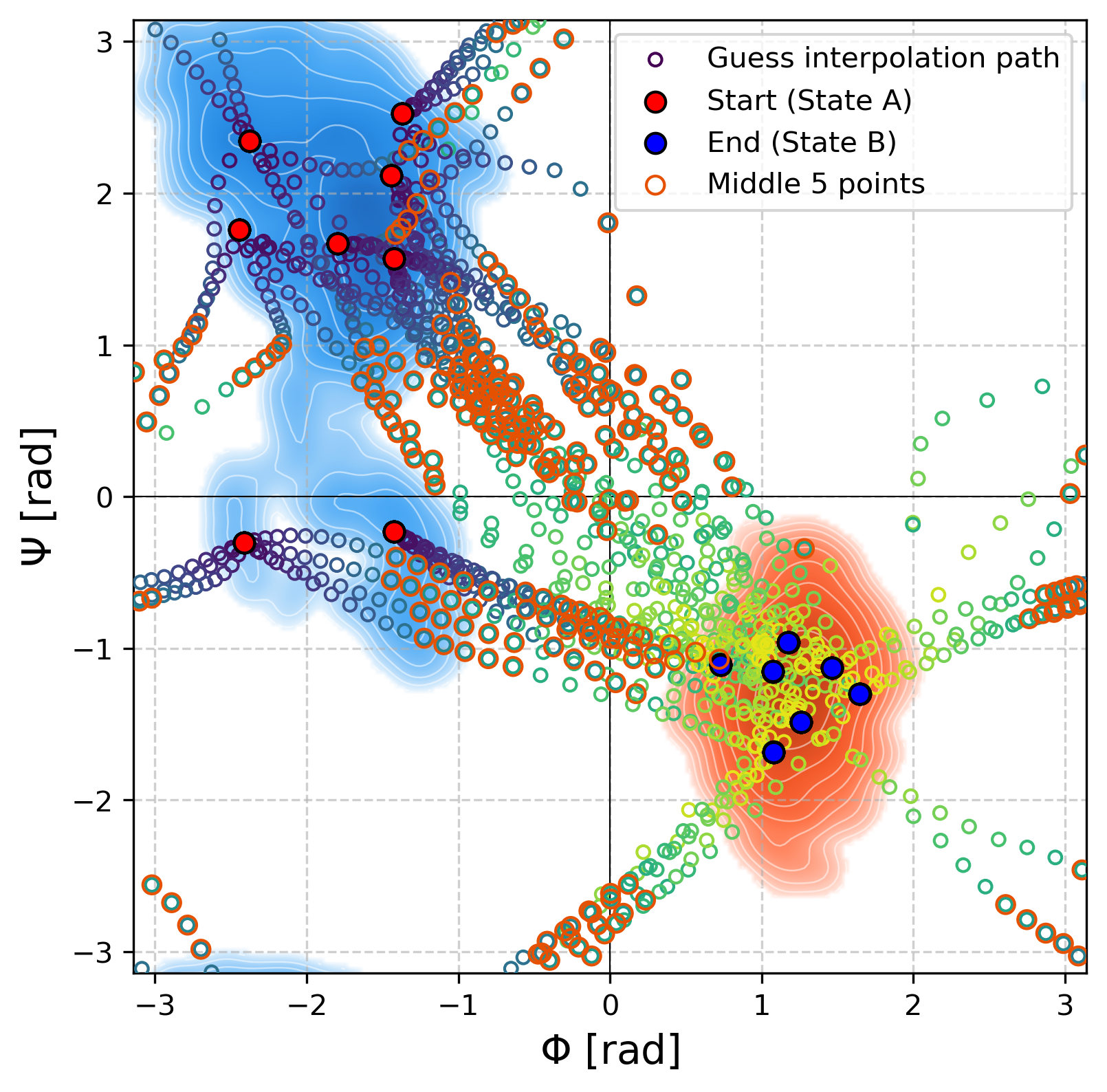}
        \caption{Non-minimized.}
        \label{fig:abl:geo_unmin}
    \end{subfigure}
    \caption{Interpolation paths using geodesic interpolation with the start and end structures used to evaluate SAA performance and the NEB baseline. We highlight in orange the 5 middle points for each path that were taken as input for SAA. The two graphs (a) and (b) correspond to whether the initial configurations are optimized with energy minimization before computing the interpolated path. As required by NEB, the minimized configuration (b) where used for NEB calculations.}
    \label{fig:abl:geo_abl} 
\end{figure}

Figure \ref{fig:abl:geo_lin_opt_neb} shows the optimized minimum energy paths obtained by running NEB starting from linear and geodesic-interpolated paths. The paths are made of 25 intermediate images. We clearly observe that NEB fails to converge the second TS from the linearly-interpolated path. This probably stems from the poor quality of these samples as highlighted in Section \ref{quality linear geodesic}. We also provide in Figure \ref{fig:abl:geo_abl} the details of the linear and geodesic interpolations. We show the different interpolated paths and the 5 selected middle frames that are used to initialize SAA for the results in Table \ref{tab:geo_lin_neb}. Additionally, we highlight that when energy minimization is not necessary (as is the case for ASTRA, unlike NEB), we can sample a broader ensemble of guesses, allowing for better coverage of the TSE.

\begin{table}[!htb]
\centering
\caption{Evaluation of TS samples generated by running SAA or NEB after initializing guess structures from classical interpolation algorithms (linear and geodesic interpolation). We report metrics associated with running the Dimer method from these structures. The "minimized" or "non-minimized" initialization corresponds to minimizing the energy of the structure before interpolating the initial guess path. Higher energy difference is better as all differences are negative.}
\label{tab:geo_lin_neb}
\sisetup{
    detect-weight=true, 
    detect-family=true,
    table-align-text-post=false 
}
\footnotesize
\begin{tabular}{lccccccc}
\toprule
\textbf{Method} & 
\textbf{\% of Comm.$\uparrow$} & 
\multirow{2}{*}{\textbf{Num. stable samples}} & 
\multirow{2}{*}{\textbf{\textbf{Conv. Rate$\uparrow$}}} & 
\multirow{2}{*}{\textbf{\textbf{Conv. Step$\downarrow$}}} & 
\textbf{Pos. RMSD$\downarrow$} & 
\textbf{Energy Diff.$\uparrow$} & \\
\cmidrule(lr){6-6} \cmidrule(lr){7-7}
&[0.4,0.6]& (out of 56) & & &(\AA) & (kcal$\cdot$mol$^{-1}$)\\
\midrule
\multicolumn{7}{l}{\textbf{Geodesic Interpolation (minimized)}} \\
SAA & 0.83 & 54 & 0.93 & 20 & 0.013 & -0.82 \\
NEB (baseline) &  & 56 & 0.98 &  & 0.07 & -0.009 \\
\midrule
\multicolumn{7}{l}{\textbf{Geodesic Interpolation (non-minimized)}} \\
SAA & 0.9 & 40 & 0.85 & 23 & 0.012 & -0.69 \\
NEB (baseline) &  & 5 & 0.0 & {--} & {--} & {--} \\
\midrule
\multicolumn{7}{l}{\textbf{Linear Interpolation (minimized)}} \\
SAA & 0.3 & 53 & 0.52 & 137 & 0.17 & -1.01 \\
NEB (baseline) &  & 50 & 0.98 &  & 0.008 & -0.025 \\

\bottomrule
\end{tabular}
\end{table}

\clearpage
\subsection{Score-Based Interpolation TS region covering}

\begin{figure}[!htb]
    \centering
    \begin{subfigure}[b]{0.48\textwidth}
        \centering
        \includegraphics[width=\textwidth]{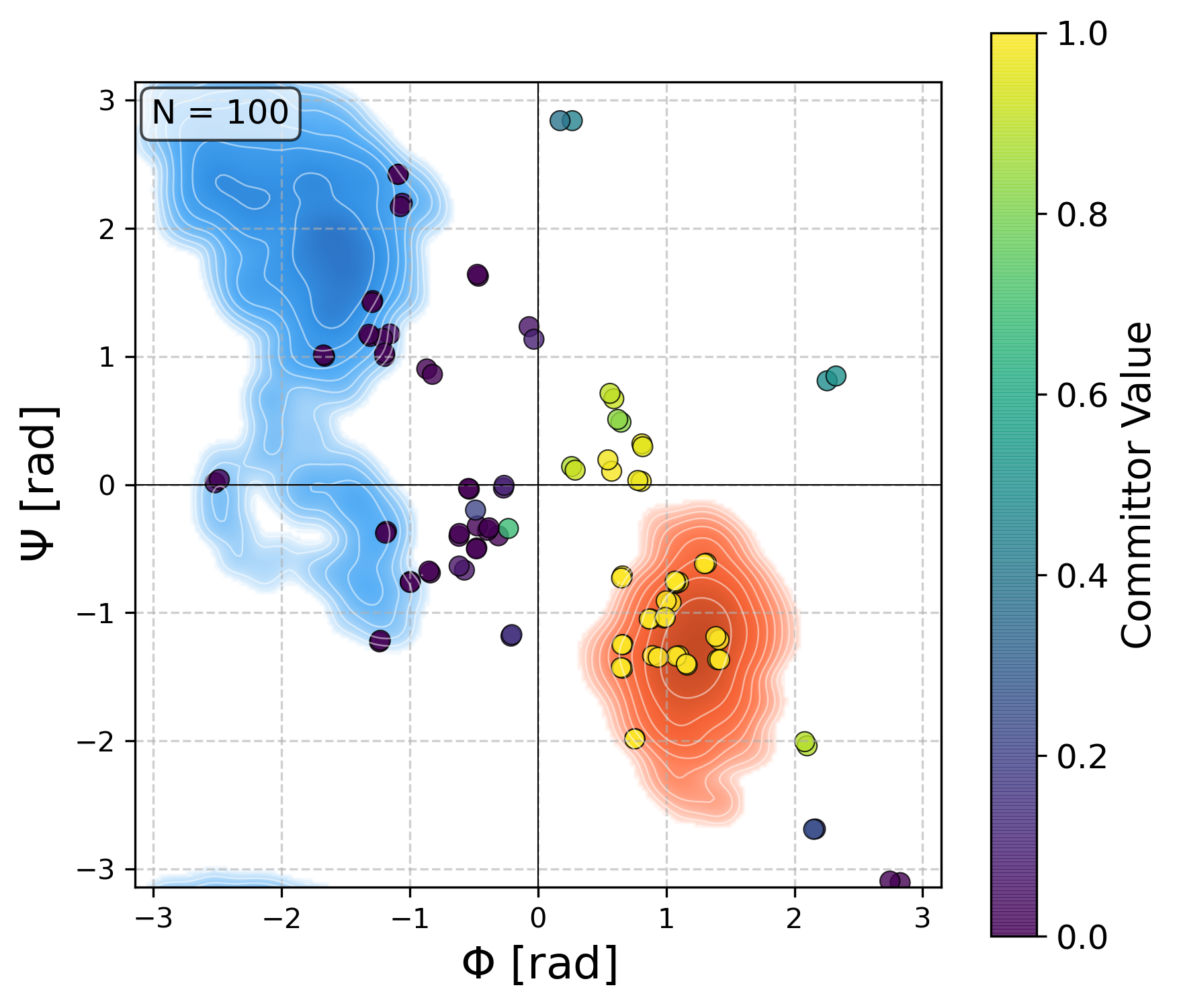}
        \caption{Guidance 1.0.}
        \label{fig:abl:and_nosaa_g1}
    \end{subfigure}
    \hfill 
    \begin{subfigure}[b]{0.48\textwidth}
        \centering
        \includegraphics[width=\textwidth]{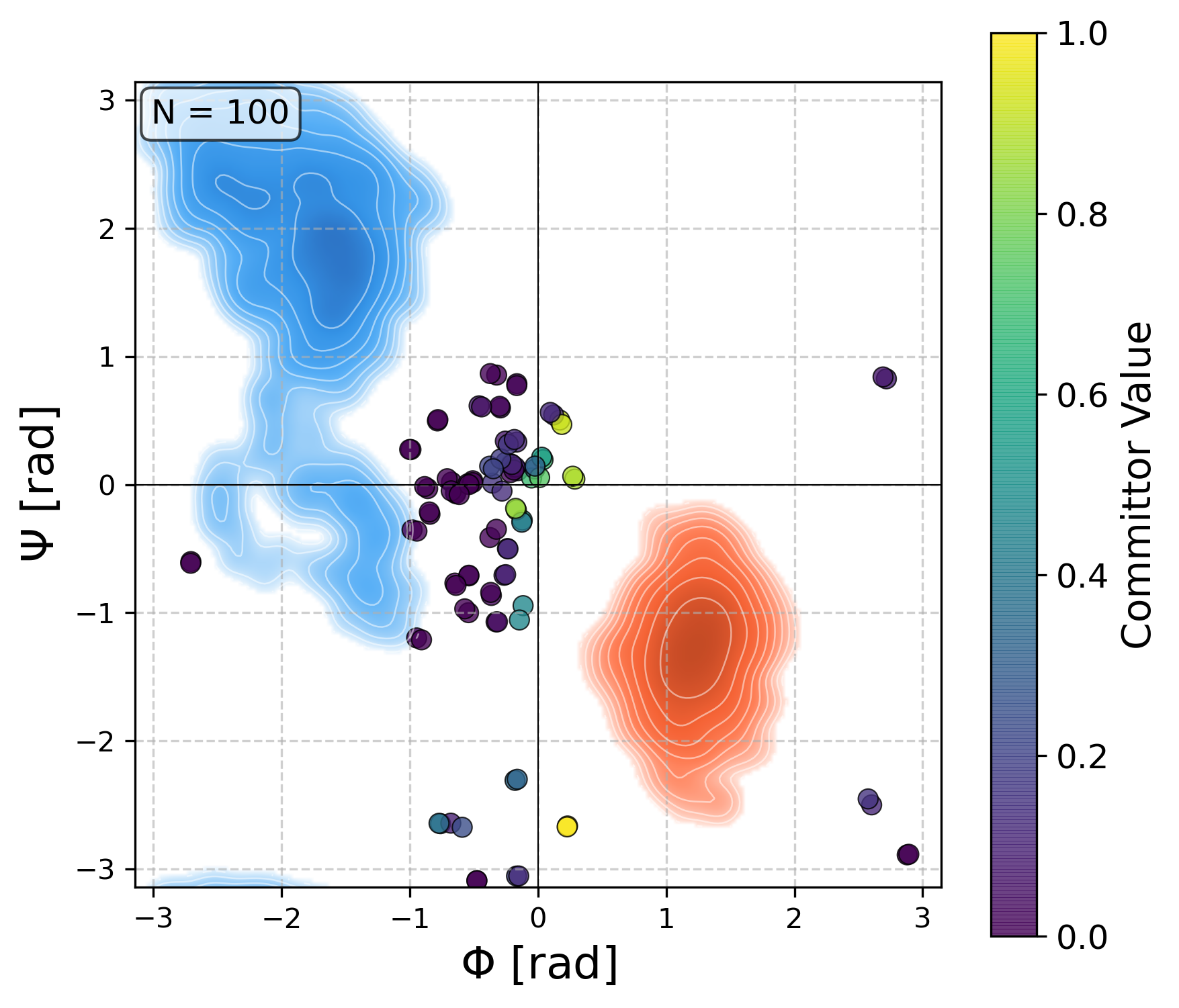}
        \caption{Guidance 2.0.}
        \label{fig:abl:and_nosaa_g2}
    \end{subfigure}
    \hfill
    \begin{subfigure}[b]{0.48\textwidth}
        \centering
        \includegraphics[width=\textwidth]{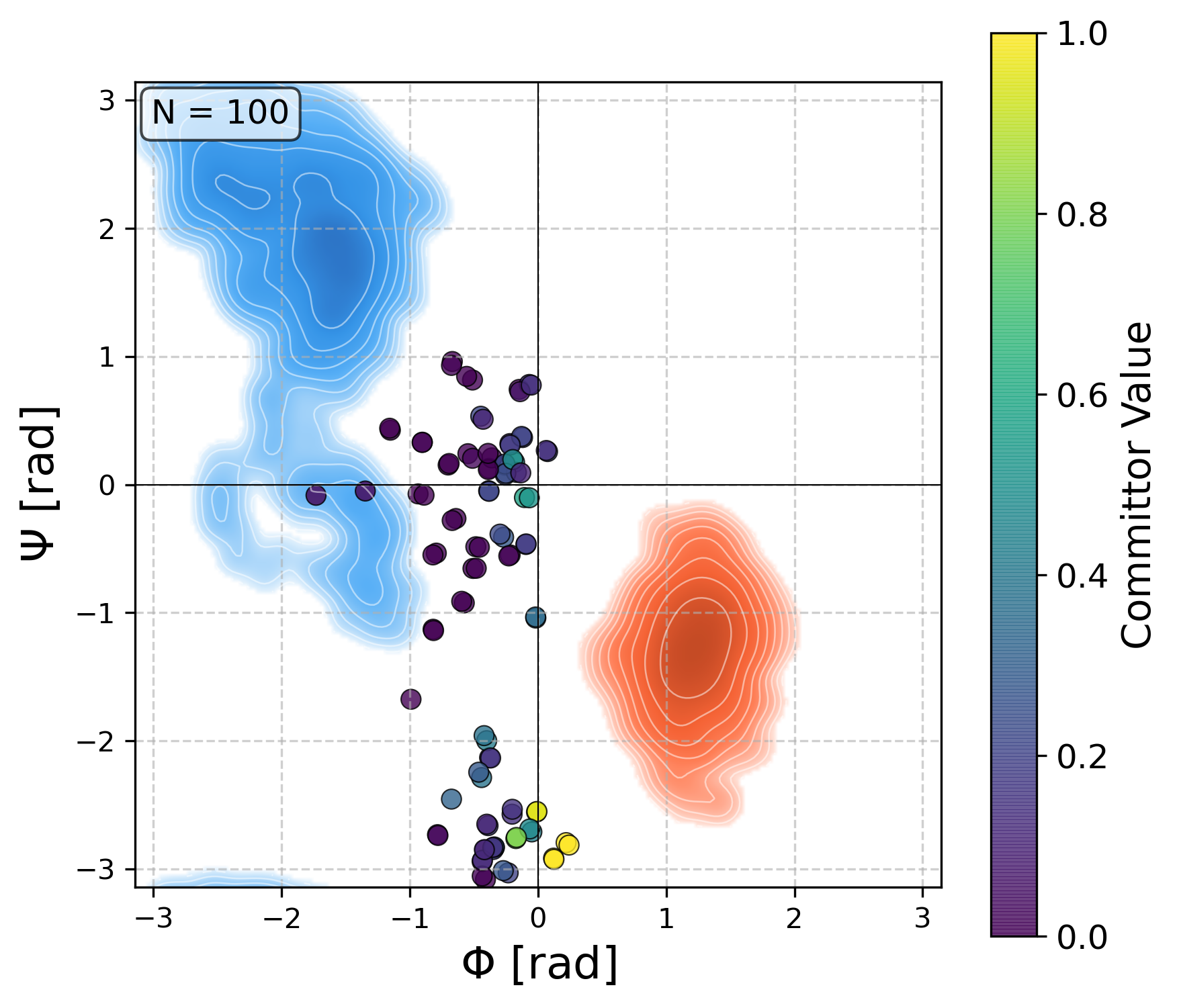}
        \caption{Guidance 3.0.}
        \label{fig:abl:and_nosaa_g3}
    \end{subfigure}
    \caption{Effect of the guidance scale when sampling 100 structures using Isodensity Interpolation to guide the reverse diffusion process.}
    \label{fig:abl:and_nosaa_guid} 
\end{figure}

\begin{figure}[!htb]
    \centering
    \begin{subfigure}[b]{0.48\textwidth}
        \centering
        \includegraphics[width=\textwidth]{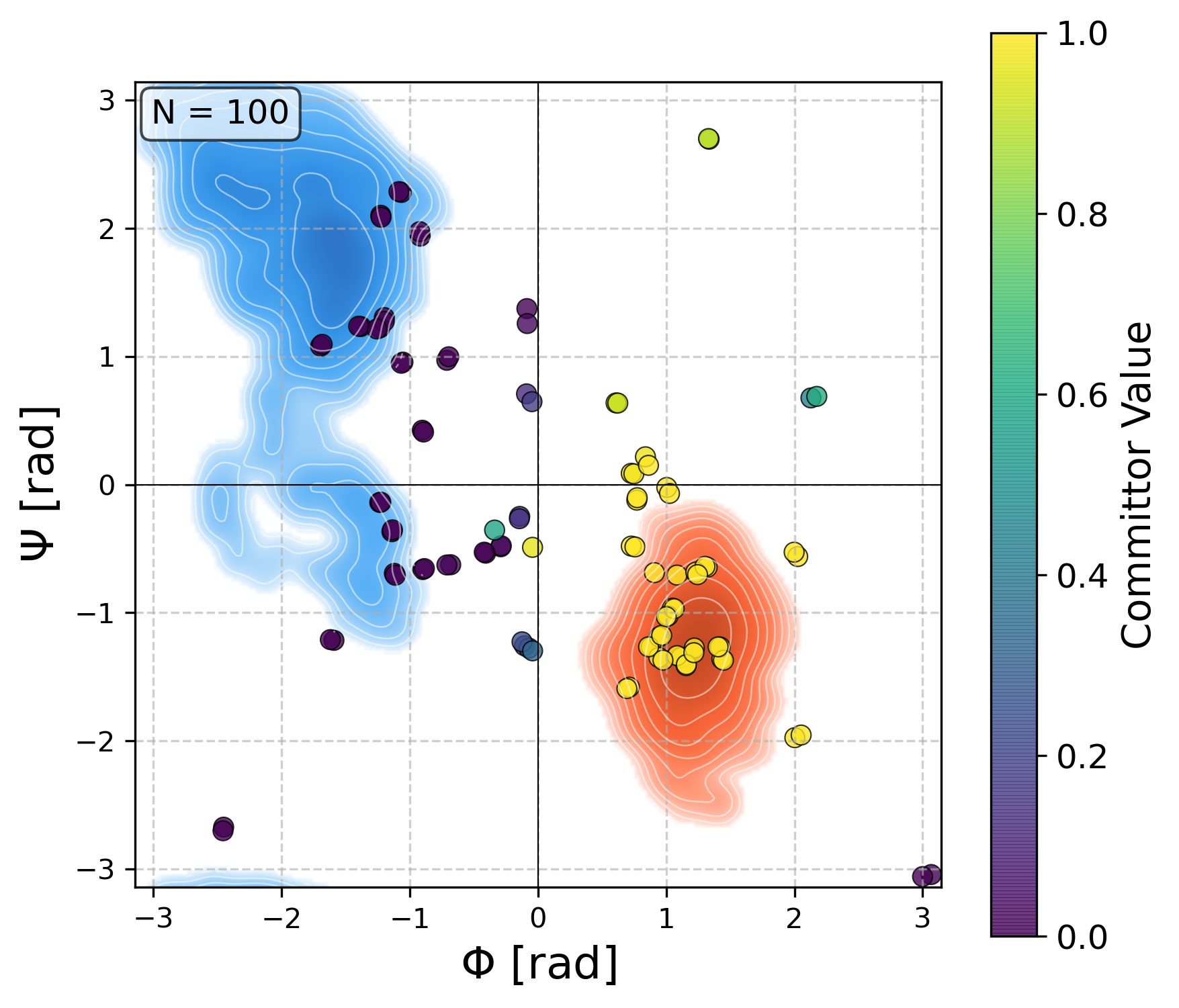}
        \caption{Guidance 1.0.}
        \label{fig:abl:sa_nosaa_g1}
    \end{subfigure}
    \hfill 
    \begin{subfigure}[b]{0.48\textwidth}
        \centering
        \includegraphics[width=\textwidth]{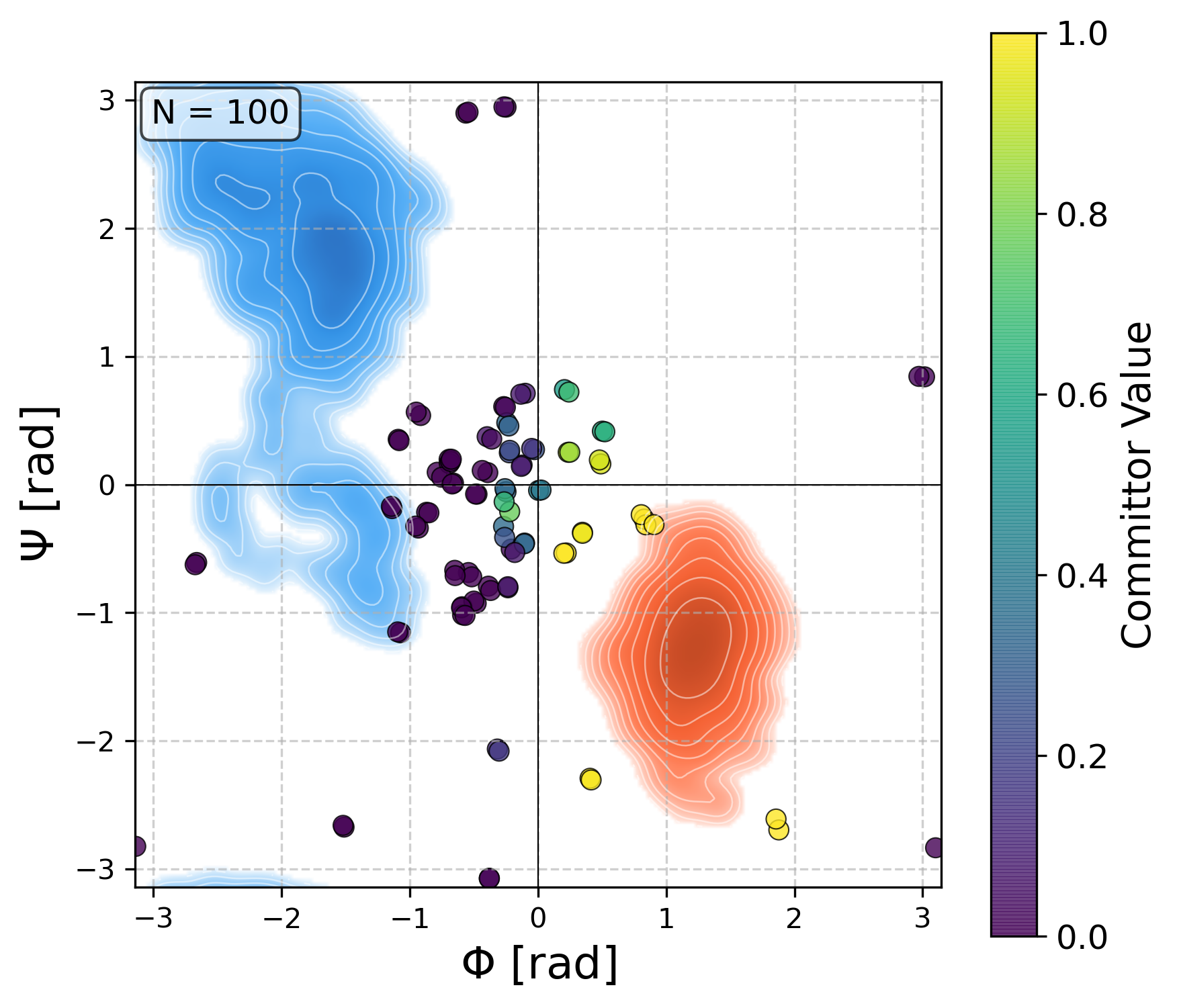}
        \caption{Guidance 2.0.}
        \label{fig:abl:sa_nosaa_g2}
    \end{subfigure}
    \hfill
    \begin{subfigure}[b]{0.48\textwidth}
        \centering
        \includegraphics[width=\textwidth]{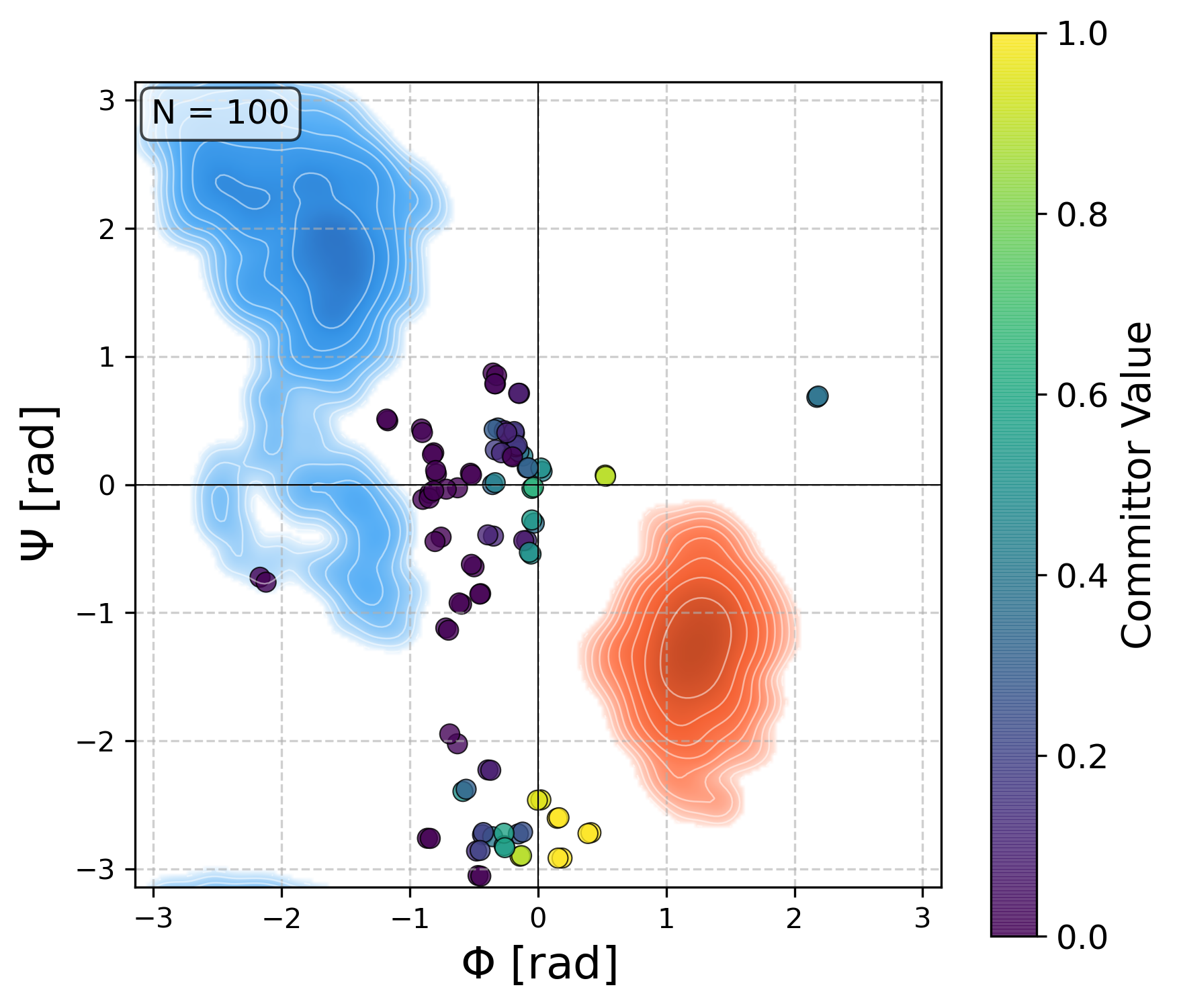}
        \caption{Guidance 3.0.}
        \label{fig:abl:sa_nosaa_g3}
    \end{subfigure}
    \caption{Effect of the guidance scale when sampling 100 structures using Simple Averaging to guide the reverse diffusion process.}
    \label{fig:abl:sa_nosaa_guid} 
\end{figure}

Figure \ref{fig:abl:and_nosaa_guid} shows the distribution of samples obtained from Isodensity Interpolation when varying the guidance scale. Figure \ref{fig:abl:sa_nosaa_guid} shows this same distribution when using Simple Averaging of scores instead of II.

\clearpage
\end{appendices}

\clearpage

\bibliography{sn-bibliography}

\end{document}